\documentclass[
  reprint,
  superscriptaddress,
  amsmath,
  amssymb,
  aip,
  jcp,
  floatfix
]{revtex4-2}

\usepackage{amsthm}
\usepackage{graphicx}
\usepackage{bm}
\usepackage[dvipsnames]{xcolor}
\usepackage{xspace}
\usepackage{multirow}
\usepackage{float}
\usepackage{algorithm}
\usepackage{algpseudocode}
\usepackage{enumitem}
\usepackage[
   colorlinks=true,
   linkcolor=NavyBlue,  % Color for internal links (Tables, Figures, Sections)
   citecolor=NavyBlue,  % Color for citations
   urlcolor=gray,       % Color for external URLs
]{hyperref}

\graphicspath{{./}}

\newcommand*{\code}[0]{\texttt{natto}\xspace}

\newcommand*{\uv}[1]{\hat{\mathbf #1}} % unit vector
\newcommand*{\avg}[1]{\left\langle #1 \right\rangle} % average over all index permutations
\newcommand*{\kt}[0]{k_{t}} % coupling-operator coefficient
\newcommand*{\Hop}[0]{\mathbf H} % harmonic operator
\newcommand*{\Kop}[0]{\mathbf K} % coupling operator

\newcommand*{\fref}[1]{Fig.~\ref{#1}}
\newcommand*{\tref}[1]{Table~\ref{#1}}
\newcommand*{\eref}[1]{Eq.~\eqref{#1}}
\newcommand*{\sref}[1]{Sec.~\ref{#1}}
\newcommand*{\aref}[1]{Appendix~\ref{#1}}
\newcommand*{\olcite}[1]{Ref.~\onlinecite{#1}}   % inline number: \cite is superscript here

\theoremstyle{definition} % For definitions (normal text)
\newtheorem{definition}{Definition}
\newtheorem{procedure}{Procedure}

\begin{document}

\title{Reusable Operators for Irreducible Cartesian Tensor Decomposition and Coupling}

\author{Mingjian Wen}
\email{mjwen@uestc.edu.cn}
\affiliation{Institute of Fundamental and Frontier Sciences, University of Electronic Science and Technology of China, Chengdu, 611731, China}
\affiliation{Frontier Center of Energy Distribution and Integration, Tianfu Jiangxi Laboratory, Chengdu, 641419, China}

\date{\today}

\begin{abstract}
   Molecular and material properties, from the polarizability to the elastic constants, are described by tensors.
   Their behavior under rotations is made explicit when a tensor is decomposed into irreducible parts that transform independently.
   In Cartesian form these parts are the symmetric and traceless irreducible Cartesian tensors (ICTs), whose decomposition and coupling underlie selection rules, orientational averages, and the use of symmetry.
   The decomposition is a textbook result at rank two, but higher rank and the intrinsic symmetry of physical tensors make it nontrivial.
   What has been lacking, unlike in the well-established spherical formalism, is a general construction for a given intrinsic symmetry, together with reusable operators that extract the ICTs and rebuild the original tensor exactly.
   Here, we develop such a construction and obtain these operators explicitly.
   These operators depend on rank and symmetry alone, and therefore each need only be built once and then applied to any tensor of that class.
   Building on them, we further obtain the Cartesian harmonics of a vector and the operators that couple two ICTs into a third, both central to equivariant machine learning.
   The construction is demonstrated on the elastic tensor, in both its second-order form of rank four and its third-order form of rank six.
   The ICTs of the rank-4 tensor also define a rotation- and scale-invariant measure of anisotropy, which we evaluate across the first-principles elastic tensors of crystalline materials from the Materials Project.
   The construction is implemented in the open-source package \code, which produces the operators in both exact symbolic and numerical form.
\end{abstract}


\maketitle

\section{Introduction}
\label{sec:intro}

Many physical quantities are tensors of rank two or higher, among them the rank-2 polarizability~\cite{buckingham1967permanent} and nuclear magnetic shielding tensor~\cite{buckingham1971asymmetry}, the rank-3 piezoelectric tensor~\cite{jerphagnon1978description}, and the rank-4 second-order elastic tensor~\cite{backus1970geometrical}.
Each decomposes into parts that transform independently under rotations, and each part can be written as a symmetric and traceless tensor, an irreducible Cartesian tensor (ICT), whose rank is its \emph{weight}~\cite{coope1965irreducible,coope1970irreducible2,coope1970irreducible3}.
For example, a rank-2 tensor decomposes as $T_{ij}=\delta_{ij}X_0+\epsilon_{ijk}X_k+X_{ij}$, where $\delta_{ij}$ is the Kronecker delta and $\epsilon_{ijk}$ the Levi-Civita symbol.
Repeated indices are summed throughout.
Here $X_0=\tfrac{1}{3}T_{kk}$ is a scalar, $X_k=\tfrac{1}{2}\epsilon_{kij}T_{ij}$ a vector, and $X_{ij}=\tfrac{1}{2}(T_{ij}+T_{ji})-\tfrac{1}{3}T_{kk}\delta_{ij}$ a symmetric traceless rank-2 tensor; these are the ICTs of weight 0, 1, and 2.
The separated parts have long played distinct roles across applications.
In spectroscopy, the weights determine which irreducible parts can interact, allowing forbidden contributions to be excluded through selection rules~\cite{harris1974theory,andrews1990symmetry}.
For molecules with random orientations, rotational averages can be reduced to scalar combinations of irreducible parts instead of evaluated orientation by orientation~\cite{morris1969averaging,bonvicini2023three,andrews1989three}.
For material property tensors, the scalar parts describe isotropic response, while the remaining parts distinguish different forms of anisotropy and help identify material symmetry~\cite{forte1996symmetry}.
The same separation is now used in equivariant machine learning, where features and predictions are organized by their weight so that rotations act correctly by construction~\cite{simeon2023tensornet,zaverkin2024higher,cheng2024cartesian,xu2025tace,xu2025cartesian3j,chen2026atomistic}.

In the spherical-tensor formalism the irreducible decomposition is long established, angular momentum theory supplying the spherical harmonics, the Wigner rotation matrices, and the Clebsch--Gordan coefficients~\cite{edmonds1996angular}.
Both formalisms carry the same irreducible content, and a Cartesian tensor can always be converted to the spherical basis, decomposed, and converted back.
We instead work directly in the Cartesian basis.
Property tensors are computed, tabulated, and consumed in Cartesian components, and their intrinsic index symmetries, such as the exchange symmetry $\alpha_{ij}=\alpha_{ji}$ of the polarizability, are statements about those components; staying in that basis avoids the conversion and its phase and normalization conventions.
A brief comparison of the two formalisms is given in supplementary Sec.~S11.

Cartesian tensor reduction has been approached from several directions, each supplying part of what a general construction needs: an independent set of operators at any rank, an orthonormal form of that set, and a way to impose a prescribed intrinsic symmetry.
One line begins with Coope and coworkers, who built the reduction from isotropic tensors, with natural projections and mapping tensors that extract the irreducible parts~\cite{coope1965irreducible,coope1970irreducible2}, and fixed Cartesian coupling tensors~\cite{coope1970irreducible3}.
Jerphagnon and coworkers carried the reduction to physical property tensors, resolving repeated weights and treating intrinsic symmetry one class at a time~\cite{jerphagnon1978description}.
In these reductions, the independent mappings are chosen by hand for each rank in turn.
Andrews and Ghoul then worked out the rank-4 decomposition in full~\cite{andrews1982irreducible}, and Lehman and Parke derived closed formulas for the Cartesian harmonics of a unit vector and for coupling two irreducible tensors~\cite{lehman1989angular}.
These formulas express the result directly in the components of the vector or tensors, and require repeated evaluation of a sum of terms, once for each input.
Bonvicini automated the selection, identifying independent operators at arbitrary rank by computation and demonstrating the method through rank five~\cite{bonvicini2024irreducible}.
Shao and coworkers obtained orthonormal decomposition matrices of the full tensor-product space, from chain contractions of Clebsch--Gordan matrices following the parentage scheme of Coope~\cite{coope1965irreducible}, by way of a change of basis to the spherical spaces~\cite{shao2025high}.
Other lines reached high rank by importing structure from elsewhere: geometric and material symmetry arguments tailored to elasticity~\cite{backus1970geometrical,cowin1992identification,forte1996symmetry}, and invariant theory giving recursive orthogonal decompositions~\cite{zheng1994theory,zheng2000irreducible,zou2001orthogonal}.

These works supply most of the ingredients of a general construction.
Independence has been settled by a matrix computation, orthonormality by invariant theory, by an existence argument or by a spherical change of basis, and intrinsic symmetry by hand one class at a time, while the harmonic and coupling operators come from derivations of their own.
What has not been done is to carry the construction through for a tensor with a prescribed intrinsic index symmetry, with the independent mappings, their orthonormal form, the symmetry-adapted combinations, and the harmonic and coupling operators all coming from a single mechanism.
Such a decomposition of the full tensor-product space might be restricted to a tensor with an intrinsic symmetry, but we are not aware that this has been done.

Here, we develop a general, linear-algebraic construction that, for any rank and prescribed intrinsic symmetry, produces a complete and independent set of reusable operators for extracting every irreducible part and reconstructing the tensor exactly.
Starting from all possible operators, an exact matrix calculation selects the independent ones.
We then retain the combinations that obey the prescribed intrinsic symmetry, and provide both an extraction--reconstruction pair and an orthonormal form.
The orthonormal form is self-dual, so one set of operators serves both directions, and the squared magnitude splits into contributions from the individual ICTs without cross terms.
The resulting operators depend only on rank, weight, and intrinsic symmetry.
They can therefore be stored as numerical arrays and reused, reducing the decomposition of any tensor in the class to contractions with fixed coefficients.
The natural projector, the mapping tensors, and the extraction and reconstruction summarized in \sref{sec:reduction:embedding} are those of Refs.~\onlinecite{coope1965irreducible,coope1970irreducible2}; the selection of the independent mappings, the orthonormal operators, the symmetry adaptation, and the implementation are developed here.

We demonstrate the construction on elastic tensors.
At rank four, it reproduces the known decomposition without using an elasticity-specific derivation~\cite{backus1970geometrical,cowin1992identification,forte1996symmetry}.
The same construction then gives the decomposition of the third-order elastic tensor at rank six, including an irreducible part of weight three that is absent at rank four.
Across first-principles elastic tensors from the Materials Project~\cite{horton2025accelerated,wen2024an}, the irreducible parts also separate the anisotropy of each material into contributions with different angular dependence.
This gives information that is not available from a single scalar measure of anisotropy.

Beyond the reduction itself, the construction provides operators for two related tasks.
Applied to repeated products of a vector, the operators produce Cartesian harmonics, which describe angular dependence in terms of ICTs.
Applied to a product of two ICTs, the same construction gives an operator that couples them into a third ICT~\cite{coope1970irreducible3,lehman1989angular}.
Both sets of operators are therefore obtained from the same decomposition rather than from separate derivations.
Such couplings are central to equivariant machine learning, and the Cartesian natural tensor networks of \olcite{chen2026atomistic} already use a condensed form of the present construction to predict properties from dipole moments to elastic constants.

\begin{figure}[tbh!]
   \centering
   \includegraphics[width=0.95\columnwidth]{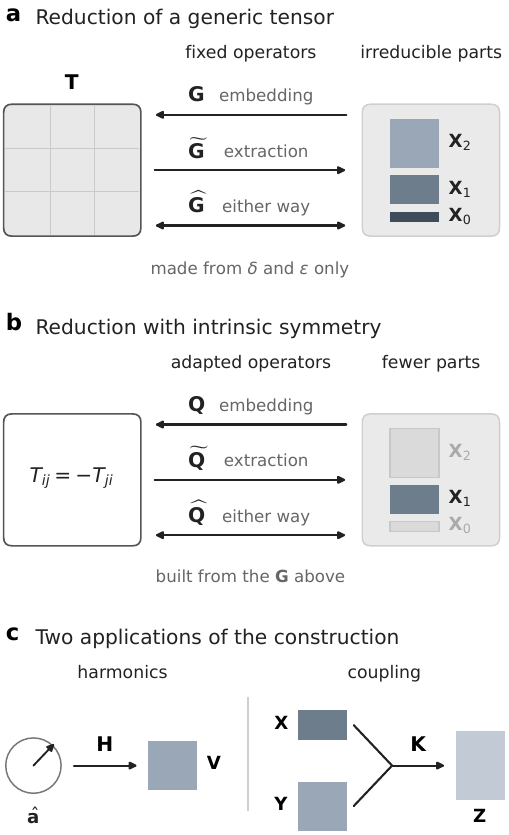}
   \caption{\textbf{One construction reduces a Cartesian tensor, adapts to its intrinsic symmetry, and yields the harmonic and coupling operators.}
   \textbf{a} Fixed operators carry a generic tensor $\mathbf T$ to its irreducible parts $\mathbf X_\ell$ and back: $\mathbf G$ embeds a part, its dual $\widetilde{\mathbf G}$ extracts one, and the orthonormal $\widehat{\mathbf G}$ does either (\sref{sec:reduction:embedding} and \sref{sec:independent:orthonormal}).
   \textbf{b} An intrinsic symmetry, here $T_{ij}=-T_{ji}$, replaces them by symmetry-adapted operators $\mathbf Q$; the weights the symmetry excludes are shown greyed (\sref{sec:arbitrary:symmetry}).
   \textbf{c} The same construction gives the harmonic operator $\Hop$, which carries a direction $\uv{a}$ to a Cartesian harmonic $\mathbf V$ (\sref{sec:nt:unit:vector}), and the coupling operator $\Kop$, which couples two ICTs into a third (\sref{sec:tp}).
   }
   \label{fig:construction:map}
\end{figure}

\fref{fig:construction:map} summarizes the construction, and the remainder of the paper is organized as follows.
\sref{sec:reduction:embedding} defines ICTs and summarizes the classical reduction and embedding operators.
\sref{sec:independent:orthonormal} counts the candidate operators, selects an independent set among them, and orthonormalizes it.
\sref{sec:arbitrary:symmetry} extends it to tensors with intrinsic symmetry, with the elastic tensor worked out in detail in \sref{sec:elastic}.
\sref{sec:nt:unit:vector} constructs the Cartesian harmonics and \sref{sec:tp} the coupling operators.
The appendices collect tables of notations and mathematical support.
The complete construction is implemented in the open-source package \code~\cite{natto}, which builds the operators in exact symbolic as well as numerical form.

\section{Foundations of Cartesian tensor reduction}
\label{sec:reduction:embedding}

This section summarizes the classical Cartesian construction for tensor reduction developed in Refs.~\cite{coope1965irreducible,coope1970irreducible2,coope1970irreducible3,jerphagnon1978description,andrews1982irreducible}.
We state the results and illustrate them by example rather than reproduce their proofs, for which the reader is referred to those references.
We work with tensors in three-dimensional Euclidean space.
A rank-$n$ Cartesian tensor $\mathbf T_n$ has components $T_{i_1i_2\dots i_n}$, where each index runs over the three Cartesian directions, giving $3^n$ components in total.
We call a tensor \emph{generic} if no relation is assumed among its components, so that all $3^n$ of them are independent; physical property tensors usually carry intrinsic symmetries that lower this count, a case treated in \sref{sec:arbitrary:symmetry}.

Throughout, boldface symbols denote tensors and indexed symbols their Cartesian components, and we simply write $\mathbf T$ when the rank is unimportant.
Einstein summation over repeated indices is implied unless otherwise stated.
Two operation symbols are used heavily: $\otimes$ denotes a tensor product, and $\odot^\ell$ a contraction over $\ell$ pairs of indices.
A compact list of all symbols and tensor operations is collected in \aref{app:notation}.

\subsection{Irreducible and isotropic Cartesian tensors}
\label{sec:natural:tensors}

\begin{definition}[Irreducible Cartesian tensor]
   A rank-$n$ tensor $\mathbf X_n$ with components $X_{i_1i_2\dots i_n}$ is an \emph{irreducible Cartesian tensor} (ICT) if it is
   (i) \emph{symmetric}, $X_{i_{\pi(1)}\dots i_{\pi(n)}} = X_{i_1\dots i_n}$ for every permutation $\pi \in S_n$ of its $n$ indices, and
   (ii) \emph{traceless}, $\delta_{i_a i_b} X_{i_1\dots i_a\dots i_b\dots i_n} = 0$ for every pair of indices $1\leq a < b \leq n$,
   where $S_n$ is the permutation group on $n$ indices and $\delta_{ij}$ is the Kronecker delta~\cite{coope1965irreducible,coope1970irreducible2,coope1970irreducible3}.
\end{definition}
Tracelessness applies for $n\geq 2$; scalars and vectors are treated as rank-0 and rank-1 ICTs.
Such tensors are also called \emph{natural} tensors, after \olcite{coope1965irreducible}, and elsewhere irreducible~\cite{coope1970irreducible2} or harmonic~\cite{forte1996symmetry} tensors.
We use \emph{irreducible Cartesian tensor} throughout.

Two properties of ICTs are central to what follows.
The first is that, in three dimensions, a rank-$n$ ICT has $2n+1$ independent components~\cite{jerphagnon1978description}, far fewer than the $3^n$ of a generic tensor; supplementary Sec.~S1 derives the count.
The second is irreducibility: under a rotation the $2n+1$ independent components of a rank-$n$ ICT transform among themselves, and no smaller set of linear combinations of them transforms only among itself~\cite{zee2016group}.
This is what makes ICTs the endpoint of the reduction: they cannot be split further into pieces that transform independently.

All the reduction and embedding operators in this paper are built from two isotropic tensors: the Kronecker delta $\bm\delta$, whose components $\delta_{ij}$ equal 1 if $i=j$ and 0 otherwise, and the Levi--Civita symbol $\bm\epsilon$, whose components $\epsilon_{ijk}$ equal $+1$ ($-1$) if $(i,j,k)$ is an even (odd) permutation of $(1,2,3)$ and 0 otherwise.
A tensor is \emph{isotropic} when its components are the same in every rotated frame, and both $\bm\delta$ and $\bm\epsilon$ are, so every operator built from them is the same in every frame.
They are moreover the elementary isotropic tensors, in that every isotropic Cartesian tensor is a linear combination of products of them, and products of two $\bm\epsilon$'s reduce to $\bm\delta$'s, so at most one $\bm\epsilon$ is ever needed.
Contracted with a pair of indices of an arbitrary tensor, $\bm\delta$ takes the trace in that pair and $\bm\epsilon$ retains only the antisymmetric part, the two rank-lowering operations on which the reduction rests.
Applied to an ICT, however, both come up empty: $\bm\delta \odot^2 \mathbf X = 0$ because an ICT is traceless and $\bm\epsilon \odot^2 \mathbf X = 0$ because it is symmetric, so neither operation lowers an ICT any further.
Supplementary Sec.~S2 collects the identities used throughout.

\subsection{Reduction spectrum}
\label{sec:reduction:spectrum}

\begin{definition}[Reduction spectrum]\label{def:reduction:spectrum}
   A generic rank-$n$ Cartesian tensor $\mathbf T_n$ decomposes into ICTs as~\cite{jerphagnon1978description,andrews1982irreducible}
   \begin{equation} \label{eq:T:as:X}
      \mathbf T_n = \sum_{\oplus\, \ell} \sum_{\oplus\, p} \mathbf X_\ell^p.
   \end{equation}
   Here, $\ell=0,1,\dots,n$ is the \emph{weight}.
   For each weight $\ell$, its \emph{multiplicity} $N_\ell$ is the number of independent ICTs of that weight.
   We call each of them a \emph{channel} of weight $\ell$, and the \emph{multiplicity index} $p=1,2,\dots,N_\ell$ labels the channels.
   The collection of ICTs $\mathbf X_\ell^p$ is the \emph{reduction spectrum} of $\mathbf T_n$.
\end{definition}

The symbol $\oplus$ signals that \eref{eq:T:as:X} is a direct sum rather than an ordinary one: ICTs of different weights have different ranks and cannot literally be added, and those of the same weight are kept as separate pieces rather than added.
Each ICT of weight $\ell$ carries $2\ell+1$ components, and the spectrum accounts for the whole rank-$n$ tensor space, $\sum_{\ell=0}^n N_\ell(2\ell+1)=3^n$.
The multiplicities $N_\ell$ are determined in \sref{sec:num:ind:nat:tensors}.

At rank two every weight occurs once, $\mathbf T_2 = \mathbf X_0 \oplus \mathbf X_1 \oplus \mathbf X_2$, and the multiplicity index is then omitted.
Multiplicity first appears at rank three, $\mathbf T_3 = \mathbf X_0 \oplus \mathbf X_1^1 \oplus \mathbf X_1^2 \oplus \mathbf X_1^3 \oplus \mathbf X_2^1 \oplus \mathbf X_2^2 \oplus \mathbf X_3$, whose seven ICTs are written out in supplementary Sec.~S6.
Both spectra can be obtained systematically through the following procedure.

\begin{procedure}\label{proc:natural:reduction}
   A weight-$\ell$ ICT $\mathbf X_\ell^p$ can be obtained from a generic rank-$n$ Cartesian tensor $\mathbf T_n$ ($\ell \leq n$) as follows~\cite{jerphagnon1978description,andrews1982irreducible}:
   \begin{enumerate}[label=(\roman*)]
      \item lower the rank of $\mathbf T_n$ to $\ell$ by contracting it with the isotropic tensors $\bm \delta$ and/or $\bm \epsilon$;
      \item symmetrize the rank-lowered tensor by averaging over all index permutations;
      \item remove traces by subtracting tensors formed from pairwise contractions.
   \end{enumerate}
   Steps (ii) and (iii) impose the two defining conditions of an ICT: symmetry and tracelessness, respectively.
\end{procedure}

The following subsections formulate the three steps as explicit operators and show how these operators perform extraction and embedding.
The rank-2 reduction serves as a running example: although simple, it contains the essential elements of the general construction.
\tref{tab:rank2:summary} summarizes its operators and results and is referenced at the relevant points below.

\begin{table*}[bth!]
   \caption{\textbf{Reduction of a generic rank-2 tensor $\mathbf T_2$, operator by operator.}
      Because each weight occurs once, $g$ denotes the single Gram coefficient at that weight.
      }
   \label{tab:rank2:summary}
   \centering
   \small
   \begin{tabular}{cccc}
      \hline
      Quantity & $\ell=0$ & $\ell=1$ & $\ell=2$ \\
      \hline
      $X_\ell$
               & $\tfrac{1}{3}T_{kk}$
               & $\tfrac{1}{2}\epsilon_{\alpha_1 i_1i_2}T_{i_1i_2}$
               & $\tfrac{1}{2}(T_{\alpha_1\alpha_2}+T_{\alpha_2\alpha_1})-\tfrac{1}{3}\delta_{\alpha_1\alpha_2}T_{kk}$                                                                  \\
      $S_n^\ell$
               & $\tfrac{1}{3}T_{kk}\delta_{i_1i_2}$
               & $\tfrac{1}{2}(T_{i_1i_2}-T_{i_2i_1})$
               & $\tfrac{1}{2}(T_{i_1i_2}+T_{i_2i_1})-\tfrac{1}{3}\delta_{i_1i_2}T_{kk}$                                                                                                \\
      \hline
      $F_{n\to \ell}$
               & $\delta_{i_1i_2}$
               & $\epsilon_{j\, i_1i_2}$
               & $1$                                                                                                                                                                    \\
      $E_{(\ell|\ell)}$
               & $1$
               & $\delta_{\alpha_1 j}$
               & $\tfrac{1}{2}(\delta_{\alpha_1i_1}\delta_{\alpha_2i_2}+\delta_{\alpha_1i_2}\delta_{\alpha_2i_1})-\tfrac{1}{3}\delta_{\alpha_1\alpha_2}\delta_{i_1i_2}$                 \\
      $G_{(\ell|n)}$
               & $\delta_{i_1i_2}$
               & $\epsilon_{\alpha_1 i_1i_2}$
               & Same as $E_{(2|2)}$                                                                                                                                                    \\
      $g$
               & $3$
               & $2$
               & $1$                                                                                                                                                                    \\
      $\widetilde G_{(\ell|n)}$
               & $\tfrac{1}{3}\delta_{i_1i_2}$
               & $\tfrac{1}{2}\epsilon_{\alpha_1 i_1i_2}$
               & Same as $E_{(2|2)}$                                                                                                                                                    \\
      \hline
   \end{tabular}
\end{table*}

\subsection{Rank-lowering tensors}
\label{sec:rank:lowering}

We take the three steps of Procedure~\ref{proc:natural:reduction} in turn, beginning with step (i).
To reach a weight-$\ell$ ICT from a generic tensor $\mathbf T_n$ ($T_{i_1i_2\dots i_n}$) of higher rank $n \geq \ell$, the rank must first be lowered from $n$ to $\ell$, which is done by contracting the surplus indices of $\mathbf T_n$ in pairs against $\bm\delta$ and, when the parity requires it, one $\bm\epsilon$.
Which indices are paired is a choice, and different choices leave different information behind.
Collecting the $\bm\delta$'s and $\bm\epsilon$ of one choice into a single \emph{rank-lowering tensor} $\mathbf F^p_{n\to \ell}$, labeled by $p$, step (i) reads~\cite{andrews1982irreducible}
\begin{equation} \label{eq:rank:lowering}
   \mathbf T'_\ell = \mathbf F^p_{n\to \ell} \odot^{m} \mathbf T_n,
\end{equation}
with $m$ the number of contracted indices.
Each $p$ gives its own rank-$\ell$ tensor $\mathbf T'_\ell$, and it is this freedom of choice that lets a single rank-$n$ tensor carry several ICTs of the same weight.
The form of $\mathbf F^p_{n\to \ell}$ depends on the parity of $n-\ell$.

When $n-\ell$ is even, the pairing is carried out by $\bm\delta$'s alone,
\begin{equation} \label{eq:F:even}
   F^p_{n\to \ell}
   = \prod_{(a,b)\,\in\,\mathcal{D}_p} \delta_{i_a i_b},
\end{equation}
where $\mathcal{D}_p$ is a set of $k = (n-\ell)/2$ disjoint pairs of indices, one pair per $\bm\delta$, so that $\delta_{i_a i_b}$ contracts with the pair $i_a, i_b$ in $\mathbf T_{i_1 \dots i_a \dots i_b \dots i_n}$.
All $m = 2k = n-\ell$ paired indices are contracted away, leaving a tensor of rank $\ell$.
Here the label $p$ is simply the label of $\mathcal{D}_p$.

As an example, for $n=3$ and $\ell=1$, $\mathbf F^p_{3\to 1}$ is a rank-2 tensor made of a single $\bm\delta$.
Then, $\mathcal{D}_1$ = $\{(1,2)\}$, $\mathcal{D}_2$ = $\{(1,3)\}$, and $\mathcal{D}_3$ = $\{(2,3)\}$, resulting in three different rank-lowering tensors,
\begin{equation} \label{eq:F:1:3}
   F^1_{3\to 1} = \delta_{i_1 i_2},
   \quad
   F^2_{3\to 1} = \delta_{i_1 i_3},
   \quad
   F^3_{3\to 1} = \delta_{i_2 i_3},
\end{equation}
one for each index of $\mathbf T_3$ left free.

When $n-\ell$ is odd, one pair is taken by $\bm\epsilon$ instead,
\begin{equation} \label{eq:F:odd}
   F^p_{n\to \ell}
   = \epsilon_{j\, i_u i_v}
   \prod_{(a,b)\,\in\,\mathcal{D}_p} \delta_{i_a i_b},
\end{equation}
with $k = (n-\ell-1)/2$ of $\bm\delta$'s left.
All $m = 2(k+1) = n-\ell+1$ paired indices are contracted away, and $\bm\epsilon$ restores one through its free index $j$, again leaving a tensor of rank $\ell$.
Here the label $p$ records two things: which pair $\bm\epsilon$ takes, and how the remaining indices are paired in $\mathcal{D}_p$.

As an example, for $n=3$ and $\ell=2$ the rank drop is one, so $k=0$ and $\mathbf F^p_{3\to 2}$ is a single $\bm\epsilon$, again one for each index of $\mathbf T_3$ left free,
\begin{equation} \label{eq:F:2:3}
   F^1_{3\to 2} = \epsilon_{j\, i_1 i_2},
   \quad
   F^2_{3\to 2} = \epsilon_{j\, i_1 i_3},
   \quad
   F^3_{3\to 2} = \epsilon_{j\, i_2 i_3}.
\end{equation}

Note that $\ell=0$ is a special case, in which $\bm\epsilon$ has no free index left over and contracts all three of its indices with $\mathbf T_n$.
Across both cases, the distinct assignments for a given rank $n$ and weight $\ell$ are indexed by $p=1,2,\dots,N_\ell^{\mathrm c}$, so $N_\ell^{\mathrm c}$ counts the distinct rank-lowering tensors.
See \aref{app:F:count} for a closed-form count of the number of distinct assignments $N_\ell^{\mathrm c}$ and more examples.

\subsection{Natural projector}
\label{sec:nat:proj}

The rank-lowered tensor $\mathbf T'_\ell$ of \eref{eq:rank:lowering} has the right rank, but is in general neither symmetric nor traceless.
Steps (ii) and (iii) of Procedure~\ref{proc:natural:reduction} remain, and are carried out together here by a single operator, fixed by the weight $\ell$ alone.
\begin{definition}[Natural projector]\label{def:natural:projector}
   The \emph{natural projector} $\mathbf E_{(\ell|\ell)}$ is the rank-$2\ell$ tensor~\cite{coope1970irreducible2}
   \begin{equation} \label{eq:natural:projector}
      E_{(\ell|\ell)}
      = \sum_{t=0}^{\lfloor \ell/2 \rfloor}
      c_t
      \avg{\delta_{\alpha i}^{\ell-2t}\delta_{\alpha\alpha}^{t}\delta_{ii}^{t}} ,
   \end{equation}
   with no sum over the repeated indices, the upper limit being $\lfloor \ell/2 \rfloor = \ell/2$ when $\ell$ is even and $(\ell-1)/2$ when $\ell$ is odd.
   Written in full, it is $\mathbf E_{(\alpha_1\alpha_2\dots\alpha_\ell|i_1i_2\dots i_\ell)}$, where the Greek indices belong to the ICT space $\mathcal{X}^\ell$ and the Roman indices belong to the Cartesian tensor space $\mathcal{T}^\ell$.
   The coefficients $c_t$ depend only on $\ell$ and $t$, and are fixed by requiring that $\mathbf E_{(\ell|\ell)}$ return a traceless tensor~\cite{coope1970irreducible2,coope1970irreducible3},
   \begin{equation} \label{eq:ct}
      c_t = (-1)^t\,\frac{(\ell!)^2\,(2\ell-2t)!}{(2\ell)!\;t!\,(\ell-t)!\,(\ell-2t)!}.
   \end{equation}
\end{definition}
The shorthand $\avg{\delta_{\alpha i}^{\ell-2t}\delta_{\alpha\alpha}^{t}\delta_{ii}^{t}}$ denotes the average of a product of $\ell$ Kronecker deltas, the exponents counting the factors of each type.
The average is taken over all $\ell!$ permutations of the $\alpha$ indices and, separately, all $\ell!$ permutations of the $i$ indices.
Many of those $(\ell!)^2$ products coincide, and each distinct product occurs the same number of times, so the average may equally be taken over the distinct products alone.
For example, when $\ell=2$, the two symmetrized products appearing in \eref{eq:natural:projector}
are
\begin{equation}
   \begin{aligned}
       & \avg{\delta_{\alpha i}^2}                 = \tfrac{1}{2}(\delta_{\alpha_1i_1}\delta_{\alpha_2i_2}+\delta_{\alpha_1i_2}\delta_{\alpha_2i_1}), \\
       & \avg{\delta_{\alpha\alpha}\delta_{ii}}  = \delta_{\alpha_1\alpha_2}\delta_{i_1i_2}.
   \end{aligned}
\end{equation}
See supplementary Sec.~S3 for more examples of the symmetrized products, and supplementary Sec.~S4 for explicit expressions for $\mathbf E_{(\ell|\ell)}$ for $\ell\leq3$.

The terms of \eref{eq:natural:projector} fall into two groups, which give the projection its two functions.
The $t=0$ term, $\avg{\delta_{\alpha i}^{\ell}}$, fully symmetrizes the tensor, as required in step (ii) of Procedure~\ref{proc:natural:reduction}, whereas the $t \geq 1$ terms remove all traces, as required in step (iii).
The $t \geq 1$ terms do so by contracting $t$ pairs of the $i$ indices, those of the tensor being projected, with $\delta_{ii}^{t}$, extracting traces, and re-expanding them with $\delta_{\alpha\alpha}^{t}$, while the remaining $\delta_{\alpha i}^{\ell-2t}$ carries the uncontracted indices over from the $i$ block to the $\alpha$ block.

The natural projector has several properties that the rest of the paper rests on.
First, \eref{eq:natural:projector} is built from $\bm\delta$'s alone, so $\mathbf E_{(\ell|\ell)}$ is isotropic (\sref{sec:natural:tensors}).

Second, by construction in \eref{eq:natural:projector} it is symmetric under any permutation of its $\alpha$ indices and, separately, of its $i$ indices, and it is unchanged when the two blocks are exchanged, so it is self-adjoint.

Third, it is traceless within each block,
\begin{equation} \label{eq:E:traceless}
   \delta_{\alpha_1\alpha_2}E_{(\alpha_1\alpha_2\dots\alpha_\ell|i_1\dots i_\ell)}
   = \delta_{i_1i_2}E_{(\alpha_1\dots\alpha_\ell|i_1i_2\dots i_\ell)} = 0,
\end{equation}
for $\ell\geq2$, because the projection must return a traceless tensor for every rank-$\ell$ tensor, the requirement that fixes the $c_t$ in \eref{eq:ct}.

Fourth, it is \emph{idempotent} under the $\ell$-fold contraction~\cite{coope1970irreducible2},
\begin{equation} \label{eq:E:idempotent}
   E_{(\alpha_1\dots\alpha_\ell|k_1\dots k_\ell)}\,E_{(k_1\dots k_\ell|i_1\dots i_\ell)}
   = E_{(\alpha_1\dots\alpha_\ell|i_1\dots i_\ell)},
\end{equation}
so that applying it twice is the same as applying it once: a tensor that is already an ICT of weight $\ell$ is its own weight-$\ell$ part, and the projection returns it unchanged.
Its trace therefore counts the components it keeps~\cite{coope1970irreducible2}, $\operatorname{tr}\mathbf E_{(\ell|\ell)} = E_{(i_1\dots i_\ell|i_1\dots i_\ell)} = 2\ell+1$, the number of independent components of an ICT of weight $\ell$.
This trace pairs each Greek index with a Roman one, and so is a different contraction from those of \eref{eq:E:traceless}, which are taken within a block and vanish.

Finally, acting on a generic rank-$\ell$ tensor $\mathbf T_\ell$ ($T_{i_1i_2\dots i_\ell}$), the natural projector extracts the ICT $\mathbf X_\ell$ of weight $\ell$,
\begin{equation}\label{eq:E:project}
   \mathbf X_\ell = \mathbf E_{(\ell|\ell)} \odot^\ell \mathbf T_\ell,
   \;
   X_{\alpha_1\dots\alpha_\ell} = E_{(\alpha_1\dots\alpha_\ell|i_1\dots i_\ell)}\,T_{i_1\dots i_\ell}.
\end{equation}
By the second property, this contraction is independent of how the $i$-index slots of the projector are paired with the indices of $\mathbf T_\ell$.

Evaluating \eref{eq:natural:projector} at $\ell=0,1,2$ gives the three natural projectors of the rank-2 reduction, listed in \tref{tab:rank2:summary}.
In particular, the $\ell=2$ result follows by substituting $c_0 = 1$ and $c_1 = -\tfrac{1}{3}$ from \eref{eq:ct}.
At $\ell=2$ there is no rank to lower, so \eref{eq:E:project} applies directly and returns the symmetric traceless $\mathbf X_2$ of \tref{tab:rank2:summary}.
At $\ell=0$ and $\ell=1$ the projector must be preceded by the rank-lowering tensors listed there, and the two are combined in the next subsection.

\subsection{Mapping tensors between ranks}
\label{sec:map:diff:ranks}

Applying $\mathbf F^p_{n\to \ell}$ and then $\mathbf E_{(\ell|\ell)}$ carries out all three steps, so the two can be combined into a single operator $\mathbf G^p_{(\ell|n)}$, the \emph{mapping tensor} between the spaces $\mathcal{X}^\ell$ and $\mathcal{T}^{n}$,
\begin{equation} \label{eq:G}
   \mathbf G^p_{(\ell|n)}
   = \mathbf E_{(\ell|\ell)} \, \mathbf F^p_{n\to \ell} .
\end{equation}
See \aref{app:G:index} for the index form of \eref{eq:G}, written out for even and odd $n-\ell$.
The resulting $\mathbf G^p_{(\ell|n)}$ is a rank-$(\ell+n)$ tensor with $\ell$ Greek indices associated with $\mathcal{X}^{\ell}$ and $n$ Roman indices associated with $\mathcal{T}^{n}$.
Contracting it with a generic rank-$n$ tensor $\mathbf T_n$ gives an ICT $\mathbf D_\ell^p=\mathbf G^p_{(\ell|n)}\odot^n\mathbf T_n$ of weight $\ell$, with components $D^p_{\alpha_1\dots\alpha_\ell}=G^p_{(\alpha_1\dots\alpha_\ell|i_1\dots i_n)}T_{i_1\dots i_n}$.

For a rank-2 tensor, combining the $F_{n\to \ell}$ and $E_{(\ell|\ell)}$ rows of \tref{tab:rank2:summary} gives the $G_{(\ell|n)}$ row listed there, completing the weight-0 and weight-1 cases left open in \sref{sec:nat:proj}.
Each weight has a single mapping tensor at this rank.
Several appear per weight from rank three on.
For example, for $n=3$ and $\ell=2$ the three rank-lowering tensors of \eref{eq:F:2:3} give three candidate mapping tensors,
\begin{equation} \label{eq:G:2:3}
   \begin{aligned}
      G^{1}_{(2|3)} & = E_{(\alpha_1\alpha_2|i_3j)}\,\epsilon_{j i_1i_2}, \\
      G^{2}_{(2|3)} & = E_{(\alpha_1\alpha_2|i_2j)}\,\epsilon_{j i_1i_3}, \\
      G^{3}_{(2|3)} & = E_{(\alpha_1\alpha_2|i_1j)}\,\epsilon_{j i_2i_3}.
   \end{aligned}
\end{equation}
Each $\bm\epsilon$ passes its free index $j$ to the projector, which also receives the one index of $\mathbf T_3$ that $\bm\epsilon$ does not take.
See \aref{app:G:index} for \eref{eq:G:2:3} expanded in $\bm\delta$ and $\bm\epsilon$, and for further examples.

\subsection{Extraction and embedding}
\label{sec:extract:reconstruct}

A complete reduction must both extract the ICTs and reconstruct $\mathbf T_n$ from them.
Here, we complete the dual pair of mapping tensors: $\mathbf G^p_{(\ell|n)}$ for embedding and $\widetilde{\mathbf G}^p_{(\ell|n)}$ for extraction.

The construction begins with the pairwise contractions of mapping tensors of the same weight.
Contracting $\mathbf G^p_{(\ell|n)}$ and $\mathbf G^q_{(\ell|n)}$ over all $n$ Roman indices leaves a rank-$2\ell$ tensor, carrying one group of $\ell$ ICT indices from each factor.
Being built from $\bm\delta$ and $\bm\epsilon$ alone, the mapping tensors are isotropic (\sref{sec:natural:tensors}), and so is their contraction.
In \eref{eq:G} the rank-lowering tensor supplies Roman indices only, so the $\ell$ Greek indices of $\mathbf G^p_{(\ell|n)}$ are the $\alpha$ indices of $\mathbf E_{(\ell|\ell)}$, which are symmetric among themselves and traceless by \eref{eq:E:traceless}.
The contraction between $\mathbf G^p_{(\ell|n)}$ and $\mathbf G^q_{(\ell|n)}$ runs over Roman indices alone and so leaves both Greek groups as they are, making the result symmetric and traceless within each of them.
Up to a scalar, the only isotropic tensor with these properties is the natural projector~\cite{coope1970irreducible2}:
\begin{equation} \label{eq:gpq}
   \mathbf G^p_{(\ell|n)} \odot^n \mathbf G^{q}_{(\ell|n)} = g_{pq} \mathbf E_{(\ell|\ell)},
\end{equation}
where the scalar $g_{pq}$ measures the overlap between $\mathbf G^p_{(\ell|n)}$ and $\mathbf G^q_{(\ell|n)}$.
The values of $g_{pq}$ form a symmetric Gram matrix $\mathbf g$, which is invertible for a linearly independent set of mapping tensors~\cite{horn2013matrix}.

The \emph{dual mapping tensors} $\widetilde{\mathbf G}^p_{(\ell|n)}$ are defined, as in \olcite{coope1970irreducible2}, by
\begin{equation} \label{eq:H}
   \widetilde{\mathbf G}^p_{(\ell|n)} = \sum_{q} (\mathbf g^{-1})_{pq} \mathbf G^q_{(\ell|n)}.
\end{equation}
Each $\widetilde{\mathbf G}^p_{(\ell|n)}$ has the same rank and index structure as $\mathbf G^p_{(\ell|n)}$.
Combining Eqs.~\eqref{eq:gpq} and \eqref{eq:H} gives the duality relation (see \aref{app:gpq} for the derivation):
\begin{equation} \label{eq:H:G:dual}
   \widetilde{\mathbf G}^p_{(\ell|n)} \odot^n \mathbf G^{q}_{(\ell|n)}
   = \delta_{pq}\mathbf E_{(\ell|\ell)}.
\end{equation}

The two sets now have complementary roles.
Acting on $\mathbf T_n$, $\widetilde{\mathbf G}^p_{(\ell|n)}$ extracts the ICT $\mathbf X_\ell^p$ of weight $\ell$ and multiplicity index $p$:
\begin{equation} \label{eq:extract}
   \mathbf X_\ell^p = \widetilde{\mathbf G}^p_{(\ell|n)} \odot^n \mathbf T_n .
\end{equation}
Conversely, $\mathbf G^p_{(\ell|n)}$ embeds $\mathbf X_\ell^p$ into $\mathcal{T}^n$:
\begin{equation} \label{eq:embed}
   \mathbf S^{\ell,p}_n = \mathbf G^p_{(\ell|n)} \odot^\ell \mathbf X_\ell^p .
\end{equation}

The embedded tensors reconstruct the original tensor through the ordinary sum
\begin{equation} \label{eq:T:as:S}
   \mathbf T_n = \sum_{\ell=0}^{n}\sum_{p=1}^{N_\ell}\mathbf S_n^{\ell,p}.
\end{equation}
Eqs.~\eqref{eq:T:as:X} and \eqref{eq:T:as:S} are the same decomposition in two forms: a direct sum of the ICTs $\mathbf X_\ell^p$, each of rank $\ell$, or an ordinary sum of their embeddings $\mathbf S_n^{\ell,p}$, each of rank $n$.
Although $\mathbf S_n^{\ell,p}$ has $3^n$ components, only $2\ell+1$ are independent, since it carries no information beyond $\mathbf X_\ell^p$.
For example, the three embedded tensors for a rank-2 tensor are provided in \tref{tab:rank2:summary}.

\section{Independent and orthonormal mapping tensors}
\label{sec:independent:orthonormal}

The reduction and embedding operators of \sref{sec:reduction:embedding} generate candidate mapping tensors $\mathbf G^p_{(\ell|n)}$.
At rank two nothing further is needed, as \tref{tab:rank2:summary} shows: each weight occurs once, its single candidate is the mapping tensor, and the familiar isotropic, antisymmetric, and symmetric traceless parts follow at once.
From rank three on a weight can occur more than once, and \eref{eq:G} may then generate more candidates than there are independent ones (\tref{tab:reduction:spectrum}): of the three weight-2 candidates in \eref{eq:G:2:3}, only two are needed.
This is the obstacle, and it surfaces in the Gram matrix $\mathbf g$ of \eref{eq:gpq}, which a linearly dependent set makes singular, so that the inverse-Gram construction of \eref{eq:H} cannot be applied to the full candidate set.
An independent subset must therefore be selected first.

The selected subset is independent but generally not orthonormal, $\mathbf g \neq \mathbf I$.
Nothing breaks: the reduction goes through the dual pair $\mathbf G^p_{(\ell|n)}$ and $\widetilde{\mathbf G}^p_{(\ell|n)}$, at the cost of building and storing two operator sets, one for extraction and one for embedding.
An orthonormal set is the more convenient basis, as an orthonormal basis usually is: it is self-dual, so a single set serves in both directions, and it leaves the ICTs on an equal footing, none scaled differently from another by the choice of mappings.

We address both in this section.
We first count the candidates and determine the multiplicity of each weight, then select an independent subset by an exact QR factorization, and finally orthonormalize the selected subset using the Gram matrix.
Everything stays within the operator formalism of \sref{sec:reduction:embedding}: the independent subset and the orthonormal set are again constant operators, precomputable for a given rank and weight and reusable for every tensor of that class.

\subsection{Number of independent mapping tensors}
\label{sec:num:ind:nat:tensors}

In \eref{eq:G} the natural projector is fixed by the weight alone, so at a given rank $n$ and weight $\ell$ the candidates differ only through the rank-lowering tensor.
Each $\mathbf F^p_{n\to \ell}$ yields one $\mathbf G^p_{(\ell|n)}$, so there are $N_\ell^{\mathrm c}$ candidates, indexed by $p=1,2,\dots,N_\ell^{\mathrm c}$.
A closed-form expression for $N_\ell^{\mathrm c}$ is derived in \aref{app:F:count}, and the values through rank six are listed in \tref{tab:reduction:spectrum}.

The number of these candidates that are linearly independent is the multiplicity $N_\ell$ of that weight in the reduction spectrum of a rank-$n$ tensor.
Following~\olcite{jerphagnon1978description}, $N_\ell$ is determined recursively using the angular-momentum addition rule: ICTs of weights $\ell_1$ and $\ell_2$ combine into ICTs of every weight $\ell$ satisfying $|\ell_1-\ell_2|\leq \ell\leq \ell_1+\ell_2$.
Since each Cartesian index has weight one, the weight content of the rank-2 reduction spectrum is $1\otimes1=0\oplus1\oplus2$.
That of rank three is $1\otimes1\otimes1=1\otimes(0\oplus1\oplus2)=0\oplus1_3\oplus2_2\oplus3$, where the subscripts denote multiplicities.
Repeating this tensor-product decomposition gives the multiplicities at higher ranks.
The resulting multiplicities are listed alongside $N_\ell^{\mathrm c}$ in \tref{tab:reduction:spectrum}.

For ranks $n\leq2$, each weight has multiplicity $N_\ell=1$.
Moreover, $N_\ell=N_\ell^{\mathrm c}$ at every weight, so every candidate mapping tensor is independent.
At higher ranks, however, $N_\ell>1$ in general, while $N_\ell^{\mathrm c}>N_\ell$ for some weights.
At the top weight $\ell=n$, $N_n=N_n^{\mathrm c}=1$ for every rank, because no rank lowering is needed there and \eref{eq:G} reduces to $\mathbf G_{(n|n)}=\mathbf E_{(n|n)}$.
A generic tensor of any rank $n$ therefore has exactly one ICT of weight $n$.

\begin{table}[bth!]
   \caption{\textbf{Numbers $N_\ell$ of independent mapping tensors by rank $n$ and weight $\ell$.}
      Parentheses give the number $N_\ell^{\mathrm c}$ of candidate mapping tensors when $N_\ell^{\mathrm c}\neq N_\ell$.}
   \label{tab:reduction:spectrum}
   \centering
   \begin{tabular}{cccccccc}
      \hline
                 & \multicolumn{7}{c}{Rank $n$}                                             \\
      \cline{2-8}
      Weight $\ell$ & 0 & 1 & 2 & 3     & 4     & 5       & 6 \\
      \hline
      0             & 1 &   & 1 & 1     & 3     & 6 (10)  & 15 \\
      1             &   & 1 & 1 & 3     & 6     & 15      & 36 (45) \\
      2             &   &   & 1 & 2 (3) & 6     & 15 (30) & 40 (45) \\
      3             &   &   &   & 1     & 3 (6) & 10      & 29 (90) \\
      4             &   &   &   &       & 1     & 4 (10)  & 15 \\
      5             &   &   &   &       &       & 1       & 5 (15) \\
      6             &   &   &   &       &       &         & 1 \\
      \hline
   \end{tabular}
\end{table}

\subsection{Selection of an independent set}
\label{sec:alg:QR}

For a given rank $n$ and weight $\ell$, we now determine the linear dependencies among the $N_\ell^{\mathrm c}$ candidate mapping tensors $\mathbf G^p_{(\ell|n)}$ and identify a subset of $N_\ell$ linearly independent ones.
A linear dependence among them is specified by a nonzero coefficient vector $\mathbf w=(w_1,w_2,\dots,w_{N_\ell^{\mathrm c}})^{\mathsf T}$ satisfying $\sum_{p=1}^{N_\ell^{\mathrm c}}w_p\mathbf G^p_{(\ell|n)}=\mathbf 0$.
Reshaping each $\mathbf G^p_{(\ell|n)}$ into a vector $\mathbf v^p$ of length $d=3^{\ell+n}$ is linear and one-to-one, so the relation becomes $\mathbf A\mathbf w=\mathbf 0$ once the $\mathbf v^p$ are stacked as the columns of a matrix $\mathbf A$.
The problem is therefore to determine the column rank of $\mathbf A$ and select $N_\ell$ independent columns.
A QR factorization of $\mathbf A$ provides this selection~\cite{golub1996matrix}, as set out in Algorithm~\ref{alg:QR}.
The entries of $\mathbf A$ are rational, since the candidates are built from $\bm\delta$ and $\bm\epsilon$ alone, and we factorize $\mathbf A=\tilde{\mathbf Q}\mathbf U$ in rational arithmetic, where the columns $\tilde{\mathbf q}_k$ of $\tilde{\mathbf Q}$ are mutually orthogonal and not normalized, and $\mathbf U$ is unit upper triangular.
A column of $\mathbf A$ that is a linear combination of the earlier ones gives $\tilde{\mathbf q}_k=\mathbf 0$, so the independent columns are those with $\lVert\tilde{\mathbf q}_k\rVert^2\neq0$.
\begin{algorithm}[H]
   \caption{Identification of linearly independent mapping tensors}
   \label{alg:QR}
   \begin{algorithmic}[1]
      \State \textbf{Input:} Candidate mapping tensors $\{\mathbf G^p_{(\ell|n)}\}_{p=1}^{N_\ell^{\mathrm c}}$
      \State \textbf{Output:} Index set $\mathcal I$ of linearly independent mapping tensors

      \For{$p = 1$ to $N_\ell^{\mathrm c}$}
      \State $\mathbf v^p \gets \operatorname{reshape}(\mathbf G^p_{(\ell|n)})$ \Comment{Length $d=3^{\ell+n}$}
      \EndFor

      \State $\mathbf A \gets [\mathbf v^1,\mathbf v^2,\dots,\mathbf v^{N_\ell^{\mathrm c}}]$ \Comment{$d\times N_\ell^{\mathrm c}$ matrix}
      \State $\tilde{\mathbf Q},\mathbf U \gets \operatorname{exactQR}(\mathbf A)$ \Comment{Rational; columns of $\tilde{\mathbf Q}$ orthogonal}
      \State $\mathcal I \gets \emptyset$

      \For{$k = 1$ to $N_\ell^{\mathrm c}$}
      \If{$\lVert\tilde{\mathbf q}_k\rVert^2\neq 0$} \Comment{Exact, in rational arithmetic}
      \State $\mathcal I \gets \mathcal I \cup \{k\}$
      \EndIf
      \EndFor

      \State \textbf{return} $\mathcal I$

   \end{algorithmic}
\end{algorithm}

From here on $\mathbf G^p_{(\ell|n)}$ denotes the selected independent set, with $p=1,2,\dots,N_\ell$.
We note that the set is not unique; any basis of the weight-$\ell$ mapping space is valid.
The three candidates of \eref{eq:G:2:3}, for instance, span a two-dimensional space, and any two of them (or two independent linear combinations) may serve.
The worked examples and \code fix one choice by the conventions of supplementary Sec.~S5, which also explains why some choice cannot be avoided.

\subsection{Orthonormal mapping tensors}
\label{sec:orthonormal:mapping}

An orthonormal choice of the mappings always exists, as noted in \olcite{coope1970irreducible2}, and we now construct it explicitly.
The mapping tensors selected in \sref{sec:alg:QR} are independent, so their Gram matrix $\mathbf g$ of \eref{eq:gpq} is positive definite as well as symmetric, and therefore has a unique symmetric positive-definite inverse square root~\cite{horn2013matrix}, which we denote by $\mathbf g^{-1/2}$.
It can be obtained by diagonalizing $\mathbf g$ and taking the inverse square roots of its eigenvalues, as detailed in \aref{app:gpq}.
The \emph{orthonormal mapping tensors} are then defined by
\begin{equation} \label{eq:G:orthonormal}
   \widehat{\mathbf G}^{p}_{(\ell|n)}
   = \sum_{q=1}^{N_\ell}(\mathbf g^{-1/2})_{pq}\mathbf G^q_{(\ell|n)}.
\end{equation}
The orthonormal set is not unique; the symmetric inverse square root selects one basis within each repeated weight.
The $\widehat{\mathbf G}^{p}_{(\ell|n)}$ satisfy the relation (derived in \aref{app:gpq})
\begin{equation} \label{eq:G:orthonormal:property}
   \widehat{\mathbf G}^{p}_{(\ell|n)}\odot^n
   \widehat{\mathbf G}^{q}_{(\ell|n)}
   = \delta_{pq}\mathbf E_{(\ell|\ell)}.
\end{equation}
They are therefore orthonormal and self-dual, so the same $\widehat{\mathbf G}^{p}_{(\ell|n)}$ performs both extraction, replacing $\widetilde{\mathbf G}^p_{(\ell|n)}$ in \eref{eq:extract}, and embedding, replacing $\mathbf G^p_{(\ell|n)}$ in \eref{eq:embed}:
\begin{equation} \label{eq:G:orthonormal:action}
   \mathbf X_\ell^p
   = \widehat{\mathbf G}^{p}_{(\ell|n)}\odot^n\mathbf T_n,
   \quad
   \mathbf S_n^{\ell,p}
   = \widehat{\mathbf G}^{p}_{(\ell|n)}\odot^\ell\mathbf X_\ell^p.
\end{equation}
Substituting \eref{eq:G:orthonormal:action} into \eref{eq:T:as:S} reconstructs $\mathbf T_n$ using only the orthonormal mapping tensors,
\begin{equation} \label{eq:G:orthonormal:complete}
   \mathbf T_n
   = \sum_{\ell=0}^{n}\sum_{p=1}^{N_\ell}
   \widehat{\mathbf G}^{p}_{(\ell|n)}\odot^\ell
   \left(\widehat{\mathbf G}^{p}_{(\ell|n)}\odot^n\mathbf T_n\right).
\end{equation}

The reduction can therefore be carried out in either of two ways: with the dual pair $\mathbf G^p_{(\ell|n)}$ and $\widetilde{\mathbf G}^p_{(\ell|n)}$ of \sref{sec:extract:reconstruct}, or with the single self-dual $\widehat{\mathbf G}^p_{(\ell|n)}$.
Both span the same weight-$\ell$ mapping space, so both recover the same total weight-$\ell$ content $\sum_p \mathbf S^{\ell,p}_n$ of $\mathbf T_n$; the individual $\mathbf X^p_\ell$ are not the same, the two differing in how the $N_\ell$ channels within a repeated weight are apportioned.
The dual pair keeps every entry rational, as $\mathbf g^{-1}$ is rational whenever $\mathbf g$ is.
The orthonormal set needs only a single operator, and by \eref{eq:G:orthonormal:property} it makes the reduction an isometry: $\lVert\mathbf T_n\rVert^2=\sum_{\ell,p}\lVert\mathbf X_\ell^p\rVert^2$, with no cross terms between channels of the same weight.
Its coefficients are in general irrational, however, since $\mathbf g^{-1/2}$ is.
Computing the mapping tensors is fast: with \code, all orthonormal mapping tensors of a rank-$n$ tensor take 10~ms at rank four and 1.5~s at rank six on one CPU core (Apple M4 Pro), and they are built once for all tensors of that rank.
The timings, and those with an intrinsic symmetry, are given in supplementary Sec.~S12.

\section{Reduction of tensors with intrinsic symmetry}
\label{sec:arbitrary:symmetry}

Having established how to select and orthonormalize mappings for generic tensors, we now extend the construction to tensors with intrinsic symmetry, sketched in \fref{fig:construction:map}b.
Physical property tensors almost always carry an intrinsic symmetry: certain permutations of their indices leave the components unchanged, or reverse their sign.
\tref{tab:reduction:physical:tensors} lists common examples with their symmetries and reduction spectra~\cite{jerphagnon1978description}.
The symmetry-free construction of \sref{sec:independent:orthonormal} cannot be applied directly, because a symmetry-constrained tensor no longer distinguishes all of its ICTs.
What survives at each weight is in general not a subset of the candidate mappings but a subspace of their span: those linear combinations that transform under the prescribed index permutations with the prescribed sign.
The multiplicity therefore drops from $N_\ell$ to the dimension of that subspace, which may be anything from $N_\ell$ down to zero.

To make this concrete, let's look at some examples.
A weight may lose all of its channels: in three dimensions an antisymmetric index pair transforms like a single vector index under rotations, so the rank-2 rotation-rate tensor, $W_{ij}=-W_{ji}$, keeps only its weight-1 channel, the weight-0 and weight-2 multiplicities of a general rank-2 tensor both falling to zero~\cite{botsis2018mechanics}.
The rank-3 piezoelectric tensor, symmetric in its last two indices, $d_{ijk}=d_{ikj}$, loses its weight-0 channel the same way, the single weight-0 channel of a general rank-3 tensor not surviving the symmetry.
A weight can also lose only some of its channels: a general rank-4 tensor has three weight-0 channels, $T_{iijj}$, $T_{ijij}$, and $T_{ijji}$, and for an elastic tensor the minor symmetry $C_{ijkl}=C_{ijlk}$ makes the last two coincide, so the weight-0 multiplicity drops from three to two.
The rest of this section builds a systematic construction for an arbitrary intrinsic symmetry, one that covers them all: the rotation-rate and piezoelectric tensors are worked through in supplementary Sec.~S7, and the elastic tensor in \sref{sec:elastic}.

\begin{table*}[tbh!]
   \caption{\textbf{Reduction spectrum of some physical tensors.}
   Intrinsic symmetries are denoted using parentheses and square brackets: indices inside parentheses are symmetric, whereas indices inside square brackets are antisymmetric, so that $(ij)$ represents $T_{ij}=T_{ji}$ and $[ij]$ represents $T_{ij}=-T_{ji}$, and nested groups are exchanged as a unit, as in $((ij)(kl))$ for the elastic tensor.
   $N_{\mathrm{ind}}$ is the number of independent components, and the entries under weight $\ell$ are the multiplicities: $N_\ell$ for the unconstrained classes, and the symmetry-restricted $N_\ell^{\mathcal S}$ of \sref{sec:arbitrary:symmetry} once an intrinsic symmetry is imposed.
   SHG: second-harmonic generation.
   }
   \label{tab:reduction:physical:tensors}
   \centering
   \small
   \begin{tabular}{cclcccccccc}
      \hline
      Rank $n$ & Symmetry                       & Example                    & $N_{\mathrm{ind}}$
               & \multicolumn{7}{c}{Weight $\ell$}                                                                                    \\
               &                                &                            &                    & 0  & 1  & 2  & 3  & 4  & 5 & 6 \\
      \hline
      0        &                                & Pressure                   & 1                  & 1                              \\
      1        & $i$                            & Spontaneous polarization   & 3                  &    & 1                         \\
      2        & $ij$                           & Optical activity           & 9                  & 1  & 1  & 1                    \\
               & $(ij)$                         & Stress and strain          & 6                  & 1  &    & 1                    \\
               & $[ij]$                         & Rotation rate              & 3                  &    & 1                         \\
      3        & $ijk$                          & Second-order optical mixing & 27                 & 1  & 3  & 2  & 1               \\
               & $i(jk)$                        & Piezoelectric effect       & 18                 &    & 2  & 1  & 1               \\
               & $(ijk)$                        & Kleinman symmetry in SHG   & 10                 &    & 1  &    & 1               \\
               & $[ij]k$                        & Hall effect                & 9                  & 1  & 1  & 1                    \\
      4        & $ijkl$                         & Third-order optical mixing & 81                 & 3  & 6  & 6  & 3  & 1          \\
               & $(ij)kl$                       & Photoelastic effect        & 54                 & 2  & 3  & 4  & 2  & 1          \\
               & $(ij)(kl)$                     & Kerr effect                & 36                 & 2  & 1  & 3  & 1  & 1          \\
               & $i(jkl)$                       & Third-harmonic generation  & 30                 & 1  & 1  & 2  & 1  & 1          \\
               & $((ij)(kl))$                   & Second-order elasticity                 & 21                 & 2  &    & 2  &    & 1          \\
               & $(ijkl)$                       & Cauchy relations           & 15                 & 1  &    & 1  &    & 1          \\
      6        & $ijklmn$                       &                            & 729                & 15 & 36 & 40 & 29 & 15 & 5 & 1 \\
               & $((ij)(kl)(mn))$               & Third-order elasticity     & 56                 & 3  &    & 3  & 1  & 2  &   & 1 \\
      \hline
   \end{tabular}
\end{table*}

An intrinsic symmetry is prescribed by a set of generators, each consisting of an index-permutation operator $\Pi_a$ and a sign $\eta_a$, $a=1,2,\dots,N_{\Pi}$:
\begin{equation} \label{eq:intrinsic:symmetry:generator}
   \Pi_a\mathbf T_n=\eta_a\mathbf T_n,
   \quad \eta_a\in\{+1,-1\}.
\end{equation}
Together, these generators define the intrinsic-symmetry class $\mathcal S$.
Only a generating set need be prescribed: a tensor invariant under the $\Pi_a$ is automatically invariant under every permutation they generate, with the sign given by the corresponding product of the $\eta_a$.
The elastic tensor $C_{ijkl}$ has $N_{\Pi}=2$ generators: a minor symmetry, exchanging the indices within the leading pair, and the major symmetry, exchanging the two pairs,
\begin{equation} \label{eq:elastic:generators}
   (\Pi_1\mathbf C)_{ijkl} = C_{jikl},\;
   (\Pi_2\mathbf C)_{ijkl} = C_{klij},\;
   \eta_1 = \eta_2 = +1 .
\end{equation}
The remaining minor symmetry, $C_{ijkl}=C_{ijlk}$, follows by composing them as $\Pi_2\Pi_1\Pi_2$.

The construction rests on one observation: an intrinsic symmetry constrains how the operators may be built, not how the ICTs behave.
Because every mapping that respects it lies in the span of the symmetry-free ones already at hand, the constraint becomes a linear condition on $N_\ell$ coefficients, and the tensor itself never enters.
The procedure below carries that out in three steps.

\begin{procedure}\label{proc:symmetry:adapted}
   At each weight $\ell$, the symmetry-adapted mapping tensors $\mathbf Q^p_{(\ell|n)}$ are obtained from the set of independent mapping tensors $\{\mathbf G^p_{(\ell|n)}\}_{p=1}^{N_\ell}$ of \sref{sec:independent:orthonormal} as follows:
   \begin{enumerate}[label=(\roman*)]
      \item write each $\mathbf Q$ as a general combination of the $\mathbf G$, \eref{eq:Q:from:symmetry:solutions}, and require it to carry the prescribed symmetry, \eref{eq:Q:symmetry};
      \item evaluate how that symmetry transforms the $\mathbf G$, \eref{eq:G:symmetry:action};
      \item select the combinations that survive, giving the $\mathbf Q$, \eref{eq:symmetry:coefficient:constraint}.
   \end{enumerate}
\end{procedure}

Starting with step~(i), each $\mathbf Q^p_{(\ell|n)}$, $p=1,2,\dots,N_\ell^{\mathcal S}$, lies in the weight-$\ell$ mapping space, so it can be written as a linear combination of the mapping tensors $\{\mathbf G^q_{(\ell|n)}\}$:
\begin{equation} \label{eq:Q:from:symmetry:solutions}
   \mathbf Q^p_{(\ell|n)}
   =\sum_{q=1}^{N_\ell}c_q^p\mathbf G^q_{(\ell|n)},
\end{equation}
where $\mathbf c^p=(c_1^p,c_2^p,\dots,c_{N_\ell}^p)^{\mathsf T}$ is a coefficient vector to be determined, as is $N_\ell^{\mathcal S}$, the number of independent symmetry-adapted mappings at weight $\ell$.
Permuting the indices of $\mathbf T_n$ permutes those of the mapping tensors and leaves the ICTs untouched, so the symmetry is imposed on the $\mathbf Q^p_{(\ell|n)}$: each must transform as $\mathbf T_n$ does in \eref{eq:intrinsic:symmetry:generator},
\begin{equation} \label{eq:Q:symmetry}
   \Pi_a\mathbf Q^p_{(\ell|n)}=\eta_a\mathbf Q^p_{(\ell|n)} .
\end{equation}

Turning to step~(ii), imposing this condition requires the action of each generator on the mapping tensors $\mathbf G^q_{(\ell|n)}$, since $\Pi_a\mathbf Q^p_{(\ell|n)}=\sum_{q}c_q^p\,\Pi_a\mathbf G^q_{(\ell|n)}$.
The permuted mapping $\Pi_a\mathbf G^q_{(\ell|n)}$ remains in the same weight-$\ell$ mapping space and can therefore be expanded uniquely in the same basis:
\begin{equation} \label{eq:G:symmetry:action}
   \Pi_a\mathbf G^q_{(\ell|n)}
   =\sum_{p=1}^{N_\ell}M^a_{pq}\mathbf G^p_{(\ell|n)}.
\end{equation}
The duals isolate its entries (\aref{app:M:mixing}),
\begin{equation} \label{eq:M:mixing}
   M^a_{pq}
   = \frac{
   \widetilde{\mathbf G}^p_{(\ell|n)}\odot^{\ell+n}\bigl(\Pi_a\mathbf G^q_{(\ell|n)}\bigr)
   }{
   2\ell+1
   }.
\end{equation}
So $\mathbf M^a$, which records how generator $a$ mixes the $N_\ell$ mappings, is built from operators already in hand.

In step~(iii), substituting \eref{eq:Q:from:symmetry:solutions} and \eref{eq:G:symmetry:action} into the requirement %$\Pi_a\mathbf Q^p_{(\ell|n)}=\eta_a\mathbf Q^p_{(\ell|n)}$
\eref{eq:Q:symmetry}
gives the linear system (see \aref{app:M:mixing})
\begin{equation} \label{eq:symmetry:coefficient:constraint}
   \bigl(\mathbf M^a-\eta_a\mathbf I\bigr)\mathbf c^p
   =\mathbf 0.
\end{equation}
In words, the surviving mappings are those the permutation leaves alone up to the prescribed sign: they are the eigenvectors of $\mathbf M^a$ at eigenvalue $\eta_a$, and $N_\ell^{\mathcal S}$ is the dimension of that eigenspace.

A weight loses a channel whenever some combination of its candidates extracts nothing from any tensor of the class, and the constraint discards exactly those combinations.
The extremes are the ones easiest to recognize: the combination can be the difference of two candidates that extract the same ICT, as $T_{ijij}$ and $T_{ijji}$ do for an elastic tensor, or a single candidate that extracts nothing on its own, as the trace $W_{ii}$ does for the rotation-rate tensor.
Both are special cases of the same rule: the combination may run over any number of candidates, which is why \eref{eq:symmetry:coefficient:constraint} is solved as a linear system rather than found by inspection.
This is not the reduction carried out in \sref{sec:alg:QR}, where candidates were cut because some were redundant for any tensor; here they stay independent in general and lose their independence only against the intrinsic symmetry.

Solving \eref{eq:symmetry:coefficient:constraint}, for example by Gaussian elimination, gives $N_\ell^{\mathcal S}$ independent coefficient vectors $\{\mathbf c^p\}_{p=1}^{N_\ell^{\mathcal S}}$, which \eref{eq:Q:from:symmetry:solutions} turns into the symmetry-adapted mapping tensors.
Which basis of solutions is taken, and how it is scaled and ordered, is again a choice, fixed in supplementary Sec.~S5.
These carry the prescribed symmetry on their Cartesian indices and span every weight-$\ell$ mapping compatible with that symmetry.
For several generators, the conditions for $\Pi_1,\Pi_2,\dots,\Pi_{N_{\Pi}}$ are imposed simultaneously, so that every $\mathbf c^p$ satisfies all the prescribed symmetries.
The combined system and its exact solution are detailed in \aref{app:M:mixing}.

The $\mathbf Q^p_{(\ell|n)}$ are again independent mapping tensors between $\mathcal X^\ell$ and $\mathcal T^n$, being combinations of the $\mathbf G^q_{(\ell|n)}$.
\sref{sec:extract:reconstruct} and \sref{sec:orthonormal:mapping} assumed no more than this, so the Gram matrix, the duals, the extraction and embedding, and the orthonormalization all carry over with the $\mathbf Q^p_{(\ell|n)}$ in place of the $\mathbf G^p_{(\ell|n)}$ and $N_\ell^{\mathcal S}$ in place of $N_\ell$.
Either route may therefore be taken: the dual pair $\mathbf Q^p_{(\ell|n)}$ and $\widetilde{\mathbf Q}^p_{(\ell|n)}$, or the self-dual orthonormal $\widehat{\mathbf Q}^p_{(\ell|n)}$, with the reconstruction of \eref{eq:G:orthonormal:complete} holding in either case.

This completes the construction: given the generators of an intrinsic symmetry, Procedure~\ref{proc:symmetry:adapted} yields the symmetry-adapted mapping tensors at every weight, and with them the reduction spectrum of the symmetry class, at any rank.
The next section applies the construction to the rank-4 elastic tensor.

\section{Application to the elastic tensor}
\label{sec:elastic}

Elastic tensors describe how a material responds to mechanical load.
A small deformation is measured by the symmetric strain $\varepsilon_{ij}$ and the internal force it generates by the symmetric stress $\sigma_{ij}$, and expanding one in powers of the other gives a hierarchy of response tensors~\cite{wallace1972thermodynamics},
\begin{equation}
   \sigma_{ij} = C_{ijkl}\,\varepsilon_{kl}
      + \tfrac{1}{2}\,C_{ijklmn}\,\varepsilon_{kl}\varepsilon_{mn} + \cdots .
\end{equation}
The leading term defines the second-order elastic tensor $C_{ijkl}$ of rank four (also called the stiffness tensor), the linear response of Hooke's law that fixes the moduli, the sound velocities, and the mechanical stability of a crystal~\cite{love1927treatise,born1954dynamical}; the next defines the third-order elastic tensor $C_{ijklmn}$ of rank six, the leading anharmonic correction.
Because the strain is symmetric, each tensor is symmetric within every index pair, and because each derives from the strain energy, each is also symmetric under permutation of its pairs, giving the symmetry classes $((ij)(kl))$ and $((ij)(kl)(mn))$ with $21$ and $56$ independent components.

The construction treats the two in the same way, one symmetry class in place of another.
Complete reductions of both, with the symmetry-restricted multiplicities, mapping bases, Gram matrices, and duals at every surviving weight, are given in supplementary Sec.~S7.
The rest of this section follows the rank-4 tensor: its decomposition, the weight fractions that partition its anisotropy, and what they reveal across computed elastic tensors.

\subsection{Reduction of the elastic tensor}
\label{sec:elastic:reduction}

The symmetry class $((ij)(kl))$ of the second-order elastic tensor $\mathbf C$ collects the minor symmetries $C_{ijkl}=C_{jikl}=C_{ijlk}$ and the major symmetry $C_{ijkl}=C_{klij}$.
Its two generators are those of \eref{eq:elastic:generators}.

Applying the construction to this symmetry class yields the reduction spectrum listed in \tref{tab:reduction:physical:tensors},
\begin{equation} \label{eq:elastic:spectrum}
   \mathbf C = 2\,\mathbf X_0 \oplus 2\,\mathbf X_2 \oplus \mathbf X_4 .
\end{equation}
The $21$ independent components are recovered by summing the dimensions of the ICTs, $2\times 1 + 2\times 5 + 1\times 9 = 21$.
Physically, the two scalars $\mathbf X_0$ capture the isotropic response (the two rotational invariants that fix the orientation-averaged bulk and shear moduli), while the weight-2 ICTs $\mathbf X_2$ and the weight-4 ICT $\mathbf X_4$ encode successively finer angular dependence of the anisotropic elastic response~\cite{jerphagnon1978description}.

We now construct the operators behind this spectrum.
The candidates follow from \eref{eq:G}: the rank-lowering tensors of \eref{eq:F:even} pair the four Cartesian indices in the three possible ways, and $E_{(0|0)}=1$, so
\begin{equation}
   \begin{aligned}
      G^1_{(0|4)} &= \delta_{i_1i_2}\delta_{i_3i_4},\\
      G^2_{(0|4)} &= \delta_{i_1i_3}\delta_{i_2i_4},\\
      G^3_{(0|4)} &= \delta_{i_1i_4}\delta_{i_2i_3} .
   \end{aligned}
\end{equation}
Algorithm~\ref{alg:QR} finds these $N_0^{\mathrm c}=3$ candidates linearly independent, so none is discarded and the multiplicity is $N_0=3$.
Each symmetry-adapted mapping is then a combination $\mathbf Q^p_{(0|4)}=\sum_q c^p_q\mathbf G^q_{(0|4)}$ of these three, \eref{eq:Q:from:symmetry:solutions}, with the coefficients still to be found.

The minor symmetry gives $\Pi_1\mathbf G^1_{(0|4)}=\mathbf G^1_{(0|4)}$, $\Pi_1\mathbf G^2_{(0|4)}=\mathbf G^3_{(0|4)}$ and $\Pi_1\mathbf G^3_{(0|4)}=\mathbf G^2_{(0|4)}$, while the pair exchange $\Pi_2$ leaves all three unchanged.
Expanding these in the basis through \eref{eq:M:mixing} gives
\begin{equation}
   \mathbf M^1 =
   \begin{pmatrix} 1&0&0\\ 0&0&1\\ 0&1&0 \end{pmatrix},
   \quad
   \mathbf M^2 = \mathbf I ,
\end{equation}
so the major symmetry imposes nothing at this weight.
With $\eta_1=\eta_2=+1$, the conditions in \eref{eq:symmetry:coefficient:constraint} reduce to $c_2=c_3$ with $c_1$ free, giving $N_0^{\mathcal S}=2$ and the sparse solution basis $\mathbf c^1=(1,0,0)^{\mathsf T}$, $\mathbf c^2=(0,1,1)^{\mathsf T}$, and hence, via \eref{eq:Q:from:symmetry:solutions},
\begin{equation} \label{eq:elastic:Q0}
   Q^1_{(0|4)} = \delta_{i_1i_2}\delta_{i_3i_4},
   \quad
   Q^2_{(0|4)} = \delta_{i_1i_3}\delta_{i_2i_4}+\delta_{i_1i_4}\delta_{i_2i_3} .
\end{equation}
The remaining steps are those of \sref{sec:extract:reconstruct}, applied to the $\mathbf Q^p_{(0|4)}$: their Gram matrix and its inverse, \eref{eq:gpq}, are
\begin{equation} \label{eq:elastic:gram}
   \mathbf g^{Q} =
   \begin{pmatrix} 9 & 6\\ 6 & 24\end{pmatrix},
   \quad
   (\mathbf g^{Q})^{-1} =
   \begin{pmatrix} \tfrac{2}{15} & -\tfrac{1}{30}\\[2pt] -\tfrac{1}{30} & \tfrac{1}{20}\end{pmatrix}.
\end{equation}
Combining the duals of \eref{eq:H} with the extraction of \eref{eq:extract}, and using \eref{eq:elastic:Q0} and \eref{eq:elastic:gram}, gives
\begin{equation} \label{eq:elastic:X0}
   X_0^1 = \tfrac{2}{15}C_{iijj}-\tfrac{1}{15}C_{ijij},
   \quad
   X_0^2 = -\tfrac{1}{30}C_{iijj}+\tfrac{1}{10}C_{ijij} .
\end{equation}
Embedding them back with \eref{eq:embed}, with the symmetry-adapted $\mathbf Q^p_{(0|4)}$ in place of the $\mathbf G^p_{(\ell|n)}$, gives the weight-0 part of $\mathbf C$:
\begin{equation}
   X_0^1\,\delta_{i_1i_2}\delta_{i_3i_4}
      + X_0^2\bigl(\delta_{i_1i_3}\delta_{i_2i_4}+\delta_{i_1i_4}\delta_{i_2i_3}\bigr),
\end{equation}
in which we recognize the classical isotropic elastic tensor of the Lam\'e decomposition~\cite{love1927treatise}.
The two scalars the construction extracts are the Lam\'e constants themselves, $\lambda=X_0^1$ and $\mu=X_0^2$.

The weight-0 sector exercises the full machinery: the candidates of \eref{eq:G}, the multiplicity from Algorithm~\ref{alg:QR}, the symmetry adaptation of Procedure~\ref{proc:symmetry:adapted}, and the Gram matrix, duals, extraction, and embedding of \sref{sec:extract:reconstruct}.
Weights 2 and 4 follow the same route, with larger matrices at each step and nothing new in kind.
The complete reduction of $\mathbf C$ is collected in supplementary Sec.~S7, which adds to the weight-0 construction above the weight-2 and weight-4 mappings, their Gram matrices and duals, and the resulting ICTs.
The third-order elastic tensor, at rank six, is reduced there by the same route.

\subsection{Weight-resolved anisotropy}
\label{sec:elastic:anisotropy}

The reduction of \sref{sec:elastic:reduction} says which ICTs an elastic tensor has; the ICTs themselves say how much of the elastic response each one carries.
Elastic anisotropy thereby becomes measurable in a resolved way: each weight has a definite angular dependence, so the response separates into an isotropic part and anisotropic parts of weight two and weight four, and the share carried by each is a physical quantity in its own right.
We can thus build a single scalar index that reports how anisotropic a crystal is, while the shares also report which angular dependence that anisotropy has.

These shares are well defined because rotations do not mix different weights, so the embeddings are mutually orthogonal in the Frobenius inner product and the squared norm of $\mathbf C$ partitions among them~\cite{jerphagnon1978description,gaith2009correlation},
\begin{equation} \label{eq:pythagoras}
   \lVert \mathbf C \rVert^2 = \sum_{\ell} \lVert \mathbf C_\ell \rVert^2 .
\end{equation}
Here $\mathbf C_\ell$ is the weight-$\ell$ part of $\mathbf C$, assembled from its ICTs by the embedding of \eref{eq:embed} as $\mathbf C_\ell = \sum_p \mathbf Q^p_{(\ell|4)} \odot^\ell \mathbf X_\ell^p$, and its norm follows from the Gram matrix of \eref{eq:gpq},
\begin{equation} \label{eq:weight:norm}
   \lVert \mathbf C_\ell \rVert^2 = \mathbf C_\ell \odot^4 \mathbf C_\ell
   = \sum_{p,q} g_{pq}\bigl(\mathbf X_\ell^p \odot^\ell \mathbf X_\ell^q\bigr) .
\end{equation}
The share of the total carried by each weight is the weight fraction
\begin{equation} \label{eq:weight:fractions}
   f_\ell = \frac{\lVert \mathbf C_\ell \rVert^2}{\lVert \mathbf C \rVert^2},
   \quad \sum_\ell f_\ell = 1 ,
\end{equation}
so $f_0$ is the isotropic share of the response, while $f_2$ and $f_4$ measure the anisotropy carried at each weight.
The anisotropic weights together carry $1-f_0$ of the response, and the anisotropy weight fractions
\begin{equation} \label{eq:relative:fractions}
   \hat f_\ell = \frac{f_\ell}{1-f_0},
   \quad \ell \neq 0,
   \quad \sum_{\ell \neq 0} \hat f_\ell = 1 ,
\end{equation}
give the weight-$\ell$ share of that, for any tensor that is not isotropic.
For $\mathbf C$ there are only two anisotropic weights, so $\hat f_2 + \hat f_4 = 1$ and either one carries the division on its own.
Being ratios, the weight fractions are dimensionless and invariant under rotation and under overall scaling of $\mathbf C$, so materials of different overall stiffness are directly comparable.
Within a repeated weight the mapping basis is fixed only up to an invertible mixing of the channels, which redistributes norm among them but leaves $\mathbf C_\ell$ unchanged, the duals of \eref{eq:H} transforming inversely.
The weight fractions are therefore the same whether the $\mathbf Q^p_{(\ell|4)}$ or the orthonormal mappings of \sref{sec:orthonormal:mapping} are used.

Because the weight-0 subspace is itself rotation invariant, $\mathbf C_0$ is the orthogonal projection of $\mathbf C$ onto the isotropic elastic tensors of \eref{eq:elastic:Q0}, so $1-f_0$ is the squared distance to isotropy, obtained without fitting.
It is the quantity compared with established anisotropy measures in \sref{sec:elastic:mp}.
The weight fractions also inherit the crystal's symmetry: because rotations do not mix weights, its point-group invariance factorizes into an independent condition on each ICT, so if the Laue class admits no invariant of weight $\ell$, meaning that no nonzero ICT of weight $\ell$ is left unchanged by all its symmetry operations, $f_\ell$ vanishes, as $f_2$ does for cubic crystals.

The weight fractions generalize the classical measure of cubic anisotropy.
The cubic classes admit no invariant at weight two~\cite{bradley1972mathematical} (supplementary Sec.~S8), so the entire anisotropy of a cubic crystal is the single weight-4 ICT and $f_4=1-f_0$; evaluating the weight-0 norm from \eref{eq:weight:norm}, with \eref{eq:elastic:X0} and \eref{eq:elastic:gram}, gives
\begin{equation}
   f_4 = \frac{6}{5}\,
   \frac{\left(C_{11}-C_{12}-2C_{44}\right)^2}{\lVert \mathbf C \rVert^2} ,
\end{equation}
which vanishes precisely when the Zener ratio $A=2C_{44}/(C_{11}-C_{12})$ equals unity~\cite{zener1948elasticity}.
A vanishing weight fraction therefore need not come from symmetry: the cubic classes admit an invariant of weight four, yet $f_4$ vanishes for a cubic crystal with $A=1$.
Together with the Lam\'e constants at weight zero in \eref{eq:elastic:X0}, the same operators thus reproduce classical elasticity at both ends of the cubic spectrum without elasticity-specific input.
The Zener ratio, however, is defined only for cubic crystals.
The weight fractions are defined for crystals of any symmetry, reproduce the cubic criterion as the special case just shown, and in addition record which weight the anisotropy occupies.

\subsection{First-principles elastic tensors}
\label{sec:elastic:mp}

Elastic anisotropy varies widely across crystalline materials, and the weight fractions give a resolved way to survey it.
We apply them to all $10\,276$ second-order elastic tensors from the Materials Project~\cite{horton2025accelerated,wen2024an,wen2023elasticity}, asking how much of the elastic response is isotropic, how the remainder divides between weights two and four, and how that picture relates to the anisotropy measures already in use.
The survey is possible at this scale because the mapping tensors are precomputed once for the $((ij)(kl))$ class, so each material costs a single contraction.

\begin{figure*}[tbh!]
   \centering
   \includegraphics[width=0.90\textwidth]{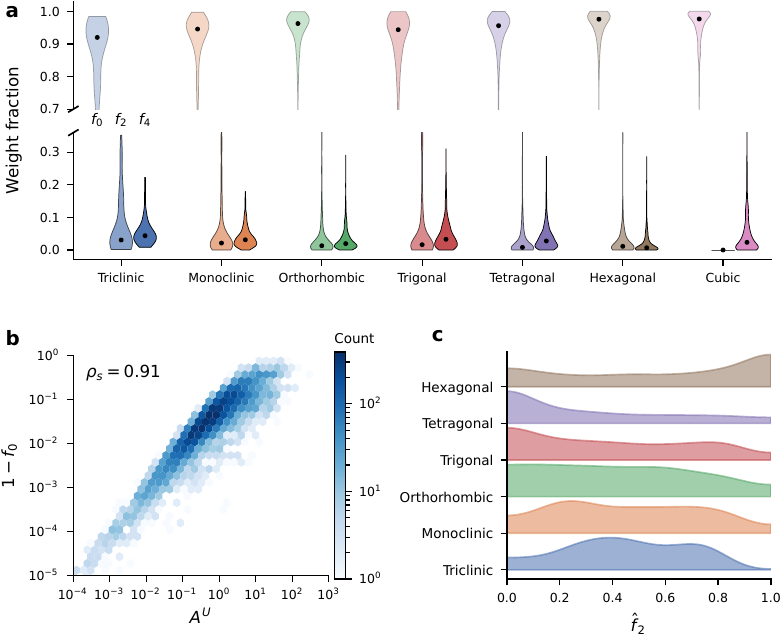}
   \caption{%
      \textbf{Weight-resolved anisotropy of second-order elastic tensors from the Materials Project.}
      \textbf{a} Distributions of the weight fractions $f_0$, $f_2$ and $f_4$ of \eref{eq:weight:fractions} by crystal system; black dots mark the medians.
      The vertical axis is broken, the two segments being drawn to a common scale.
      \textbf{b} The squared distance to isotropy $1-f_0$ against the universal anisotropy index $A^{U}$, binned by material count, with $\rho_s$ the Spearman rank correlation.
      \textbf{c} Distributions of the anisotropy weight fraction $\hat f_2$ of \eref{eq:relative:fractions} by crystal system, each curve scaled to a common height.
      Cubic entries are omitted, $\hat f_2$ vanishing identically for them.
   }
   \label{fig:mp:anisotropy}
\end{figure*}

\fref{fig:mp:anisotropy}a resolves the three weight fractions by crystal system.
The isotropic part dominates throughout, its mean fraction of the squared norm being $0.88$ for triclinic crystals and $0.95$ for hexagonal and cubic ones, so, measured by the norm, the anisotropic part of a real crystal is relatively small.
Both anisotropic weights are populated in every class but the cubic one, where $f_2$ vanishes identically, the selection rule of \sref{sec:elastic:anisotropy} recovered although the operators are built for the $((ij)(kl))$ class alone and nothing about cubic symmetry enters them.

The weight fractions measure how the squared norm of $\mathbf C$ is distributed over the weights, and $f_0$ close to unity means that the norm is almost entirely carried by the isotropic part.
This, however, does not rule out a strong directional dependence of the mechanical response, for example of the shear or Young's modulus.
As shown in \sref{sec:elastic:reduction}, the isotropic part $\mathbf C_0$ is the Lam\'e tensor with the Lam\'e constants $\lambda$ and $\mu$ of \eref{eq:elastic:X0}, equivalently a bulk modulus $K=\lambda+\tfrac{2}{3}\mu$ and a shear modulus $G=\mu$~\cite{love1927treatise}, which for these $\lambda$ and $\mu$ are the Voigt averages of the moduli of the crystal~\cite{hill1952elastic}.
Both are the same in every direction, so they carry no directional information, yet they contribute to the squared norm of the isotropic part.
A material with a large bulk modulus can therefore have $f_0$ close to unity even when its moduli, such as the shear or Young's modulus, vary strongly with direction.
Applied to the compliance, the inverse of $\mathbf C$, the same reduction gives the Reuss averages~\cite{hill1952elastic} in place of the Voigt ones for the isotropic part.

We next compare $1-f_0$ with the measures of elastic anisotropy already in use.
All of them are built from the mismatch between the Voigt and Reuss bounds on the bulk and shear moduli, a gap that closes only for an isotropic material.
Writing $K_V,G_V$ and $K_R,G_R$ for those bounds, they are the universal index~\cite{ranganathan2008universal} $A^U = 5G_V/G_R + K_V/K_R - 6$, the log-Euclidean index~\cite{kube2016elastic} $A^L = ([\ln(K_V/K_R)]^2 + 5[\ln(G_V/G_R)]^2)^{1/2}$, and the Chung--Buessem shear and bulk anisotropies~\cite{chung1967elastic} $A_G = (G_V-G_R)/(G_V+G_R)$ and $A_B = (K_V-K_R)/(K_V+K_R)$.
Because $f_0$ is a norm fraction rather than a bound mismatch, the comparison sets two ways of quantifying the same departure from isotropy against each other.
\fref{fig:mp:anisotropy}b makes it for $A^U$, and \tref{tab:anisotropy:correlation} for all four.
The rank correlation with $A^U$, $A^L$ and $A_G$ is uniformly high, and resolving it by crystal system leaves it between $0.89$ and $0.95$ in every group, so it does not arise from pooling symmetries.
Rank correlation is the appropriate comparison, the measures differing widely in scale; the lower raw Pearson value for $A^U$ reflects its unbounded range rather than a genuinely weaker association.
Since these measures encode established practice, the agreement validates $f_0$ rather than distinguishing it.
The bulk anisotropy $A_B$ is the exception, and the fault lies with it rather than with $f_0$: cubic symmetry forces $K_V=K_R$, so $A_B$ vanishes for the $41\%$ of the database that is cubic, although those crystals are not isotropic.

A scalar index reports one number, and the weight fractions add a second dimension to it.
\fref{fig:mp:anisotropy}c shows the distribution of the anisotropy weight fraction $\hat f_2$ of \eref{eq:relative:fractions} for each crystal system, the cubic entries being omitted because the selection rule fixes $\hat f_2=0$ for them.
Pooled, the fifth to ninety-fifth percentiles run from $0.02$ to $0.94$, and no crystal system covers less than three quarters of that interval, so how the anisotropy divides between weights two and four is a property of the material rather than of its symmetry.
The established measures fix the magnitude of the departure from isotropy, on which $f_0$ agrees with them, while $f_2$ and $f_4$ record which weight that departure occupies.

\begin{table}[tbh!]
   \caption{\textbf{Correlation of $1-f_0$ with established scalar measures of elastic anisotropy.}
      The measures $A^U$, $A^L$, $A_G$ and $A_B$ are defined in the text.
      Spearman is the rank correlation; the Pearson columns are computed on the raw values and on their logarithms.
      Values of $A_B$ below $10^{-12}$ in magnitude are rounding noise and are set to zero, which applies to the $4\,217$ cubic crystals, for which $A_B=0$ by symmetry; these are excluded from the logarithm, and all other quantities are positive for every crystal.
   }
   \label{tab:anisotropy:correlation}
   \centering
   \begin{tabular}{lccc}
      \hline
      Measure & Spearman & Pearson & Pearson (log) \\
      \hline
      $A^U$   & 0.91    & 0.56   & 0.93 \\
      $A^L$   & 0.91    & 0.76   & 0.94 \\
      $A_G$   & 0.90    & 0.76   & 0.93 \\
      $A_B$   & 0.41    & 0.73   & 0.61 \\
      \hline
   \end{tabular}
\end{table}

\section{Cartesian harmonics}
\label{sec:nt:unit:vector}

The construction now turns from property tensors to the ICTs themselves: building them from a vector here, and coupling two of them in \sref{sec:tp}.
The ICTs built from a vector $\mathbf a$ are the \emph{Cartesian harmonics}, one for each weight $n$; their components are homogeneous polynomials of degree $n$ in the Cartesian coordinates that satisfy Laplace's equation, that is, solid harmonics.
The reduction of the preceding sections produces them with no new machinery, and yields a single operator for each weight.
The construction below uses the unit vector $\uv{a}=\mathbf a/a$, of length $a=|\mathbf a|$, which fixes the scale; homogeneity restores the general case, $\mathbf V_n(\mathbf a)=a^n\mathbf V_n(\uv{a})$.

The rank-$n$ polyadic $\uv{a}^{\otimes n}$ has components $\hat a_{i_1}\hat a_{i_2}\cdots \hat a_{i_n}$.
Full symmetry is one of the intrinsic symmetries treated in \sref{sec:arbitrary:symmetry}, so the reduction developed in \sref{sec:reduction:embedding} to \sref{sec:arbitrary:symmetry} applies to it unchanged.
Two properties of the polyadic simplify that reduction: $\uv{a}$ has unit length, and $\uv{a}^{\otimes n}$ is fully symmetric.

Unit length disposes of the rank lowering.
A harmonic of weight $\ell<n$ is built from $\uv{a}^{\otimes \ell}$ directly, and $\hat a_i\hat a_i=1$ makes the weight-$\ell$ part of $\uv{a}^{\otimes n}$ proportional to it, so nothing is lost by taking the polyadic of the weight one wants.
Step (i) of Procedure~\ref{proc:natural:reduction} therefore never enters, and the mapping tensor of \eref{eq:G} reduces to the projector alone, $\mathbf G_{(n|n)}=\mathbf E_{(n|n)}$.

Full symmetry disposes of the symmetrization and simplifies what is left.
Step (ii) of Procedure~\ref{proc:natural:reduction}, the average over index permutations, is already carried out in $\uv{a}^{\otimes n}$, so the projector has only the traces of step (iii) to remove, and it does that in one step: the weight-$n$ part of $\uv{a}^{\otimes n}$ is $\mathbf E_{(n|n)}\odot^{n}\uv{a}^{\otimes n}$.
The projector itself also simplifies, because each of its terms averages over all $n!$ permutations of the Greek indices and, separately, over all $n!$ permutations of the Roman ones, and the Roman permutations only change which index of $\uv{a}^{\otimes n}$ each $\bm\delta$ meets.
They therefore contribute equally, their average may be dropped, and $\delta_{ii}^{t}$ comes outside it, leaving
\begin{equation} \label{eq:E:symmetric}
   E_{(n|n)} \cong
   \sum_{t=0}^{\lfloor n/2 \rfloor} c_t\,
   \avg{\delta_{\alpha i}^{\,n-2t}\delta_{\alpha\alpha}^{t}}\,
   \delta_{ii}^{t},
\end{equation}
with the coefficients $c_t$ of \eref{eq:natural:projector} unchanged.
Here $\cong$ marks an equality that holds when both sides are contracted with a fully symmetric tensor, not a generic tensor.
\aref{app:E:symmetric} gives the detailed steps.

What the reduction does not fix is the normalization.
The projector $\mathbf E_{(n|n)}$ is scaled so as to return a weight-$n$ tensor unchanged, whereas harmonics are conventionally scaled by the Legendre polynomial they generate~\cite{coope1965irreducible,coope1970irreducible2,lehman1989angular}.
Rescaling \eref{eq:E:symmetric} by $C=(2n-1)!!/n!$ gives the \emph{harmonic operator}
\begin{equation} \label{eq:H:unit:vector}
   H_{(n|n)} =
   C \sum_{t=0}^{\lfloor n/2 \rfloor} c_t\,
   \avg{\delta_{\alpha i}^{\,n-2t}\delta_{\alpha\alpha}^{t}}\,
   \delta_{ii}^{t}.
\end{equation}

Contracting it with the polyadic returns the Cartesian harmonic of weight $n$, $\mathbf V_n=\Hop_{(n|n)}\odot^{n}\uv{a}^{\otimes n}$, which has the Legendre property
\begin{equation} \label{eq:legendre:normalization}
   \mathbf V_n\odot^{n}\uv{b}^{\otimes n}=P_n(\uv{a}\cdot\uv{b})
\end{equation}
for every unit vector $\uv{b}$, where $P_n$ is the Legendre polynomial of degree $n$, and in particular $\mathbf V_n\odot^{n}\uv{a}^{\otimes n}=P_n(1)=1$.
\aref{app:normalization} verifies \eref{eq:legendre:normalization}.

The $2n+1$ independent components of $\mathbf V_n$ span the degree-$n$ solid harmonics; \aref{app:E:symmetric} writes out $\Hop_{(2|2)}$ and $\Hop_{(3|3)}$ and the harmonics they produce.
With this normalization, \eref{eq:H:unit:vector} agrees with the Cartesian harmonic tensor of \olcite{lehman1989angular}, and supplementary Sec.~S9 gives the conversion between the two conventions.
Although mathematically equivalent, the two differ in form.
The classical result is an expression for the harmonic of a given direction, with $\uv{a}$ written inside it, whereas $\Hop_{(n|n)}$ depends only on $n$ and meets the vector only through a contraction.
Bond directions, surface normals, gradients, and other vector descriptors are the elementary building blocks of orientational order parameters in materials modeling and of equivariant machine learning models built on irreducible Cartesian tensors~\cite{zaverkin2024higher,cheng2024cartesian,xu2025tace}, and the construction applies to all of them directly.

\section{Coupling of irreducible Cartesian tensors}
\label{sec:tp}

Given two ICTs $\mathbf X_{\ell_1}$ and $\mathbf Y_{\ell_2}$, their product $\mathbf P_{n}=\mathbf X_{\ell_1}\otimes\mathbf Y_{\ell_2}$ is a tensor of rank $n=\ell_1+\ell_2$, symmetric within its first $\ell_1$ indices and within its last $\ell_2$ but not across the two groups.
That is an intrinsic symmetry in the sense of \sref{sec:arbitrary:symmetry}, so the reduction developed there applies to $\mathbf P_n$ unchanged.
Its irreducible parts $\mathbf Z_{\ell_3}$ have weights
$|\ell_1-\ell_2|\leq \ell_3\leq \ell_1+\ell_2$,
following the same selection rule as Clebsch--Gordan coupling in the spherical formalism~\cite{edmonds1996angular}.
For each weight triple, the reduction yields a single operator that produces $\mathbf Z_{\ell_3}$ from the pair by contraction.

As in \sref{sec:nt:unit:vector}, two properties simplify the reduction, here narrowing the choices within its steps rather than removing steps outright: being ICTs, $\mathbf X_{\ell_1}$ and $\mathbf Y_{\ell_2}$ are traceless, and each is symmetric.
A contraction taken inside either of them vanishes, so only contractions between the two survive, and since each is symmetric, all such contractions give the same result.
Step (i) of Procedure~\ref{proc:natural:reduction} is therefore left with a single candidate instead of the $N_\ell^{\mathrm c}$ of \sref{sec:rank:lowering}, and step (iii) with a single series.
Step (ii) is not free, since $\mathbf P$ is symmetric only within each group, and it is carried out by $\mathbf E_{(\ell_3|\ell_3)}$ along with the trace removal.

Write $\alpha$ for the indices of $\mathbf X_{\ell_1}$, $\beta$ for those of $\mathbf Y_{\ell_2}$, and $\gamma$ for the free indices of $\mathbf Z_{\ell_3}$.
Write $L=\ell_1+\ell_2+\ell_3$ and let $L_i=\lfloor L/2\rfloor-\ell_i$ be the \emph{triangle numbers} of the weight triple.
Each $L_i$ counts the contractions between the two groups other than $i$: $L_1$ those between $\mathbf Y_{\ell_2}$ and $\mathbf Z_{\ell_3}$, $L_2$ those between $\mathbf X_{\ell_1}$ and $\mathbf Z_{\ell_3}$, and $L_3$ those between $\mathbf X_{\ell_1}$ and $\mathbf Y_{\ell_2}$.
By the two properties above, rank lowering keeps only the $L_3$ contractions between $\mathbf X_{\ell_1}$ and $\mathbf Y_{\ell_2}$, so the rank-lowering tensors $\mathbf F^p_{n\to \ell_3}$ of \sref{sec:rank:lowering} reduce to a single candidate.
It takes one of the two forms of \sref{sec:rank:lowering}, $\bm\epsilon$ entering exactly when the rank drop $n-\ell_3$ is odd, which is the parity of $L=n+\ell_3$:
\begin{equation} \label{eq:tp:F}
   F_{n\to \ell_3} =
   \begin{cases}
      \delta_{\alpha\beta}^{\,L_3},
       & L \text{ even},   \\[2pt]
      \epsilon_{\gamma\alpha\beta}\,\delta_{\alpha\beta}^{\,L_3},
       & L \text{ odd},
   \end{cases}
\end{equation}
every $\bm\delta$, and the $\bm\epsilon$ of the odd case, taking one index from each group.

The natural projector $\mathbf E_{(\ell_3|\ell_3)}$ of \eref{eq:natural:projector} simplifies for the same reason.
The symmetrization of step (ii) collapses, every surviving permutation of the Roman indices returning the same tensor.
For the trace removal of step (iii), the projector's factor $\delta_{ii}^{t}$ takes $t$ traces of $\mathbf P$, of which only the cross traces survive, each contracting $\mathbf X_{\ell_1}$ with $\mathbf Y_{\ell_2}$ once more, so the series is indexed by the number $t$ of these.
As in \eref{eq:E:symmetric}, $\delta_{ii}^{t}$ then comes outside the average, leaving the average over the free $\gamma$ indices alone.

Composing the two through \eref{eq:G} gives a single mapping tensor $\mathbf G_{(\ell_3|n)}=\mathbf E_{(\ell_3|\ell_3)}\,\mathbf F_{n\to \ell_3}$, which reduces $\mathbf P$ to $\mathbf Z_{\ell_3}$.
The composition carries $c_t$ of \eref{eq:ct} into
\begin{equation} \label{eq:ct:tp}
   \kt = (-1)^t\,
   \frac{(2\ell_3-2t-1)!!}{(2\ell_3-1)!!}\,
   \frac{L_2!\,L_1!}{(L_2-t)!\,(L_1-t)!\,t!},
\end{equation}
the counterpart of $c_t$ for the coupling, derived in supplementary Sec.~S9.1.
With the weight-$\ell_3$ mapping space one dimensional, the selection and orthonormalization of \sref{sec:independent:orthonormal} are not needed, and extraction and embedding differ only by the single Gram-matrix element $g_{11}$.

Rescaling $\mathbf G_{(\ell_3|n)}$ by a scalar $C$ gives the \emph{coupling operator} $\Kop_{(\ell_3|\ell_1,\ell_2)}=C\,\mathbf G_{(\ell_3|n)}$ for the weights $(\ell_1,\ell_2,\ell_3)$.
When $L$ is even,
\begin{equation} \label{eq:tp:even:H}
   K_{(\ell_3|\ell_1,\ell_2)} =
   C
   \sum_{t=0}^{\min(L_2,L_1)}
   \kt\,
   \avg{\delta_{\alpha\gamma}^{\,L_2-t}
      \delta_{\beta\gamma}^{\,L_1-t}
      \delta_{\gamma\gamma}^{\,t}}\,
   \delta_{\alpha\beta}^{\,L_3+t},
\end{equation}
and when $L$ is odd,
\begin{equation} \label{eq:tp:odd:H}
   K_{(\ell_3|\ell_1,\ell_2)} =
   C
   \sum_{t=0}^{\min(L_2,L_1)}
   \kt
   \avg{\epsilon_{\alpha\beta\gamma}\,
      \delta_{\alpha\gamma}^{\,L_2-t}
      \delta_{\beta\gamma}^{\,L_1-t}
      \delta_{\gamma\gamma}^{\,t}}
   \delta_{\alpha\beta}^{\,L_3+t},
\end{equation}
where the averages are those of \eref{eq:natural:projector}, read as in \eref{eq:E:symmetric}: only the average over $\gamma$ survives, and the factors $\delta_{\alpha\beta}$, which carry the contractions between $\mathbf X_{\ell_1}$ and $\mathbf Y_{\ell_2}$, stand outside it.
Applied to the pair, it returns the weight-$\ell_3$ ICT,
\begin{equation} \label{eq:tp:apply}
   \mathbf Z_{\ell_3}
   =
   \left(\Kop_{(\ell_3|\ell_1,\ell_2)}\odot^{\ell_2}\mathbf Y_{\ell_2}\right)
   \odot^{\ell_1}\mathbf X_{\ell_1}.
\end{equation}
The two contractions are independent, so the rank-$(\ell_1 + \ell_2)$ product $\mathbf P$ is never formed.

Here $C$ is a scaling factor, as in \sref{sec:nt:unit:vector}, and the one thing the construction leaves open.
Its value depends on the three weights and differs between the even and odd cases.
We fix it by the two conditions of \olcite{lehman1989angular}: for even $L$, that two harmonics of the same direction couple to the harmonic of weight $\ell_3$ of that direction; for odd $L$, where that coupling vanishes as the directions merge, that its rate of vanishing is unity.
See \aref{app:normalization} for the conditions in full, which give, for even $L$,
\begin{equation} \label{eq:C:constant}
   C =
   \frac{\ell_1!\,\ell_2!\,(2\ell_3-1)!!}
   {(2L_2-1)!!\,(2L_1-1)!!\,(2L_3-1)!!\,(L/2)!}
\end{equation}
and, for odd $L$,
\begin{equation} \label{eq:D:constant}
   C =
   \frac{2\,\ell_3\,\ell_1!\,\ell_2!\,(2\ell_3-1)!!}
   {(2L_2+1)!!\,(2L_1+1)!!\,(2L_3+1)!!\,\left(\frac{L+1}{2}\right)!},
\end{equation}
with $(-1)!!=1$.

The two operators in Eqs.~\eqref{eq:tp:even:H} and \eqref{eq:tp:odd:H} agree with the coupling obtained previously, normalization included~\cite{lehman1989angular}.
The coupling of two irreducible tensors into a given weight is unique up to a factor, so they also agree, up to normalization, with the fixed Cartesian coupling tensors of \olcite{coope1970irreducible3}.
What the derivation here adds is the route and the form.
The coupling follows from \eref{eq:natural:projector}, the operator that reduces an arbitrary tensor, rather than from a derivation specific to this problem.
It is also delivered as a tensor that depends only on the three weights and meets $\mathbf X_{\ell_1}$ and $\mathbf Y_{\ell_2}$ only through contractions: the sum that produces it is carried out once, before $\mathbf X_{\ell_1}$ and $\mathbf Y_{\ell_2}$ are given, and the operator is reused for every pair at those weights.
In the expression of \olcite{lehman1989angular}, in contrast, $\mathbf X_{\ell_1}$ and $\mathbf Y_{\ell_2}$ appear inside every term of the sum, so it is carried out after they are given, for each pair.
The form used in \olcite{lehman1989angular} and its relation to ours are discussed in supplementary Sec.~S9.
The same coupling appears in the spherical formalism as the Clebsch--Gordan coefficients and, in the form symmetric in the three weights, as the $3j$ symbols; supplementary Sec.~S10 gives the Cartesian counterpart of the latter, and supplementary Sec.~S11 compares the two formalisms.

The cost of building and applying the coupling operators is given in supplementary Sec.~S12.

\section{Conclusion}
\label{sec:conclusion}

We have presented a constructive Cartesian reduction of high-rank tensors with prescribed intrinsic symmetry into ICTs.
The construction separates three tasks that are often entangled in hand-derived decompositions: constructing all symmetry-free candidate mappings from invariant tensors, resolving repeated weights by QR selection and Gram orthonormalization, and reinstating intrinsic symmetry through symmetry-adapted mapping tensors that reconstruct the original Cartesian tensor.
This gives an explicit reduction--reconstruction pair rather than only a representation-theoretic spectrum.

The same construction also gives Cartesian operators for two further operations.
The harmonic operators build the Cartesian harmonics from repeated products of a vector, and the coupling operators couple two ICTs into a third.
These are the Cartesian counterparts of the spherical harmonics and of the Clebsch--Gordan coefficients, and both are operations at which a Cartesian treatment would otherwise convert to the spherical formalism.
Keeping them in Cartesian form removes the conversion along with the phase and normalization conventions that come with it.

The elastic tensor puts the construction to work both where the answer is already known and where it is not.
At rank four it recovers the decomposition established in the elasticity literature~\cite{backus1970geometrical,cowin1992identification,forte1996symmetry}, with no elasticity-specific input.
At rank six it gives a systematic treatment of the third-order elastic tensor, previously handled case by case~\cite{norris1991symmetry,auffray2013matrix,auffray2021explicit}.
The same operators also make anisotropy measurable rather than merely classifiable.
Distinct weights carry distinct angular dependence, so a material's departure from isotropy separates into parts that a single scalar index cannot tell apart.
Nothing in this is specific to elasticity, the same weight fractions resolving the angular content of any property tensor the construction reduces, among them the piezoelectric and photoelastic tensors.

More broadly, the weight-resolved decomposition provides a design principle for equivariant machine learning.
Predicting a tensor through its ICTs and embedding them back makes the output equivariant by construction, and enforces its intrinsic symmetry exactly.
The partition of the norm across weights also gives a weight-wise normalization of the learning targets.
Part of the scheme developed here is already in use in \olcite{chen2026atomistic}.

Everything the construction produces shares one character.
The natural projector, the rank-lowering and mapping tensors, and the harmonic and coupling operators are all constant isotropic arrays: built from $\bm\delta$ and $\bm\epsilon$ alone, and fixed by the ranks, weights, and intrinsic symmetry at hand rather than by the data they act on.
Each can therefore be pre-evaluated once, in exact symbolic or in numerical form, and stored as an array, then reused for every tensor of that rank and symmetry.
Applying it is a single tensor contraction, the operation an array library provides through, for example, matrix multiplication or \texttt{einsum}.
The operators are fast to build, and the reduction and the tensor product that apply them are fast to evaluate (supplementary Sec.~S12).
Reduction, reconstruction, formation of harmonics, and coupling are then the same operation with different stored arrays, which makes the construction straightforward to implement and use.
They are what the open-source package \code~\cite{natto} builds, stores, and applies.

\section*{Supplementary material}
The supplementary material provides additional mathematical details and worked examples supporting the construction presented in the main text.
It includes (1) the number of independent components of an irreducible Cartesian tensor, (2) the isotropic tensor identities used throughout, (3) the counts of distinct symmetrized products, (4) an explicit form of the natural projector, (5) the conventions that fix the channels of a repeated weight, (6) worked examples of the reduction for generic tensors and (7) for tensors with intrinsic symmetry, (8) the character counts behind the multiplicities and the crystal-system content, (9) the conversion between operator conventions, (10) the relation between the coupling operator and the Cartesian $3j$ tensor, and (11) a comparison of the Cartesian and spherical formalisms.

\begin{acknowledgments}
   This work is supported by startup funding from the University of Electronic Science and Technology of China (UESTC) and uses computational resources provided by the Center for High-Performance Computing (HPC) at UESTC.
   It also uses computational resources provided by the Hefei Advanced Computing Center.
\end{acknowledgments}

\section*{Author declarations}

\subsection*{Conflict of interest}
The author has no conflicts to disclose.

\subsection*{Author contributions}
M.W.: Conceptualization; Methodology; Formal analysis; Software; Investigation; Visualization; Writing -- original draft; Writing -- review \& editing.

\section*{Data availability}
The data that support the findings of this study are available within the article and its supplementary material.
The \code{} package implementing the irreducible Cartesian tensor algorithms described in this work is openly available at \url{https://github.com/wengroup/natto}~\cite{natto}.

\appendix

\section{Notation}
\label{app:notation}

\tref{tab:notation} lists the symbols used throughout, and \tref{tab:operations} lists the tensor operations.

\begin{table*}[htb!]
   \caption{Symbols.}
   \label{tab:notation}
   \centering
   \footnotesize
   \begin{tabular}{ccl}
      \hline
      Symbol                               & Indicial notation                                           & Explanation \\
      \hline
      $\bm\delta$                          & $\delta_{ij}$                                               & Kronecker delta \\
      $\bm\epsilon$                        & $\epsilon_{ijk}$                                            & Levi--Civita symbol \\
      $\mathcal{T}^n$                      &                                                             & Rank-$n$ Cartesian tensor space \\
      $\mathcal{X}^\ell$                   &                                                             & Weight-$\ell$ ICT space \\
      $\mathbf a$ and $\uv{a}$            & $a_i$ and $\hat a_i$                                                       & Vector and the corresponding unit vector \\
      $\mathbf T$ or $\mathbf T_n$         & $T_{i_1i_2\dots i_n}$                                       & Rank-$n$ tensor in $\mathcal{T}^n$ \\
      $\mathbf U$ or $\mathbf U_n$         & $U_{i_1i_2\dots i_n}$                                       & Rank-$n$ symmetric tensor in $\mathcal{T}^n$ \\
      $\mathbf X$ or $\mathbf X_\ell$      & $X_{\alpha_1\alpha_2\dots \alpha_\ell}$                     & Weight-$\ell$ ICT in $\mathcal{X}^\ell$ \\
      $\mathbf X_\ell^p$                   & $X_{\alpha_1\alpha_2\dots \alpha_\ell}^{p}$                 & Weight-$\ell$ ICT in $\mathcal{X}^\ell$ with multiplicity index $p$ \\
      $\mathbf S_n^{\ell,p}$               & $S_{i_1i_2\dots i_n}^{\ell,p}$                              & Embedding of $\mathbf X_\ell^p$ in $\mathcal{T}^n$ \\
      $\mathbf E_{(\ell|\ell)}$            & $E_{(\alpha_1\dots \alpha_\ell|i_1\dots i_\ell)}$           & Natural projector, $\mathcal{T}^\ell\to\mathcal{X}^\ell$ \\
      $\mathbf F^p_{n\to \ell}$            &                                                             & Rank-lowering tensor, $\mathcal{T}^n\to\mathcal{T}^\ell$ \\
      $\mathbf G^p_{(\ell|n)}$             & $G^p_{(\alpha_1\dots \alpha_\ell|i_1\dots i_n)}$            & Mapping tensor, embeds $\mathcal{X}^\ell\to\mathcal{T}^n$ \\
      $\widetilde{\mathbf G}^p_{(\ell|n)}$ & $\widetilde G^p_{(\alpha_1\dots \alpha_\ell|i_1\dots i_n)}$ & Dual mapping tensor, extracts $\mathcal{T}^n\to\mathcal{X}^\ell$ \\
      $\widehat{\mathbf G}^p_{(\ell|n)}$   & $\widehat G^p_{(\alpha_1\dots \alpha_\ell|i_1\dots i_n)}$   & Orthonormal mapping tensor, self-dual, $\mathcal{T}^n\leftrightarrow\mathcal{X}^\ell$ \\
      $\mathbf Q^p_{(\ell|n)}$             & $Q^p_{(\alpha_1\dots \alpha_\ell|i_1\dots i_n)}$            & Symmetry-adapted mapping tensor, embeds $\mathcal{X}^\ell\to\mathcal{T}^n$ \\
      $\widetilde{\mathbf Q}^p_{(\ell|n)}$ & $\widetilde Q^p_{(\alpha_1\dots \alpha_\ell|i_1\dots i_n)}$ & Dual symmetry-adapted mapping tensor, extracts $\mathcal{T}^n\to\mathcal{X}^\ell$ \\
      $\widehat{\mathbf Q}^p_{(\ell|n)}$   & $\widehat Q^p_{(\alpha_1\dots \alpha_\ell|i_1\dots i_n)}$   & Orthonormal symmetry-adapted mapping tensor, self-dual, $\mathcal{T}^n\leftrightarrow\mathcal{X}^\ell$ \\
      $\Hop_{(n|n)}$                       & $H_{(\alpha_1\dots \alpha_n|i_1\dots i_n)}$                 & Harmonic operator \\
      $\mathbf V_n$                        & $V_{i_1i_2\dots i_n}$                                      & Cartesian harmonic of weight $n$ \\
      $\Kop_{(\ell_3|\ell_1,\ell_2)}$      & $K_{(\gamma\dots|\alpha\dots,\beta\dots)}$                  & Cartesian coupling operator \\
      $\mathbf g$                          & $g_{pq}$                                                    & Gram matrix of the $\mathbf G^p_{(\ell|n)}$ \\
      $\mathcal S$                         &                                                             & Intrinsic symmetry class of $\mathbf T_n$ \\
      $\Pi_a$, $\eta_a$                    &                                                             & Index-permutation operator and sign of generator $a$ \\
      $N_\ell^{\mathrm c}$                                &                                                             & Number of candidate mappings at weight $\ell$, for a rank-$n$ tensor \\
      $N_\ell$                             &                                                             & Multiplicity at weight $\ell$: the number of independent ICTs, or channels \\
      $N_\ell^{\mathcal S}$                &                                                             & Multiplicity at weight $\ell$ restricted to the symmetry class $\mathcal S$ \\
      \hline
   \end{tabular}
\end{table*}

\begin{table*}[htb!]
   \caption{Operations and relations.}
   \label{tab:operations}
   \centering
   \small
   \begin{tabular}{cll}
      \hline
      Operation                & Example                                                                               & Explanation \\
      \hline
      $\otimes$                & $\mathbf T \otimes\mathbf S \rightarrow T_{i_1\dots i_n}S_{j_1\dots j_n}$          & Tensor product \\
      $\otimes^n$              & $\mathbf a^{\otimes n}\rightarrow a_{i_1}\cdots a_{i_n}$                              & $n$-fold tensor product \\
      $\odot^n$                & $\mathbf T \odot^n \mathbf S \rightarrow T_{i_1\dots i_n} S_{i_1\dots i_n}$ & $n$-fold contraction \\
      $\cong$                  &                                                                                       & Equality under the specified conditions \\
      $\avg{\cdot} $            &                                                                                       & Average over index permutations \\
      $\lfloor \cdot \rfloor $ &                                                                                       & Floor function \\
      \hline
   \end{tabular}
\end{table*}

\section{Counting rank-lowering tensors}
\label{app:F:count}

The distinct rank-lowering tensors $\mathbf F^p_{n\to \ell}$ correspond to the distinct ways of assigning the indices of $\mathbf T_n$ to the $\bm\delta$ and $\bm\epsilon$ factors in \eref{eq:F:even} and \eref{eq:F:odd}, and counting those assignments gives $N_\ell^{\mathrm c}$.
It is also the number of candidate mapping tensors, as summarized in \sref{sec:num:ind:nat:tensors}.

The number of ways to form $k$ disjoint unordered pairs from $N$ labeled indices is
\begin{equation}
   P(N,k)
   = \frac{1}{k!}\prod_{s=0}^{k-1}\binom{N-2s}{2}
   = \frac{N!}{(N-2k)!\,k!\,2^k}.
\end{equation}
The product chooses the pairs successively, and the factor $1/k!$ removes the ordering of the $k$ pairs.

For even $n-\ell$, the $\bm\delta$ factors of \eref{eq:F:even}, $k = (n-\ell)/2$ of them, draw their pairs from all $n$ indices, so
\begin{equation} \label{eq:Nc:even}
   N_\ell^{\mathrm c}
   = P(n,k)
   = \frac{n!}{\ell! \, \left( \frac{n-\ell}{2} \right)! \, 2^{(n-\ell)/2}} .
\end{equation}
Read as two choices in turn, the same count is $N_\ell^{\mathrm c}=\binom{n}{\ell}(n-\ell-1)!!$: which $\ell$ indices are left unpaired, these being the ones the rank-$\ell$ result keeps, and how the remaining $n-\ell$ are paired among themselves.
For $\mathbf F^p_{3\to 1}$ of \eref{eq:F:1:3} this is $3\times1=3$, the three choices of the contracted pair.
As another example, for $n=5$ and $\ell=1$ it is $5\times3=15$, and leaving $i_5$ unpaired gives the three
\begin{equation}
   \delta_{i_1 i_2}\delta_{i_3 i_4},\;
   \delta_{i_1 i_3}\delta_{i_2 i_4},\;
   \delta_{i_1 i_4}\delta_{i_2 i_3},
\end{equation}
the other twelve following from the four other choices of unpaired index.

For odd $n-\ell$ and $\ell>0$, $\mathbf F^p_{n\to \ell}$ of \eref{eq:F:odd} contains one $\bm\epsilon$ and $k=(n-\ell-1)/2$ of $\bm\delta$ factors.
The two indices of $\mathbf T_n$ contracted with $\bm\epsilon$ can be chosen in $\binom{n}{2}$ ways; exchanging them changes only the sign and does not produce a distinct candidate.
The $k$ unordered $\bm\delta$ pairs are then selected from the remaining $n-2$ indices in $P(n-2,k)$ ways.
Therefore,
\begin{equation} \label{eq:Nc:odd}
      N_\ell^{\mathrm c} = \binom{n}{2}P(n-2,k)
        = \frac{n!}{(\ell-1)! \, \left( \frac{n-\ell-1}{2} \right)! \, 2^{(n-\ell+1)/2}} .
\end{equation}
After these contractions, $\ell-1$ indices of $\mathbf T_n$ remain; together with the free index of $\bm\epsilon$, they form the $\ell$ indices of the rank-lowered tensor.

The $\bm\epsilon$ and $\bm\delta$ factors are not interchangeable.
For example, at $n=4$ and $\ell=1$ the candidates $F^p_{4\to 1} = \epsilon_{j i_3 i_4} \delta_{i_1 i_2}$ and $F^q_{4\to 1} = \epsilon_{j i_1 i_2} \delta_{i_3 i_4}$ use the same two pairs but assign them to different isotropic tensors, and are therefore distinct candidates.

The remaining case is $\ell=0$ with $n$ odd, where $\bm\epsilon$ contracts all three of its indices with $\mathbf T_n$ rather than supplying a free index:
\begin{equation}
   F^p_{n\to 0} = \epsilon_{i_w i_u i_v} \prod_{(a,b)\,\in\,\mathcal{D}_p} \delta_{i_a i_b} ,
\end{equation}
where $k = (n-3)/2$ is the number of $\bm\delta$ factors.
The rank of $\mathbf F^p_{n\to 0}$ is $3+2k=n$, and all $n$ indices are contracted with $\mathbf T_n$ in \eref{eq:rank:lowering}, leaving a scalar.

There are $\binom{n}{3}$ choices for the indices contracted with $\bm\epsilon$ and $P(n-3,k)$ pairings of the remaining indices.
Permuting the three $\bm\epsilon$ indices changes at most the sign and does not produce a distinct candidate.
Thus
\begin{equation}
   N_0^{\mathrm c}
   = \binom{n}{3} P(n-3,k)
   = \frac{n!}{6 \cdot \left( \frac{n-3}{2} \right)! \, 2^{(n-3)/2}} .
\end{equation}

For $n=3$ there is a single such tensor, the rank-3 $F^1_{3\to 0} = \epsilon_{i_1 i_2 i_3}$.
For $n=5$, $k=1$ and $N_0^{\mathrm c}=\binom{5}{3}=10$:
\begin{equation}
   \begin{gathered}
   F^{1}_{5\to 0} = \epsilon_{i_3 i_4 i_5}\delta_{i_1 i_2},\; F^{2}_{5\to 0} = \epsilon_{i_2 i_4 i_5}\delta_{i_1 i_3},\\
   F^{3}_{5\to 0} = \epsilon_{i_2 i_3 i_5}\delta_{i_1 i_4},\; F^{4}_{5\to 0} = \epsilon_{i_2 i_3 i_4}\delta_{i_1 i_5},\\
   F^{5}_{5\to 0} = \epsilon_{i_1 i_4 i_5}\delta_{i_2 i_3},\; F^{6}_{5\to 0} = \epsilon_{i_1 i_3 i_5}\delta_{i_2 i_4},\\
   F^{7}_{5\to 0} = \epsilon_{i_1 i_3 i_4}\delta_{i_2 i_5},\; F^{8}_{5\to 0} = \epsilon_{i_1 i_2 i_5}\delta_{i_3 i_4},\\
   F^{9}_{5\to 0} = \epsilon_{i_1 i_2 i_4}\delta_{i_3 i_5},\; F^{10}_{5\to 0} = \epsilon_{i_1 i_2 i_3}\delta_{i_4 i_5},
   \end{gathered}
\end{equation}
one for each choice of the three indices taken by $\bm\epsilon$, the remaining pair going to $\bm\delta$.

\section{Explicit form of the mapping tensors}
\label{app:G:index}

\eref{eq:G} writes the mapping tensor compactly as $\mathbf G^p_{(\ell|n)} = \mathbf E_{(\ell|\ell)}\mathbf F^p_{n\to \ell}$, leaving implicit which index of $\mathbf T_n$ meets which slot of the projector.
This appendix writes that out, in the two cases the rank drop allows, even and odd $n-\ell$.

When $n-\ell$ is even, let $i_{s_1},\dots,i_{s_\ell}$ be the indices of $\mathbf T_n$ that do not occur in the pairs $\mathcal D_p$ of \eref{eq:F:even}.
These $\ell$ indices occupy the $i$-index slots of $\mathbf E_{(\ell|\ell)}$, giving
\begin{equation} \label{eq:G:index}
   G^p_{(\ell|n)}
   =
   E_{(\alpha_1\dots\alpha_\ell|i_{s_1}\dots i_{s_\ell})}
   \prod_{(a,b)\,\in\,\mathcal{D}_p} \delta_{i_a i_b} .
\end{equation}
For example, take $n=3$ and $\ell=1$, where each $\mathcal D_p$ of \eref{eq:F:1:3} is a single pair and the one remaining index goes to the projector, $E_{(\alpha_1|i)}=\delta_{\alpha_1i}$.
The three choices give
\begin{equation} \label{eq:G:1:3}
   \begin{aligned}
      G^1_{(1|3)} & = E_{(\alpha_1|i_3)}\delta_{i_1i_2} = \delta_{\alpha_1i_3}\delta_{i_1i_2}, \\
      G^2_{(1|3)} & = E_{(\alpha_1|i_2)}\delta_{i_1i_3} = \delta_{\alpha_1i_2}\delta_{i_1i_3}, \\
      G^3_{(1|3)} & = E_{(\alpha_1|i_1)}\delta_{i_2i_3} = \delta_{\alpha_1i_1}\delta_{i_2i_3}.
   \end{aligned}
\end{equation}

When $n-\ell$ is odd and $\ell>0$, let $i_{s_1},\dots,i_{s_{\ell-1}}$ be the indices of $\mathbf T_n$ that occur in neither the pairs $\mathcal D_p$ nor the $\bm\epsilon$ pair $(i_u,i_v)$ of \eref{eq:F:odd}.
These $\ell-1$ indices occupy $\ell-1$ of the $i$-index slots of $\mathbf E_{(\ell|\ell)}$, while its remaining $i$-index slot is contracted with the free index $j$ of $\bm\epsilon$:
\begin{equation} \label{eq:G:index:odd}
   G^p_{(\ell|n)}
   =
   E_{(\alpha_1\dots\alpha_\ell|i_{s_1}\dots i_{s_{\ell-1}}j)}
   \epsilon_{j\, i_u i_v}
   \prod_{(a,b)\,\in\,\mathcal{D}_p} \delta_{i_a i_b} ,
\end{equation}
where $j$ is summed.
The three mapping tensors of \eref{eq:G:2:3}, for $n=3$ and $\ell=2$, are the three choices of the index retained by $\mathbf E_{(2|2)}$ in this form.
Substituting $\mathbf E_{(2|2)}$ from \tref{tab:rank2:summary} expands them into $\bm\delta$ and $\bm\epsilon$ alone,
\begin{equation} \label{eq:G:2:3:expanded}
   \begin{aligned}
      G^{1}_{(2|3)} & = \tfrac{1}{2}\left(\delta_{\alpha_1i_3}\epsilon_{\alpha_2i_1i_2}+\delta_{\alpha_2i_3}\epsilon_{\alpha_1i_1i_2}\right)-\tfrac{1}{3}\delta_{\alpha_1\alpha_2}\epsilon_{i_1i_2i_3}, \\
      G^{2}_{(2|3)} & = \tfrac{1}{2}\left(\delta_{\alpha_1i_2}\epsilon_{\alpha_2i_1i_3}+\delta_{\alpha_2i_2}\epsilon_{\alpha_1i_1i_3}\right)+\tfrac{1}{3}\delta_{\alpha_1\alpha_2}\epsilon_{i_1i_2i_3}, \\
      G^{3}_{(2|3)} & = \tfrac{1}{2}\left(\delta_{\alpha_1i_1}\epsilon_{\alpha_2i_2i_3}+\delta_{\alpha_2i_1}\epsilon_{\alpha_1i_2i_3}\right)-\tfrac{1}{3}\delta_{\alpha_1\alpha_2}\epsilon_{i_1i_2i_3},
   \end{aligned}
\end{equation}
the middle sign differing because $\epsilon_{i_2i_1i_3}=-\epsilon_{i_1i_2i_3}$ while $\epsilon_{i_3i_1i_2}=\epsilon_{i_1i_2i_3}$.
By construction in \eref{eq:natural:projector}, $\mathbf E_{(\ell|\ell)}$ is symmetric under every permutation of its $i$ indices, so the order in which its $i$-index slots receive the remaining indices and $j$ does not affect the mapping tensor.

In the remaining case, $n-\ell$ odd with $\ell=0$, the projector is $E_{(0|0)}=1$ and the mapping tensor is the rank-lowering tensor itself, $G^p_{(0|n)}=F^p_{n\to0}$, which \aref{app:F:count} writes out.

At rank two each weight has a single mapping tensor, and the two cases above reproduce the $G_{(\ell|n)}$ row of \tref{tab:rank2:summary}, the $\ell=2$ entry because a zero rank drop leaves $\mathcal D_p$ empty and \eref{eq:G:index} returns the projector alone.

\section{Gram matrix, duality, and orthonormality}
\label{app:gpq}

This appendix evaluates the Gram coefficient $g_{pq}$ of \eref{eq:gpq}, derives the duality relation \eref{eq:H:G:dual} and the orthonormality relation \eref{eq:G:orthonormal:property}, and gives the eigendecomposition used to compute $\mathbf g^{-1/2}$.

Taking the trace of both sides of \eref{eq:gpq} and rearranging gives
\begin{equation} \label{eq:gpq:trace}
   g_{pq}
   = \frac{\operatorname{tr}\bigl(\mathbf G^p_{(\ell|n)}\odot^{n}\mathbf G^q_{(\ell|n)}\bigr)}{\operatorname{tr}\mathbf E_{(\ell|\ell)}}
   = \frac{\mathbf G^p_{(\ell|n)}\odot^{\ell+n}\mathbf G^q_{(\ell|n)}}{2\ell+1}.
\end{equation}
In the second equality the trace of a tensor carrying two groups of $\ell$ ICT-space indices is the contraction of the first group with the second, which adds $\ell$ contractions to the $n$ of \eref{eq:gpq} and closes all $\ell+n$ indices of the two mapping tensors.
The value $\operatorname{tr}\mathbf E_{(\ell|\ell)}=2\ell+1$ comes from \sref{sec:nat:proj}.
Exchanging $p$ and $q$ leaves the numerator unchanged, so $g_{pq}=g_{qp}$ and $\mathbf g=[g_{pq}]$ is symmetric.

The duality relation \eref{eq:H:G:dual} follows directly.
Using \eref{eq:H} and \eref{eq:gpq},
\begin{equation}
   \begin{aligned}
      \widetilde{\mathbf G}^p_{(\ell|n)} \odot^n \mathbf G^q_{(\ell|n)}
       & = \sum_r (\mathbf g^{-1})_{pr}
      \left(\mathbf G^r_{(\ell|n)} \odot^n \mathbf G^q_{(\ell|n)}\right) \\
       & = \sum_r (\mathbf g^{-1})_{pr}g_{rq}\mathbf E_{(\ell|\ell)}     \\
       & = (\mathbf g^{-1}\mathbf g)_{pq}\mathbf E_{(\ell|\ell)}         \\
       & = \delta_{pq}\mathbf E_{(\ell|\ell)}.
   \end{aligned}
\end{equation}

The inverse square root used in \eref{eq:G:orthonormal} is obtained from the eigendecomposition
\begin{equation}
   \mathbf g=\mathbf W\bm\Lambda\mathbf W^{\mathsf T},
   \quad
   \bm\Lambda=\operatorname{diag}(\lambda_1,\dots,\lambda_{N_\ell}),
\end{equation}
where $\mathbf W$ is orthogonal and all eigenvalues $\lambda_p$ are positive.
The symmetric positive-definite inverse square root is then
\begin{equation}
   \begin{aligned}
      \mathbf g^{-1/2}  & = \mathbf W\bm\Lambda^{-1/2}\mathbf W^{\mathsf T},                  \\
      \bm\Lambda^{-1/2} & = \operatorname{diag}(\lambda_1^{-1/2},\dots,\lambda_{N_\ell}^{-1/2}).
   \end{aligned}
\end{equation}

The orthonormality relation \eref{eq:G:orthonormal:property} follows by substituting \eref{eq:G:orthonormal} and applying \eref{eq:gpq}:
\begin{equation}
   \begin{aligned}
      & \widehat{\mathbf G}^{p}_{(\ell|n)} \odot^n
      \widehat{\mathbf G}^{q}_{(\ell|n)}                              \\
       & = \sum_{r,s}
      (\mathbf g^{-1/2})_{pr}(\mathbf g^{-1/2})_{qs}
      \left(\mathbf G^r_{(\ell|n)} \odot^n \mathbf G^s_{(\ell|n)}\right) \\
       & = \sum_{r,s}
      (\mathbf g^{-1/2})_{pr}g_{rs}(\mathbf g^{-1/2})_{qs}
      \mathbf E_{(\ell|\ell)}                                            \\
       & = (\mathbf g^{-1/2}\mathbf g\mathbf g^{-1/2})_{pq}
      \mathbf E_{(\ell|\ell)}                                            \\
       & = \delta_{pq}\mathbf E_{(\ell|\ell)}.
   \end{aligned}
\end{equation}
Here the third equality uses the symmetry of $\mathbf g^{-1/2}$, and the last uses $\mathbf g^{-1/2}\mathbf g\mathbf g^{-1/2}=\mathbf I$.

\section{Mixing matrices and symmetry constraints}
\label{app:M:mixing}

This appendix derives \eref{eq:M:mixing} for the mixing coefficients $M^a_{pq}$ of \sref{sec:arbitrary:symmetry}, the constraint \eref{eq:symmetry:coefficient:constraint} that one generator imposes, and the stacked system for all $N_{\Pi}$ generators.
As in the main text, a fixed weight $\ell$ is understood throughout and is suppressed in the notation for $\mathbf M^a$.

To extract the $M^a_{pq}$ of \eref{eq:G:symmetry:action}, contract that equation over the $n$ Cartesian indices with the dual mapping tensor $\widetilde{\mathbf G}^p_{(\ell|n)}$ and use the duality relation \eref{eq:H:G:dual},
\begin{equation}
   \widetilde{\mathbf G}^p_{(\ell|n)}\odot^n\bigl(\Pi_a\mathbf G^q_{(\ell|n)}\bigr)
   =\sum_{r=1}^{N_\ell}M^a_{rq}\,\delta_{pr}\mathbf E_{(\ell|\ell)}
   =M^a_{pq}\mathbf E_{(\ell|\ell)}.
\end{equation}
The $\delta_{pr}$ that \eref{eq:H:G:dual} supplies collapses the sum to its $r=p$ term.
Taking the trace of both sides as in \eref{eq:gpq:trace} and dividing by $\operatorname{tr}\mathbf E_{(\ell|\ell)}=2\ell+1$ gives \eref{eq:M:mixing}.

Write $\mathbf Q=\sum_{q}c_q\mathbf G^q_{(\ell|n)}$ for a general combination of the mapping tensors, as in \eref{eq:Q:from:symmetry:solutions}, and impose on it the requirement $\Pi_a\mathbf Q=\eta_a\mathbf Q$ that $\mathbf Q$ transform as $\mathbf T_n$ does in \eref{eq:intrinsic:symmetry:generator}.
On the left $\Pi_a$ acts term by term, and \eref{eq:G:symmetry:action} replaces each $\Pi_a\mathbf G^q_{(\ell|n)}$ by $\sum_p M^a_{pq}\mathbf G^p_{(\ell|n)}$, so
\begin{equation}
   \Pi_a\left(\sum_{q=1}^{N_\ell}c_q\mathbf G^q_{(\ell|n)}\right)
   =\sum_{p=1}^{N_\ell}\bigl(\mathbf M^a\mathbf c\bigr)_p\,\mathbf G^p_{(\ell|n)},
\end{equation}
while the right-hand side is simply $\eta_a\sum_p c_p\mathbf G^p_{(\ell|n)}$.
Equating the two and collecting the coefficient of each $\mathbf G^p_{(\ell|n)}$ gives
\begin{equation}
   \sum_{p=1}^{N_\ell}
   \left[\bigl(\mathbf M^a\mathbf c\bigr)_p-\eta_a c_p\right]
   \mathbf G^p_{(\ell|n)}=\mathbf 0.
\end{equation}
The $\mathbf G^p_{(\ell|n)}$ are independent as mappings on a general tensor, by the selection of \sref{sec:independent:orthonormal}, so each bracket must vanish separately.
Thus $(\mathbf M^a-\eta_a\mathbf I)\mathbf c=\mathbf0$; taking $\mathbf c=\mathbf c^p$ gives \eref{eq:symmetry:coefficient:constraint} for each symmetry-adapted mapping.

Imposing all $N_{\Pi}$ generators at once gives the stacked system
\begin{equation}
   \mathbf A^{(\ell)}\mathbf c=\mathbf 0,
   \quad
   \mathbf A^{(\ell)}
   =\begin{bmatrix}
      \mathbf M^1-\eta_1\mathbf I \\
      \vdots                      \\
      \mathbf M^{N_{\Pi}}-\eta_{N_{\Pi}}\mathbf I
   \end{bmatrix}.
\end{equation}
The admissible coefficient vectors are therefore the null space of $\mathbf A^{(\ell)}$, and the symmetry-restricted multiplicity is
\begin{equation}
   N_\ell^{\mathcal S}=N_\ell-\operatorname{rank}\mathbf A^{(\ell)}.
\end{equation}
The entries of $\mathbf G^p_{(\ell|n)}$, $\widetilde{\mathbf G}^p_{(\ell|n)}$, and hence $\mathbf M^a$ and $\mathbf A^{(\ell)}$, are rational, so the null space is obtained exactly by Gaussian elimination over the rationals.
Its $N_\ell^{\mathcal S}$ basis vectors define the symmetry-adapted mappings through \eref{eq:Q:from:symmetry:solutions}; any other basis of the same null space gives the same mapping subspace.

If $\mathbf A^{(\ell)}$ has full column rank, the only solution is $\mathbf c=\mathbf 0$ and weight $\ell$ is absent from the reduction spectrum of the symmetry class.
If every $\mathbf M^a=\eta_a\mathbf I$, then $\mathbf A^{(\ell)}=\mathbf 0$ and the symmetry imposes no restriction at that weight, so $N_\ell^{\mathcal S}=N_\ell$.

\section{Projector for a polyadic}
\label{app:E:symmetric}

This appendix gives the steps leading from \eref{eq:natural:projector} to \eref{eq:E:symmetric} for a polyadic $\mathbf a^{\otimes n}$, and writes out the harmonic operators and the harmonics they give at low rank.
Only full symmetry is used, so the result holds for any fully symmetric $\mathbf U_n$, and the length of $\mathbf a$ is left free; \sref{sec:nt:unit:vector} sets $a=1$.
Write $\Delta_t$ for the object inside the angle brackets of \eref{eq:natural:projector}, taken at $\ell=n$ since the weight extracted here equals the rank, before either average is applied:
\begin{equation}
   \begin{aligned}
      \Delta_t
       & = \delta_{\alpha i}^{\,n-2t}\delta_{\alpha\alpha}^{t}\delta_{ii}^{t}    \\
       & = \prod_{k=1}^{n-2t}\delta_{\alpha_k i_k}
      \prod_{k=1}^{t}\delta_{\alpha_{n-2k+1}\alpha_{n-2k+2}}\,
      \delta_{i_{n-2k+1}i_{n-2k+2}}.
   \end{aligned}
\end{equation}
The first product carries $n-2t$ indices from Roman to Greek, and the second contracts the remaining $2t$ Greek indices in pairs and the remaining $2t$ Roman indices in the matching pairs.
The average in \eref{eq:natural:projector} runs over the $(n!)^2$ pairs of Greek and Roman permutations of $\Delta_t$.

Nothing so far restricts the argument, \eref{eq:natural:projector} holding for any rank-$n$ tensor.
Contracting with $\mathbf a^{\otimes n}$ is what makes the Roman average redundant.
Permuting the Roman indices of $\Delta_t$ only changes which index of $\mathbf a^{\otimes n}$ each $\bm\delta$ meets.
Since $\mathbf a^{\otimes n}$ is unchanged by any permutation of its indices, all $n!$ Roman permutations contribute equally, and their average is any one of them.
Fixing the Roman indices in the order written above leaves an average over the $n!$ Greek permutations alone, and the $t$ factors $\delta_{ii}$ carry no Greek index and so stand outside it, giving \eref{eq:E:symmetric}.
Rescaling it by $C=(2n-1)!!/n!$ forms the harmonic operator of \eref{eq:H:unit:vector}.

Examples at low rank make the operator explicit.
The coefficients $c_t$ are those of \eref{eq:ct}, which depend on $n$ as well as $t$ and give $c_0=1$ at every $n$.
For $n=0$ and $n=1$ only the $t=0$ term is present and $C=1$, so $\Hop_{(0|0)}=1$ and $\Hop_{(1|1)}=\delta_{\alpha_1i_1}$.
The first two nontrivial cases are
\begin{equation}
   \begin{aligned}
      \Hop_{(2|2)}
       & = \tfrac{3}{2}\left(\avg{\delta_{\alpha i}^2}
      -\tfrac{1}{3}\delta_{\alpha_1\alpha_2}\delta_{i_1i_2}\right)                     \\
       & \cong \tfrac{3}{2}\left(\delta_{\alpha_1i_1}\delta_{\alpha_2i_2}
      -\tfrac{1}{3}\delta_{\alpha_1\alpha_2}\delta_{i_1i_2}\right)
   \end{aligned}
\end{equation}
for $n=2$, with $C=3/2$ and $c_1=-1/3$, and
\begin{equation}
   \begin{aligned}
      \Hop_{(3|3)}
       & = \tfrac{5}{2}\left(\avg{\delta_{\alpha i}^3}
      -\tfrac{3}{5}\avg{\delta_{\alpha i}\delta_{\alpha\alpha}}\delta_{i_2i_3}\right)  \\
       & \cong \tfrac{5}{2}\Bigl(\delta_{\alpha_1i_1}\delta_{\alpha_2i_2}\delta_{\alpha_3i_3} \\
       & \quad -\tfrac{1}{5}\bigl(\delta_{\alpha_1i_1}\delta_{\alpha_2\alpha_3}
      +\delta_{\alpha_2i_1}\delta_{\alpha_1\alpha_3}
      +\delta_{\alpha_3i_1}\delta_{\alpha_1\alpha_2}\bigr)\delta_{i_2i_3}\Bigr)
   \end{aligned}
\end{equation}
for $n=3$, with $C=5/2$ and $c_1=-3/5$, the $\tfrac{1}{3}$ of the three-term average combining with $c_1$ to give the $\tfrac{1}{5}$ shown.
The second line of each drops the Greek average on the $t=0$ term, which a single representative product replaces once the argument is symmetric, hence the $\cong$ of \eref{eq:E:symmetric}; the $t\geq1$ terms keep theirs, the result being symmetric in the $\alpha$ only after the average is taken.

Contracting these operators with $\mathbf a^{\otimes n}$ sends each $\delta_{\alpha i}$ to $a_\alpha$ and each $\delta_{ii}$ to $a^2=\mathbf a\cdot\mathbf a$, so they return the Cartesian harmonics, which are $V(\mathbf a)=1$ and $V_i(\mathbf a)=a_i$ at $n=0$ and $n=1$, and
\begin{equation}
   \begin{aligned}
      V_{ij}(\mathbf a)  & = \tfrac{3}{2}\left(a_i a_j - \tfrac{1}{3}a^2\delta_{ij}\right), \\
      V_{ijk}(\mathbf a) & = \tfrac{5}{2}\left(a_i a_j a_k - \tfrac{1}{5}a^2\bigl(\delta_{ij}a_k+\delta_{jk}a_i+\delta_{ki}a_j\bigr)\right),
   \end{aligned}
\end{equation}
one factor of $a^2$ for each $\delta_{ii}$.
The entries are homogeneous polynomials of degree two and three, each satisfying Laplace's equation.

Writing $a_i=a\hat a_i$ pulls a factor $a^n$ out of every term,
\begin{equation}
   V_{ij}(\mathbf a)
   = \tfrac{3}{2}\left(a^2\,\hat a_i\hat a_j - \tfrac{1}{3}a^2\delta_{ij}\right)
   = a^2\,V_{ij}(\uv{a}),
\end{equation}
and likewise for $V_{ijk}$, leaving the unit-vector harmonics of \sref{sec:nt:unit:vector} and the scaling $\mathbf V_n(\mathbf a)=a^n\mathbf V_n(\uv{a})$ stated there.

\section{Normalization of the harmonic and coupling operators}
\label{app:normalization}

Contracting \eref{eq:E:symmetric} with $\uv{a}^{\otimes n}$ sends each $\delta_{ii}$ to unity and each $\delta_{\alpha i}$ to $\hat a_\alpha$.
Contracting the result with a second unit vector $\uv{b}$ replaces each $\hat a_\alpha$ by $u=\uv{a}\cdot\uv{b}$ and each $\delta_{\alpha\alpha}$ by unity, so every term of the average reduces to $u^{\,n-2t}$,
\begin{equation}
   \mathbf V_n \odot^{n}\uv{b}^{\otimes n}
   = C\sum_{t=0}^{\lfloor n/2 \rfloor} c_t\;u^{\,n-2t}
   = C\,\frac{n!}{(2n-1)!!}\,P_n(u).
\end{equation}
The second equality follows on writing the coefficients $c_t$ of \eref{eq:ct} in double-factorial form,
\begin{equation}
   c_t = (-1)^t\,\frac{(2n-2t-1)!!}{(2n-1)!!}\,\frac{n!}{2^t\,t!\,(n-2t)!},
\end{equation}
since $(-1)^t(2n-2t-1)!!/[2^t\,t!\,(n-2t)!]$ is the coefficient of $u^{\,n-2t}$ in the Legendre polynomial $P_n$, leaving the common factor $n!/(2n-1)!!$.
The projector therefore delivers $P_n$ up to a single overall factor, and $C=(2n-1)!!/n!$ removes it, giving \eref{eq:legendre:normalization} and, at $\uv{b}=\uv{a}$, the unit normalization $\mathbf V_n\odot^{n}\uv{a}^{\otimes n}=P_n(1)=1$.

The coupling constants of \sref{sec:tp} likewise rest on the conditions of \olcite{lehman1989angular}.
The constants of \eref{eq:C:constant} and \eref{eq:D:constant} are fixed by two conditions on the ICT $\mathbf Z_{\ell_3}$ that \eref{eq:tp:even:H} and \eref{eq:tp:odd:H} produce from $\mathbf X_{\ell_1}$ and $\mathbf Y_{\ell_2}$.
$C$ is fixed for even $L$ by requiring that when $\mathbf X_{\ell_1}$ and $\mathbf Y_{\ell_2}$ are constructed from a unit vector $\uv{a}$ as in \eref{eq:H:unit:vector},
$\mathbf Z_{\ell_3}$ obtained from \eref{eq:tp:even:H} is the same as the ICT constructed from the same unit vector $\uv{a}$ using \eref{eq:H:unit:vector}.
The second condition fixes $C$ for odd $L$: when $\mathbf X_{\ell_1}$ is constructed from a unit vector $\uv{a}$ and $\mathbf Y_{\ell_2}$ from another unit vector $\uv{b}$,
\begin{equation}
   \lim_{\uv{b}\to\uv{a}}  \frac{ | \mathbf Z_{\ell_3} \odot^{\ell_3-1} \uv{a}^{\otimes \ell_3-1} |} {|\uv{a}\times\uv{b}|} = 1,
\end{equation}
where $|\cdot|$ denotes the norm of a vector.

%\bibliography{main.bib}
%aipnum4-2.bst 2019-01-14 (MD) hand-edited version of apsrev4-1.bst

\end{document}

\title{Supplementary material for:
   Reusable Operators for Irreducible Cartesian Tensor Decomposition and Coupling}

\author{Mingjian Wen}
\email{mjwen@uestc.edu.cn}
\affiliation{Institute of Fundamental and Frontier Sciences, University of Electronic Science and Technology of China, Chengdu, 611731, China}
\affiliation{Frontier Center of Energy Distribution and Integration, Tianfu Jiangxi Laboratory, Chengdu, 641419, China}

\date{\today}

\maketitle

\makeatletter
\renewcommand*{\l@subsubsection}[2]{}
\renewcommand*{\l@f@section}{\addpenalty{\@secpenalty}\addvspace{0.05em plus\p@}\bf}
{\footnotesize\linespread{0.85}\selectfont\@starttoc{toc}}
\makeatother
\clearpage

Equation, section, and table numbers prefixed with ``S'' refer to this document;
all other cross-references point to the main text.

\section{Independent components of an irreducible Cartesian tensor}
\label{si:ict:count}

A rank-$n$ ICT has $2n+1$ independent components, the count quoted in Sec.~II\,A\maintag{}.
The two conditions of the definition give it in turn.

Take symmetry first.
A symmetric component is unchanged by any reordering of its indices, so all that distinguishes one component from another is how many of the $n$ indices are $x$, how many are $y$, and how many are $z$.
Each component therefore corresponds to a triple $(a,b,c)$ of non-negative integers with $a+b+c=n$, and counting components means counting triples.
Choosing $a$ leaves $b$ anywhere between $0$ and $n-a$, with $c=n-a-b$ then fixed, so each $a$ admits $n-a+1$ triples and the total is
\begin{equation} \label{eq:si:count:sym}
   \sum_{a=0}^{n}(n-a+1)
   = (n+1)+n+\dots+1
   = \frac{(n+1)(n+2)}{2} .
\end{equation}
At $n=2$ this gives the six components $T_{xx}$, $T_{yy}$, $T_{zz}$, $T_{xy}$, $T_{xz}$ and $T_{yz}$ of a symmetric matrix.

Tracelessness next.
Contracting any pair of indices of a symmetric tensor returns the same symmetric tensor of rank $n-2$, so demanding that the trace vanish is one condition for each component of that tensor.
Their number is \eref{eq:si:count:sym} with $n$ replaced by $n-2$,
\begin{equation} \label{eq:si:count:trace}
   \frac{(n-2+1)(n-2+2)}{2} = \frac{n(n-1)}{2} .
\end{equation}
None of these conditions follows from the others, because every symmetric rank-$(n-2)$ tensor arises as the trace of some symmetric rank-$n$ tensor, so the trace may be set to zero one component at a time.

Subtracting the conditions from the components leaves
\begin{equation} \label{eq:si:count:ict}
   \frac{(n+1)(n+2)}{2}-\frac{n(n-1)}{2}
   = \frac{4n+2}{2}
   = 2n+1 ,
\end{equation}
far fewer than the $3^n$ components of a generic rank-$n$ tensor.

\section{Isotropic tensor identities}
\label{si:iso}

Every operator in this work is built from two isotropic tensors, the Kronecker delta $\bm\delta$ and the Levi--Civita symbol $\bm\epsilon$.
This section records what makes them isotropic and collects the contractions used throughout.

For any orthogonal $\mathbf R\in\mathrm O(3)$, contracting the two with copies of $\mathbf R$ gives the identities~\cite{horn2013matrix,gurtin2010mechanics}
\begin{equation} \label{eq:detid}
   R_{ia}R_{jb}\,\delta_{ab} = \delta_{ij},
   \qquad
   R_{ia}R_{jb}R_{kc}\,\epsilon_{abc} = \det(\mathbf R)\,\epsilon_{ijk} .
\end{equation}
Both are therefore invariant under $\mathrm{SO}(3)$, which is what makes them isotropic.
The array $\epsilon_{ijk}$ itself has the same components in every coordinate system, including after inversion.
What \eref{eq:detid} records is the result of contracting $\bm\epsilon$ with three copies of $\mathbf R$: it gives $\det(\mathbf R)\,\epsilon_{ijk}$, which is $+\epsilon_{ijk}$ for a rotation and $-\epsilon_{ijk}$ for an improper $\mathbf R$.
Contracting $\bm\delta$ with two copies of $\mathbf R$, by contrast, gives back $\delta_{ij}$ for every $\mathbf R$.
This is why contracting polar tensors with one $\bm\epsilon$ gives axial tensors.
For example, take the cross product $c_i=\epsilon_{ijk}a_jb_k$ of two polar vectors, which transform as $a'_j=R_{jm}a_m$ and $b'_k=R_{kn}b_n$.
Since $\epsilon$ is unchanged,
\begin{equation} \label{eq:cross:product}
   c'_i=\epsilon_{ijk}\,R_{jm}R_{kn}\,a_mb_n
   =\det(\mathbf R)\,R_{ip}\,\epsilon_{pmn}\,a_mb_n
   =\det(\mathbf R)\,R_{ip}\,c_p,
\end{equation}
where the second equality follows from \eref{eq:detid} by contracting both sides with $R_{jm}R_{kn}$ and using the orthogonality $R_{jm'}R_{jm}=\delta_{m'm}$ together with $\det(\mathbf R)^2=1$.
A polar vector would transform as $c'_i=R_{ip}c_p$, without the factor $\det(\mathbf R)$, so $\mathbf c$ is an axial vector.
For inversion, $\mathbf R=-\mathbf I$ and $\det(\mathbf R)=-1$, so $\mathbf c$ is unchanged whereas a polar vector would change sign.

Products of two Levi--Civita symbols reduce to Kronecker deltas,
\begin{equation} \label{eq:ee:1}
   \epsilon_{ijk}\epsilon_{pqr}
   = \delta_{ip}(\delta_{jq}\delta_{kr} - \delta_{jr}\delta_{kq})
   + \delta_{iq}(\delta_{jr}\delta_{kp} - \delta_{jp}\delta_{kr})
   + \delta_{ir}(\delta_{jp}\delta_{kq} - \delta_{jq}\delta_{kp}),
\end{equation}
so the identity removes Levi--Civita symbols in pairs: a product of an even number of them is a sum of products of $\bm\delta$, and a product of an odd number reduces to a single $\bm\epsilon$ times such a sum.
No operator built from the two therefore carries more than one $\bm\epsilon$, which is why the rank-lowering tensors of Eq.~(3)\maintag{} and Eq.~(5)\maintag{} have either none or one.
Contracting over shared indices gives
\begin{equation} \label{eq:ee:2}
   \epsilon_{ijk}\epsilon_{pqk} =
   \delta_{ip}\delta_{jq}  - \delta_{iq}\delta_{jp},
\end{equation}
\begin{equation} \label{eq:ee:3}
   \epsilon_{ijk}\epsilon_{pjk} = 2 \delta_{ip} .
\end{equation}

\section{Counting the distinct symmetrized products}
\label{si:symmetrization}

Every operator in this work carries an average over index permutations, so evaluating one means counting the distinct terms it contains.
This section does that counting.

The average $\avg{X}$ runs over the distinct arrays obtained by permuting the free indices of $X$.
Writing $\{X\}$ for their sum, the convention of \olcite{lehman1989angular}, the two differ only by that count $N$,
\begin{equation} \label{eq:si:avg:brace}
   \avg{X} = \frac{1}{N}\{X\},
   \qquad
   N = \frac{n!}{|\mathcal H|},
\end{equation}
where $\mathcal H\subset S_n$ is the subgroup of permutations that leave $X$ unchanged and $|\mathcal H|$ is its order, the number of such permutations.
The work is therefore the same in both conventions: identify $\mathcal H$, count its elements, and divide.
The subsections below carry this out, with worked examples, for the three products that occur in this work: those of the general reduction, of the harmonic operators, and of the coupling operators.
All three are products of Kronecker deltas with at most one Levi--Civita symbol.
\sref{si:brace} uses these counts to convert the operators of Sec.~VI\maintag{} and Sec.~VII\maintag{} into the brace form.

\subsection{General reduction}

The products symmetrized in the natural projector of Eq.~(7)\maintag{}, and hence in every mapping tensor built from it, are products of Kronecker deltas.
Permutations of the free indices coincide whenever the product is left unchanged, and for a product of Kronecker deltas there are three sources of that:
\begin{enumerate}[label=(\roman*)]
   \item the two indices of a single delta may be swapped, $\delta_{i_1i_2}=\delta_{i_2i_1}$;
   \item two deltas of the same kind may be exchanged, $\delta_{i_1i_2}\delta_{i_3i_4}=\delta_{i_3i_4}\delta_{i_1i_2}$;
   \item two mixed deltas may likewise be exchanged, $\delta_{\alpha_1i_1}\delta_{\alpha_2i_2}=\delta_{\alpha_2i_2}\delta_{\alpha_1i_1}$, which moves a Greek and a Roman index together.
\end{enumerate}
Each source divides the permutations by the size of the corresponding subgroup, and the counts below are that division carried out.

A product of $t$ deltas of one kind carries $2t$ indices and
\begin{equation} \label{eq:si:count:delta}
   N = \frac{(2t)!}{2^{t}\,t!}
\end{equation}
distinct terms, the $2^{t}$ coming from (i) and the $t!$ from (ii).
For example, $t=2$ gives three terms,
\begin{equation} \label{eq:si:avg:dd}
   \avg{\delta_{ii}^{2}}
   = \tfrac{1}{3}\left(
   \delta_{i_1i_2}\delta_{i_3i_4}
   + \delta_{i_1i_3}\delta_{i_2i_4}
   + \delta_{i_1i_4}\delta_{i_2i_3}
   \right),
\end{equation}
the remaining permutations reproducing these three.

The product averaged in Eq.~(7)\maintag{} is
\begin{equation} \label{eq:si:avg:E}
   \avg{\delta_{\alpha i}^{\,\ell-2t}\delta_{\alpha\alpha}^{t}\delta_{ii}^{t}},
\end{equation}
which carries all three kinds of delta, and whose permutations act on the $\ell$ Greek and the $\ell$ Roman indices separately.
Sources (i) and (ii) act on the $t$ Greek deltas and on the $t$ Roman ones, and (iii) on the $\ell-2t$ mixed deltas, so the subgroup has order $(2^{t}t!)^2(\ell-2t)!$ and
\begin{equation} \label{eq:si:count:E}
   N = \frac{(\ell!)^2}{(2^{t}\,t!)^2\,(\ell-2t)!} ,
\end{equation}
distinct terms.

For example, at $\ell=1$ only $t=0$ occurs and the average is a single term, $\avg{\delta_{\alpha i}}=\delta_{\alpha_1i_1}$.

At $\ell=2$ the two cases are
\begin{equation} \label{eq:si:avg:E:ell2}
   \avg{\delta_{\alpha i}^{2}}
   = \tfrac{1}{2}\left(
   \delta_{\alpha_1i_1}\delta_{\alpha_2i_2}
   + \delta_{\alpha_1i_2}\delta_{\alpha_2i_1}
   \right),
   \qquad
   \avg{\delta_{\alpha\alpha}\delta_{ii}}
   = \delta_{\alpha_1\alpha_2}\delta_{i_1i_2},
\end{equation}
the second a single term because every permutation leaves it unchanged.

At $\ell=3$ the two cases have $N=6$,
\begin{equation} \label{eq:si:avg:E:ell3:t0}
   \begin{aligned}
        \avg{\delta_{\alpha i}^3}
       & = \tfrac{1}{6}( \delta_{\alpha_1i_1} \delta_{\alpha_2i_2} \delta_{\alpha_3i_3}
      + \delta_{\alpha_1i_1} \delta_{\alpha_2i_3} \delta_{\alpha_3i_2}
      + \delta_{\alpha_1i_2} \delta_{\alpha_2i_1} \delta_{\alpha_3i_3}               \\
       & \quad + \delta_{\alpha_1i_3} \delta_{\alpha_2i_1} \delta_{\alpha_3i_2}
      + \delta_{\alpha_1i_2} \delta_{\alpha_2i_3} \delta_{\alpha_3i_1}
      + \delta_{\alpha_1i_3} \delta_{\alpha_2i_2} \delta_{\alpha_3i_1}),
   \end{aligned}
\end{equation}
and $N=9$,
\begin{equation} \label{eq:si:avg:E:ell3:t1}
   \begin{aligned}
        \avg{\delta_{\alpha i}\delta_{\alpha\alpha}\delta_{ii}}
        &= \tfrac{1}{9} (\delta_{\alpha_1i_1} \delta_{\alpha_2\alpha_3} \delta_{i_2i_3}
      + \delta_{\alpha_1i_2} \delta_{\alpha_2\alpha_3} \delta_{i_1i_3}
      + \delta_{\alpha_1i_3} \delta_{\alpha_2\alpha_3} \delta_{i_1i_2}               \\
       & \quad + \delta_{\alpha_2i_1} \delta_{\alpha_1\alpha_3} \delta_{i_2i_3}
      + \delta_{\alpha_2i_2} \delta_{\alpha_1\alpha_3} \delta_{i_1i_3}
      + \delta_{\alpha_2i_3} \delta_{\alpha_1\alpha_3} \delta_{i_1i_2}               \\
       &\quad + \delta_{\alpha_3i_1} \delta_{\alpha_1\alpha_2} \delta_{i_2i_3}
      + \delta_{\alpha_3i_2} \delta_{\alpha_1\alpha_2} \delta_{i_1i_3}
      + \delta_{\alpha_3i_3} \delta_{\alpha_1\alpha_2} \delta_{i_1i_2}).
   \end{aligned}
\end{equation}

Substituting these averages into Eq.~(7)\maintag{} gives $\mathbf E_{(2|2)}$ and $\mathbf E_{(3|3)}$, assembled in \eref{eq:si:E:ell2} and \eref{eq:si:E:ell3}.

\subsection{Harmonic operators}

The average in Eq.~(46)\maintag{} is
\begin{equation} \label{eq:si:avg:uv}
   \avg{\delta_{\alpha i}^{\,n-2t}\delta_{\alpha\alpha}^{t}},
\end{equation}
the factor $\delta_{ii}^{t}$ standing outside it.
It carries $n$ Greek indices, $2t$ of them on the $t$ deltas $\delta_{\alpha\alpha}$ and $n-2t$ on the deltas $\delta_{\alpha i}$.
Both index groups are permuted in Eq.~(7)\maintag{}, which is why $(\ell!)^2$ appears in \eref{eq:si:count:E}.
Here the Roman indices meet a symmetric argument, by the convention of Eq.~(45)\maintag{}, so permuting them produces nothing new and only the $n!$ Greek permutations are counted.
Sources (i) and (ii) act on the $t$ factors $\delta_{\alpha\alpha}$ and (iii) on the $n-2t$ factors $\delta_{\alpha i}$, giving
\begin{equation} \label{eq:si:count:delta:sym}
   N = \frac{n!}{2^{t}\,t!\,(n-2t)!} ,
\end{equation}
the three factors in the denominator coming from (i), (ii), and (iii) in turn, and \eref{eq:si:count:delta} is the special case $n=2t$ in which no index is left over.

For example, at $t=0$ the count is $N=1$ for every $n$: with the Roman indices equivalent, all $n!$ Greek permutations give the same array, and $\avg{\delta_{\alpha i}^{\,n}}=\delta_{\alpha_1i_1}\delta_{\alpha_2i_2}\cdots\delta_{\alpha_ni_n}$.

For $n=3$ and $t=1$ it gives $N=3$,
\begin{equation} \label{eq:si:avg:H:n3}
   \avg{\delta_{\alpha i}\delta_{\alpha\alpha}}
   = \tfrac{1}{3}\left(\delta_{\alpha_1i_1}\delta_{\alpha_2\alpha_3}
   + \delta_{\alpha_2i_1}\delta_{\alpha_1\alpha_3}
   + \delta_{\alpha_3i_1}\delta_{\alpha_1\alpha_2}\right).
\end{equation}
The Roman index carried by $\delta_{\alpha i}$ has been taken to be $i_1$ here, the others sitting in the $\delta_{ii}$ outside the average.
Any other choice differs by a permutation of the Roman indices alone, which acts identically on a symmetric argument, as the $\cong$ of Eq.~(45)\maintag{} records.

\subsection{Coupling operators}

The average in Eq.~(50)\maintag{} is
\begin{equation} \label{eq:si:avg:tp}
   \avg{\delta_{\alpha\gamma}^{\,L_2-t}
   \delta_{\beta\gamma}^{\,L_1-t}
   \delta_{\gamma\gamma}^{\,t}},
\end{equation}
the factor $\delta_{\alpha\beta}^{\,L_3+t}$ standing outside it.
Here $\alpha$ labels the $\ell_1$ indices of $\mathbf X_{\ell_1}$, $\beta$ the $\ell_2$ indices of $\mathbf Y_{\ell_2}$, and $\gamma$ the $\ell_3$ indices of the coupled ICT; the triangle numbers $L_1$ and $L_2$ of Sec.~VII\maintag{} count the contractions that $\mathbf Y_{\ell_2}$ and $\mathbf X_{\ell_1}$ make with those $\ell_3$ indices, and $t$ counts the traces taken among them.
The three kinds of factor thus use up all $\ell_3$ indices $\gamma$, $L_2-t$ of them on the deltas $\delta_{\alpha\gamma}$, $L_1-t$ on the $\delta_{\beta\gamma}$ and $2t$ on the $t$ deltas $\delta_{\gamma\gamma}$, so that $L_1+L_2=\ell_3$ when $L$ is even.
As in \eref{eq:si:avg:uv}, and by the convention of Eq.~(45)\maintag{}, only the $\gamma$ are permuted.

Permuting the $\alpha$ indices among themselves leaves the average unchanged, because they are contracted with the symmetric $\mathbf X_{\ell_1}$, and the same holds for the $\beta$ with $\mathbf Y_{\ell_2}$.
Which $\gamma$ sits on which $\delta_{\alpha\gamma}$ is therefore immaterial, and only the split of the $\gamma$ among the three kinds of factor matters.
Counting the splits in three steps,
\begin{enumerate}[label=(\arabic*)]
   \item choose $L_2-t$ of the $\ell_3$ indices for the $\delta_{\alpha\gamma}$, in $\binom{\ell_3}{L_2-t}$ ways;
   \item choose $L_1-t$ of the remaining $\ell_3-L_2+t$ for the $\delta_{\beta\gamma}$, in $\binom{\ell_3-L_2+t}{L_1-t}$ ways;
   \item distribute the last $2t$ over the $t$ deltas $\delta_{\gamma\gamma}$, in $(2t)!/(2^{t}t!)$ ways by \eref{eq:si:count:delta}.
\end{enumerate}
The product of the three is
\begin{equation} \label{eq:si:count:tp}
   N = \frac{\ell_3!}{(L_2-t)!\,(L_1-t)!\,2^{t}\,t!},
\end{equation}
which is the count \sref{si:brace} uses to convert Eq.~(50)\maintag{} into its brace form.

For example, the weights $(\ell_1,\ell_2,\ell_3)=(3,1,2)$, for which $(L_1,L_2,L_3)=(0,2,1)$, put at $t=0$ both indices $\gamma$ on deltas $\delta_{\alpha\gamma}$, since $L_1-t=0$ leaves no $\delta_{\beta\gamma}$ and $t=0$ no $\delta_{\gamma\gamma}$.
There is one way to do that, $N=1$,
\begin{equation} \label{eq:si:avg:tp:a}
   \avg{\delta_{\alpha\gamma}^2}
   = \delta_{\alpha_1\gamma_1}\delta_{\alpha_2\gamma_2}.
\end{equation}
The weights $(1,1,2)$, for which $(L_1,L_2,L_3)=(1,1,0)$, instead put one $\gamma$ on a $\delta_{\alpha\gamma}$ and the other on a $\delta_{\beta\gamma}$.
The two choices of which index goes where give $N=2$,
\begin{equation} \label{eq:si:avg:tp:b}
   \avg{\delta_{\alpha\gamma}\delta_{\beta\gamma}}
   = \tfrac{1}{2}\left(\delta_{\alpha_1\gamma_1}\delta_{\beta_1\gamma_2}
   + \delta_{\alpha_1\gamma_2}\delta_{\beta_1\gamma_1}\right).
\end{equation}
For $(3,1,4)$, with $(L_1,L_2,L_3)=(1,3,0)$, the term $t=1$ puts two of the four $\gamma$ on deltas $\delta_{\alpha\gamma}$ and the other two on a single $\delta_{\gamma\gamma}$, in $N=6$ ways,
\begin{equation}
   \begin{aligned}
      \avg{\delta_{\alpha\gamma}^2\delta_{\gamma\gamma}}
       & = \tfrac{1}{6}\big(
      \delta_{\alpha_1\gamma_1}\delta_{\alpha_2\gamma_2}\delta_{\gamma_3\gamma_4} + \delta_{\alpha_1\gamma_1}\delta_{\alpha_2\gamma_3}\delta_{\gamma_2\gamma_4} + \delta_{\alpha_1\gamma_1}\delta_{\alpha_2\gamma_4}\delta_{\gamma_2\gamma_3} \\
       & \quad + \delta_{\alpha_1\gamma_2}\delta_{\alpha_2\gamma_3}\delta_{\gamma_1\gamma_4} + \delta_{\alpha_1\gamma_2}\delta_{\alpha_2\gamma_4}\delta_{\gamma_1\gamma_3} + \delta_{\alpha_1\gamma_3}\delta_{\alpha_2\gamma_4}\delta_{\gamma_1\gamma_2}
      \big).
   \end{aligned}
\end{equation}

The odd case of Eq.~(51)\maintag{} works the same way, with one extra step.
The $\bm\epsilon$ takes one of the $\ell_3$ indices $\gamma$, in $\ell_3$ ways.
The remaining $\ell_3-1$ are then split over $\delta_{\alpha\gamma}$, $\delta_{\beta\gamma}$ and $\delta_{\gamma\gamma}$ by the three steps above, now with $L_1+L_2=\ell_3-1$, in $(\ell_3-1)!/[(L_2-t)!\,(L_1-t)!\,2^{t}\,t!]$ ways.
Multiplying the two,
\begin{equation} \label{eq:si:count:tp:odd}
   N = \ell_3\,\frac{(\ell_3-1)!}{(L_2-t)!\,(L_1-t)!\,2^{t}\,t!}
   = \frac{\ell_3!}{(L_2-t)!\,(L_1-t)!\,2^{t}\,t!},
\end{equation}
which is \eref{eq:si:count:tp} again, so the count holds in both parities.

For example, the weights $(2,2,3)$, for which $(L_1,L_2,L_3)=(1,1,0)$, give at $t=0$ the count $N=3\times2=6$,
\begin{equation}
   \begin{aligned}
      \avg{\epsilon_{\alpha\beta\gamma} \delta_{\alpha\gamma}
      \delta_{\beta\gamma}}
       & = \tfrac{1}{6}\big(
      \epsilon_{\alpha_1\beta_1\gamma_1} \delta_{\alpha_2\gamma_2} \delta_{\beta_2\gamma_3} + \epsilon_{\alpha_1\beta_1\gamma_1} \delta_{\alpha_2\gamma_3}\delta_{\beta_2\gamma_2} + \epsilon_{\alpha_1\beta_1\gamma_2} \delta_{\alpha_2\gamma_1}\delta_{\beta_2\gamma_3} \\
       & \quad + \epsilon_{\alpha_1\beta_1\gamma_2} \delta_{\alpha_2\gamma_3}\delta_{\beta_2\gamma_1} + \epsilon_{\alpha_1\beta_1\gamma_3} \delta_{\alpha_2\gamma_1}\delta_{\beta_2\gamma_2} + \epsilon_{\alpha_1\beta_1\gamma_3}  \delta_{\alpha_2\gamma_2}\delta_{\beta_2\gamma_1}
      \big).
   \end{aligned}
\end{equation}

\section{Explicit form of the natural projector}
\label{si:Ejj}

Eq.~(7)\maintag{} defines the natural projector as a sum over $t$ of averages, weighted by the coefficients $c_t$ of Eq.~(8)\maintag{}.
This section gives those coefficients in two further forms, and the projector itself for $\ell\leq3$, using the averages counted in \sref{si:symmetrization}.

Every factor of Eq.~(8)\maintag{} cancels at $t=0$, so $c_0=1$ for every $\ell$.
Written with binomial coefficients, the general term is
\begin{equation} \label{eq:si:ct:binomial}
   c_t = (-1)^t\,\frac{(\ell!)^2}{(2\ell)!}\,\binom{\ell}{t}\binom{2\ell-2t}{\ell},
\end{equation}
the form given by \olcite{coope1970irreducible2}.
The ratio of consecutive terms in Eq.~(8)\maintag{} gives each coefficient from the one before,
\begin{equation} \label{eq:si:ct:recursion}
   c_t = -\frac{(\ell-2t+2)(\ell-2t+1)}{2t(2\ell-2t+1)}\,c_{t-1},
   \quad c_0 = 1.
\end{equation}
For example, $c_1 = -\tfrac{1}{3}$ for $\ell=2$ and $c_1 = -\tfrac{3}{5}$ for $\ell=3$.

At $\ell=0$ and $\ell=1$ the sum has the single term $t=0$, and the projector is $E_{(0|0)}=1$ and $E_{(1|1)}=\delta_{\alpha_1i_1}$.

At $\ell=2$ the two averages are those of \eref{eq:si:avg:E:ell2}, and with $c_1=-\tfrac{1}{3}$,
\begin{equation} \label{eq:si:E:ell2}
   E_{(2|2)} = \avg{\delta_{\alpha i}^2} - \tfrac{1}{3} \avg{\delta_{\alpha\alpha}\delta_{ii}}.
\end{equation}

At $\ell=3$ the two averages are those of \eref{eq:si:avg:E:ell3:t0} and \eref{eq:si:avg:E:ell3:t1}, and with $c_1=-\tfrac{3}{5}$,
\begin{equation} \label{eq:si:E:ell3}
   E_{(3|3)}
   = \avg{\delta_{\alpha i}^3}
   - \tfrac{3}{5}\avg{\delta_{\alpha i}\delta_{\alpha\alpha}\delta_{ii}}.
\end{equation}

\section{Conventions}
\label{si:conventions}

The construction determines the weight-$\ell$ mapping space of a rank-$n$ tensor class, but not a basis of it.
Where a weight occurs more than once, any $N_\ell$ independent mapping tensors spanning that space serve, in any order, and none is distinguished, so which operator carries channel $p$ is a matter of choice.
The worked examples in this work and the \code package make that choice by the conventions below.

Rank three and weight two shows how open the choice is.
The construction gives three candidates $\mathbf G^1_{(2|3)}$, $\mathbf G^2_{(2|3)}$ and $\mathbf G^3_{(2|3)}$, Eq.~(14)\maintag{}, but they are not independent, satisfying \eref{eq:si:rank3:dep}, so any two of them span the weight-2 mapping space and $N_2=2$.
Which two is open: each of the three pairs serves, and so does a pair of combinations such as $\mathbf G^1_{(2|3)}+\mathbf G^2_{(2|3)}$ and $\mathbf G^1_{(2|3)}-\mathbf G^3_{(2|3)}$.
Whichever is taken, $\mathbf T_3$ has the same weight-2 content; only its split into the two ICTs $\mathbf X^1_2$ and $\mathbf X^2_2$ differs.

\subsection{Selecting the mapping tensors}
\label{si:conventions:select}

A candidate is specified by the pairs of positions its $\bm\delta$ factors contract and, for odd $n-\ell$, the pair its $\bm\epsilon$ takes, Eq.~(3)\maintag{} and Eq.~(5)\maintag{}.
The candidates are numbered by their $\bm\delta$ pairs, then by their $\bm\epsilon$ pair, both in ascending order, so that $p=1$ contracts the earliest positions.
At rank three this gives Eq.~(4)\maintag{} and Eq.~(6)\maintag{}: $\delta_{i_1i_2}$, $\delta_{i_1i_3}$, $\delta_{i_2i_3}$ at weight one, and $\epsilon_{j\,i_1i_2}$, $\epsilon_{j\,i_1i_3}$, $\epsilon_{j\,i_2i_3}$ at weight two.
Each candidate is taken with coefficient one and with its $\bm\epsilon$ written $\epsilon_{j\,i_ui_v}$ with $u<v$, or $\epsilon_{i_wi_ui_v}$ with $w<u<v$ at weight zero; the reverse order would reverse the sign of the candidate and of every mapping built from it.

The candidates are scanned in this order, and each is kept when it is independent of those already kept; the scan stops once $N_\ell$ are kept.
The selected mappings are therefore the earliest independent candidates, and $p$ numbers them in that order.
The order thus decides which candidates are kept, not merely how they are labeled: a different order would in general keep a different set.
At rank three and weight two the scan keeps $\mathbf G^1_{(2|3)}$ and $\mathbf G^2_{(2|3)}$, and at rank four and weight three, where six candidates carry three channels, it keeps the three whose $\bm\epsilon$ takes the first position.
In \code{} the squared lengths $\lVert\tilde{\mathbf q}_k\rVert^2$ are computed from the Gram matrix $\mathbf g$, as the pivots of the $\mathbf L\mathbf D\mathbf L^{\mathsf T}$ factorization of $\mathbf A^{\mathsf T}\mathbf A=(2\ell+1)\,\mathbf g$, without evaluating $\mathbf A$.

\subsection{Symmetry-adapted mappings}
\label{si:conventions:symmetry}

Each symmetry-adapted mapping is a combination $\mathbf Q^p=\sum_q c^p_q\mathbf G^q$ of the selected mappings, Eq.~(27)\maintag{}, and Eq.~(31)\maintag{} fixes the space in which the coefficient vectors $\mathbf c^p$ lie, not a basis of it.
The basis taken is the reduced row echelon one, in the order of the $\mathbf G^q$, so that each $\mathbf Q^p$ has coefficient one on the last mapping it contains.
Unlike the order of the candidates in \sref{si:conventions:select}, which decides which mappings are selected, the order of the $\mathbf Q^p$ leaves the set of them untouched and only assigns the labels $p$; it is fixed here so that channel $p$ of a symmetry class means the same everywhere.
To number them, list for each $\mathbf Q^p$ the labels $q$ of the mappings $\mathbf G^q$ it contains, in increasing order, so that $\mathbf G^1+\mathbf G^2+\mathbf G^5$ has the list $(1,2,5)$.
The $\mathbf Q^p$ are then ordered by these lists as words are ordered in a dictionary, by the first entry, then the second, and so on: $(1,2,5)$ comes before $(1,3,4)$, the first entries agreeing and $2<3$.
No two lists coincide, so the order has no ties.
The piezoelectric tensor at weight one, \sref{si:piezo}, gives $\mathbf Q^1=\mathbf G^1+\mathbf G^2$ and $\mathbf Q^2=\mathbf G^3$, and the elastic tensor at weight two, \sref{si:elastic:operators}, numbers first the mapping that contains the first candidate, $\mathbf G^{(12)}_{(2|4)}$.

\section{Worked examples: generic tensors}
\label{si:examples:reduction}

The two lowest ranks at which the construction has anything to do are worked through here: rank two, where every weight occurs once, and rank three, where a repeated weight first forces a selection.

\subsection{Rank-2 tensors}
\label{si:example:rank2}

A generic rank-2 tensor has the reduction spectrum $\mathbf T_2 = \mathbf X_0 \oplus \mathbf X_1 \oplus \mathbf X_2$, three ICTs of dimensions $1+3+5=9$, in which every weight occurs once, so the Gram matrix is $1\times1$ at every weight and the multiplicity index is dropped.
Table~I\maintag{} lists the operators; this subsection writes out the mapping tensors and computes the Gram coefficients $g$ given there.

Each $g$ follows from Eq.~(15)\maintag{}, which at $n=2$ reads $\mathbf G_{(\ell|2)}\odot^2\mathbf G_{(\ell|2)}=g\,\mathbf E_{(\ell|\ell)}$.
The three mapping tensors are $G_{(0|2)}=\delta_{i_1i_2}$, $G_{(1|2)}=\epsilon_{\alpha_1i_1i_2}$ and $G_{(2|2)}=E_{(\alpha_1\alpha_2|i_1i_2)}$.
Contracting each with a second copy over $i_1$ and $i_2$, and writing that copy's ICT-space indices $\mu$ to keep them apart from the first's $\alpha$,
\begin{equation} \label{eq:si:g:rank2}
   \begin{aligned}
      G_{(0|2)}\odot^2 G_{(0|2)}
       & = \delta_{i_1i_2}\,\delta_{i_1i_2} = \delta_{i_1i_1} = 3\,E_{(0|0)},                                  \\
      G_{(1|2)}\odot^2 G_{(1|2)}
       & = \epsilon_{\alpha_1i_1i_2}\,\epsilon_{\mu_1i_1i_2} = 2\,\delta_{\alpha_1\mu_1} = 2\,E_{(1|1)},       \\
      G_{(2|2)}\odot^2 G_{(2|2)}
       & = E_{(\alpha_1\alpha_2|i_1i_2)}\,E_{(\mu_1\mu_2|i_1i_2)} = E_{(\alpha_1\alpha_2|\mu_1\mu_2)},
   \end{aligned}
\end{equation}
by the trace of the identity in three dimensions, the Levi--Civita identity of \eref{eq:ee:3}, and the idempotency of Eq.~(11)\maintag{}.
Hence $g=3$, $2$ and $1$ at weights $0$, $1$ and $2$, and the duals of Eq.~(16)\maintag{} carry the reciprocals $\tfrac{1}{3}$, $\tfrac{1}{2}$ and $1$, as Table~I\maintag{} lists.

\subsection{Rank-3 tensors}
\label{si:example:rank3}

A generic rank-3 tensor has the reduction spectrum
$\mathbf T_3 = \mathbf X_0 \oplus \mathbf X_1^1 \oplus \mathbf X_1^2 \oplus \mathbf X_1^3 \oplus \mathbf X_2^1 \oplus \mathbf X_2^2 \oplus \mathbf X_3$,
seven ICTs in which weight $1$ occurs three times and weight $2$ twice, of dimensions $1+3\times3+2\times5+7=27$.
This subsection gives one explicit choice of the seven, weight by weight, with the selected mappings and their Gram matrices collected in \tref{tab:n3}.

At weight $0$ the rank drop $n-\ell=3$ is odd with $\ell=0$, the last case of Appendix~C\maintag{}, so the mapping tensor is the rank-lowering tensor itself, and Appendix~B\maintag{} gives the single candidate $G_{(0|3)}=F_{3\to0}=\epsilon_{i_1i_2i_3}$.
Its Gram coefficient is $g=\epsilon_{i_1i_2i_3}\epsilon_{i_1i_2i_3}=2\,\delta_{i_1i_1}=6$ by \eref{eq:ee:3}, so
\begin{equation} \label{eq:si:rank3:X0}
   X_0 = \tfrac{1}{6}\epsilon_{i_1i_2i_3}T_{i_1i_2i_3}.
\end{equation}

At weight $1$ the rank drop $n-\ell=2$ is even, and Eq.~(C1)\maintag{} applied to the three pairings of Eq.~(3)\maintag{} gives the three candidates of Eq.~(C2)\maintag{}.
Algorithm~1\maintag{} finds all three independent, so all are kept, and they contract $\mathbf T_3$ to its three traces $T_{kk\alpha_1}$, $T_{k\alpha_1k}$ and $T_{\alpha_1kk}$.
The duals of Eq.~(16)\maintag{} combine these through the inverse of the Gram matrix of \tref{tab:n3},
\begin{equation} \label{eq:si:rank3:X1}
   \begin{pmatrix}
      X_{\alpha_1}^1 \\ X_{\alpha_1}^2 \\ X_{\alpha_1}^3
   \end{pmatrix}
   = \tfrac{1}{10}
   \begin{pmatrix}
      4  & -1 & -1 \\
      -1 & 4  & -1 \\
      -1 & -1 & 4
   \end{pmatrix}
   \begin{pmatrix}
      T_{kk\alpha_1} \\ T_{k\alpha_1k} \\ T_{\alpha_1kk}
   \end{pmatrix}.
\end{equation}

At weight $2$ the three candidates of Eq.~(14)\maintag{} are not independent.
Their expanded forms in Eq.~(C4)\maintag{} differ only in which index carries the $\bm\delta$ and in the sign of the last term, and satisfy
\begin{equation} \label{eq:si:rank3:dep}
   \mathbf G^1_{(2|3)}-\mathbf G^2_{(2|3)}+\mathbf G^3_{(2|3)} = \mathbf 0 .
\end{equation}
The selection of \sref{si:conventions:select} finds this without inspection, returning a two-dimensional span and keeping $\mathbf G^1_{(2|3)}$ and $\mathbf G^2_{(2|3)}$ as $p=1,2$, which give the candidate ICTs $\mathbf D^p_2$ of Sec.~II\,E\maintag{},
\begin{equation} \label{eq:si:rank3:D2}
   \begin{aligned}
      D^1_{\alpha_1\alpha_2}
       & = \left[\tfrac{1}{2}\left(\delta_{\alpha_1 i_3}\epsilon_{\alpha_2 i_1i_2}+\delta_{\alpha_2 i_3}\epsilon_{\alpha_1 i_1i_2}\right)
      -\tfrac{1}{3}\delta_{\alpha_1\alpha_2}\epsilon_{i_1i_2i_3}\right]T_{i_1i_2i_3}, \\
      D^2_{\alpha_1\alpha_2}
       & = \left[\tfrac{1}{2}\left(\delta_{\alpha_1 i_2}\epsilon_{\alpha_2 i_1i_3}+\delta_{\alpha_2 i_2}\epsilon_{\alpha_1 i_1i_3}\right)
      +\tfrac{1}{3}\delta_{\alpha_1\alpha_2}\epsilon_{i_1i_2i_3}\right]T_{i_1i_2i_3},
   \end{aligned}
\end{equation}
which combine as
\begin{equation} \label{eq:si:rank3:X2}
   \begin{pmatrix}
      X_{\alpha_1\alpha_2}^1 \\ X_{\alpha_1\alpha_2}^2
   \end{pmatrix}
   = \tfrac{1}{3}
   \begin{pmatrix}
      2  & -1 \\
      -1 & 2
   \end{pmatrix}
   \begin{pmatrix}
      D^1_{\alpha_1\alpha_2} \\ D^2_{\alpha_1\alpha_2}
   \end{pmatrix}.
\end{equation}
Other choices of two independent mappings in Eq.~(14)\maintag{} give equivalent weight-2 multiplicity bases; \sref{si:conventions} explains why one must nonetheless be made.

At weight $3$ the rank drop is zero, so $\mathcal D_p$ is empty in Eq.~(C1)\maintag{} and the mapping is the natural projector itself, with $g=1$ by Eq.~(11)\maintag{}.
The $t=0$ term of Eq.~(7)\maintag{} fully symmetrizes $\mathbf T_3$,
\begin{equation} \label{eq:si:rank3:U}
   U_{\alpha_1\alpha_2\alpha_3}
   = \tfrac{1}{6}\sum_{\pi\in S_3}
   \delta_{\alpha_1 i_{\pi(1)}}\delta_{\alpha_2 i_{\pi(2)}}\delta_{\alpha_3 i_{\pi(3)}}
   T_{i_1i_2i_3},
\end{equation}
and its $t=1$ term, weighted by $c_1=-\tfrac{3}{5}$ as in \eref{eq:si:E:ell3}, removes the traces of that,
\begin{equation} \label{eq:si:rank3:X3}
   X_{\alpha_1\alpha_2\alpha_3}
   = U_{\alpha_1\alpha_2\alpha_3}
   -\tfrac{1}{5}\left(
   \delta_{\alpha_1\alpha_2}U_{\alpha_3\lambda\lambda}
   +\delta_{\alpha_1\alpha_3}U_{\alpha_2\lambda\lambda}
   +\delta_{\alpha_2\alpha_3}U_{\alpha_1\lambda\lambda}
   \right).
\end{equation}

\begin{table}[tbh!]
   \caption{Reduction of a generic rank-3 tensor $\mathbf T_3$, weight by weight.
      $N_\ell$ is the multiplicity of weight $\ell$.
      At weight two the first two candidates, $\mathbf G^1_{(2|3)}$ and $\mathbf G^2_{(2|3)}$, are kept as $p=1,2$ (\sref{si:conventions:select}).}
   \label{tab:n3}
   \centering
   \small
   \begin{tabular}{ccccc}
      \hline
      Quantity & $\ell=0$ & $\ell=1$ & $\ell=2$ & $\ell=3$ \\
      \hline
      $N_\ell$
               & $1$ & $3$ & $2$ & $1$                                                          \\
      \hline
      $G^p_{(\ell|3)}$
               & $\epsilon_{i_1i_2i_3}$
               & $\begin{gathered}
                      \delta_{\alpha_1 i_3}\delta_{i_1i_2} \\
                      \delta_{\alpha_1 i_2}\delta_{i_1i_3} \\
                      \delta_{\alpha_1 i_1}\delta_{i_2i_3}
                  \end{gathered}$
               & $\begin{gathered}
                      E_{(\alpha_1\alpha_2|i_3j)}\epsilon_{ji_1i_2} \\
                      E_{(\alpha_1\alpha_2|i_2j)}\epsilon_{ji_1i_3}
                  \end{gathered}$
               & $E_{(\alpha_1\alpha_2\alpha_3|i_1i_2i_3)}$                                     \\
      \hline
      $\mathbf g$
               & $6$
               & $\begin{pmatrix}3&1&1\\1&3&1\\1&1&3\end{pmatrix}$
               & $\begin{pmatrix}2&1\\1&2\end{pmatrix}$
               & $1$                                                                            \\
      $\mathbf g^{-1}$
               & $\tfrac{1}{6}$
               & $\tfrac{1}{10}\begin{pmatrix}4&-1&-1\\-1&4&-1\\-1&-1&4\end{pmatrix}$
               & $\tfrac{1}{3}\begin{pmatrix}2&-1\\-1&2\end{pmatrix}$
               & $1$                                                                            \\
      \hline
   \end{tabular}
\end{table}

\section{Worked examples: tensors with intrinsic symmetry}
\label{si:symmetry:examples}

Four tensors with intrinsic symmetry are reduced here, in order of rank.
Each gives the surviving weights, a basis of symmetry-adapted mappings, the Gram data the duals need, and the resulting ICTs, except where the main text has already supplied them.

\subsection{Rotation-rate tensor}
\label{si:rotation}

The rotation-rate tensor $\mathbf W$ of a velocity field, written $\mathbf T$ below so that the general notation applies unchanged, is
\begin{equation} \label{eq:si:rot:def}
   T_{i_1i_2}=W_{i_1i_2}=\tfrac12\left(
   \frac{\partial v_{i_1}}{\partial x_{i_2}}
   -\frac{\partial v_{i_2}}{\partial x_{i_1}}\right),
   \quad T_{i_1i_2}=-T_{i_2i_1},
\end{equation}
so its symmetry class is $[i_1i_2]$, with three independent components, a single generator $(\Pi\mathbf T)_{i_1i_2}=T_{i_2i_1}$, and sign $\eta=-1$.
By Eq.~(31)\maintag{} a mapping tensor survives only if it carries that same sign, and the three rank-2 mappings of Table~I\maintag{} transform as
\begin{equation} \label{eq:si:rot:selection}
   \delta_{i_2i_1}=+\delta_{i_1i_2},
   \quad
   \epsilon_{\alpha_1i_2i_1}=-\epsilon_{\alpha_1i_1i_2},
   \quad
   E_{(\alpha_1\alpha_2|i_2i_1)}
   =+E_{(\alpha_1\alpha_2|i_1i_2)},
\end{equation}
so only the weight-1 mapping qualifies: $N_1^{\mathcal S}=1$ and $N_0^{\mathcal S}=N_2^{\mathcal S}=0$, and $\mathbf T$ reduces to a single weight-1 ICT, the rotation-rate entry of Table~III\maintag{}.

With a single surviving mapping, the dual of Eq.~(16)\maintag{} and the orthonormal mapping of Eq.~(21)\maintag{} are scalar rescalings of it,
\begin{equation} \label{eq:si:rot:gram}
   Q_{(\alpha_1|i_1i_2)}=\epsilon_{\alpha_1i_1i_2},
   \quad g^Q=2,
   \quad
   \widetilde Q_{(\alpha_1|i_1i_2)}
   =\tfrac12\epsilon_{\alpha_1i_1i_2},
   \quad
   \widehat Q_{(\alpha_1|i_1i_2)}
   =\tfrac{1}{\sqrt2}\epsilon_{\alpha_1i_1i_2},
\end{equation}
the Gram coefficient being that of \eref{eq:si:g:rank2}.
Writing $w_i=\epsilon_{ijk}\,\partial v_k/\partial x_j$ for the vorticity, the extraction of Eq.~(18)\maintag{} and the reconstruction of Eq.~(19)\maintag{}, both with the orthonormal mapping, give
\begin{equation} \label{eq:si:rot:extract}
   X_{\alpha_1}
   =\widehat Q_{(\alpha_1|i_1i_2)}T_{i_1i_2}
   =-\tfrac{1}{\sqrt2}w_{\alpha_1},
   \quad
   T_{i_1i_2}
   =\widehat Q_{(\alpha_1|i_1i_2)}X_{\alpha_1}
   =-\tfrac12\epsilon_{i_1i_2k}w_k.
\end{equation}

\subsection{Piezoelectric tensor}
\label{si:piezo}

The piezoelectric tensor $\mathbf d$, likewise written $\mathbf T$ below, satisfies $T_{i_1i_2i_3}=T_{i_1i_3i_2}$ and belongs to the class $i_1(i_2i_3)$, with $18$ independent components.
Everything but the symmetry is already in hand: the candidate mappings and their Gram matrices are those of a generic rank-3 tensor, \sref{si:example:rank3}, and only the symmetry step of Sec.~IV\maintag{} remains.
The sole generator $\Pi_1$ exchanges $i_2$ and $i_3$, with $\eta_1=+1$, so a mapping tensor survives only if it is symmetric under that exchange.
Testing the candidates weight by weight, as below, leaves
\begin{equation} \label{eq:si:piezo:spectrum}
   \mathbf T=2\,\mathbf X_1\oplus\mathbf X_2\oplus\mathbf X_3,
   \quad 18=2\times3+5+7.
\end{equation}

At weight zero \tref{tab:n3} gives the one candidate $\mathbf G_{(0|3)}=\epsilon_{i_1i_2i_3}$, whose sign the generator reverses, so Eq.~(29)\maintag{} gives the $1\times1$ matrix $\mathbf M^1=(-1)$.
The coefficient condition of Eq.~(31)\maintag{} then reads $(-1-\eta_1)c_1=-2c_1=0$, leaving only $c_1=0$, so $N_0^{\mathcal S}=0$ and weight zero disappears.

At weight one the generator leaves the third of the candidates listed in \tref{tab:n3}, $\mathbf G^3_{(1|3)}=\delta_{\alpha_1i_1}\delta_{i_2i_3}$, unchanged and exchanges the other two, so Eq.~(29)\maintag{} gives
\begin{equation} \label{eq:app:piezo:M}
   \mathbf M^1
   =\begin{pmatrix}
      0 & 1 & 0 \\
      1 & 0 & 0 \\
      0 & 0 & 1
   \end{pmatrix}.
\end{equation}
The coefficient condition $(\mathbf M^1-\mathbf I)\mathbf c=\mathbf 0$ of Eq.~(31)\maintag{} reduces to the single relation $c_1=c_2$, with $c_3$ free.
Choosing each free variable in turn gives the sparse solution basis $\mathbf c^1=(1,1,0)^{\mathsf T}$ and $\mathbf c^2=(0,0,1)^{\mathsf T}$, so Eq.~(27)\maintag{} yields
\begin{equation} \label{eq:app:piezo:Q}
   \mathbf Q^1_{(1|3)}=\mathbf G^1_{(1|3)}+\mathbf G^2_{(1|3)},
   \quad
   \mathbf Q^2_{(1|3)}=\mathbf G^3_{(1|3)}.
\end{equation}
Both are symmetric under $i_2\leftrightarrow i_3$ and together span the compatible weight-1 space; any other two independent solutions of the same condition would do as well, and these are the ones the conventions of \sref{si:conventions:symmetry} select.

Since $\mathbf Q^p_{(1|3)}=\sum_q c^p_q\mathbf G^q_{(1|3)}$, Eq.~(15)\maintag{} gives $g^Q_{pq}=\sum_{rs}c^p_rc^q_sg_{rs}$ from the weight-1 Gram matrix $\mathbf g$ of \tref{tab:n3}.
The result is not diagonal, so the two mappings mix when the dual basis is formed,
\begin{equation} \label{eq:si:piezo:gram}
   \mathbf g^{Q}=\begin{pmatrix}8&2\\2&3\end{pmatrix},
   \quad
   (\mathbf g^{Q})^{-1}
   =\begin{pmatrix}\tfrac3{20}&-\tfrac1{10}\\-\tfrac1{10}&\tfrac25\end{pmatrix}.
\end{equation}
With $A_{\alpha_1}=T_{\alpha_1kk}$ and $B_{\alpha_1}=T_{kk\alpha_1}$, the two mappings extract $2\mathbf B$ and $\mathbf A$, which the duals of Eq.~(16)\maintag{} combine into
\begin{equation} \label{eq:si:piezo:X1}
   \mathbf X_1^1=\tfrac3{10}\mathbf B-\tfrac1{10}\mathbf A,
   \quad
   \mathbf X_1^2=\tfrac25\mathbf A-\tfrac15\mathbf B.
\end{equation}

At weight two the selection has already reduced the three candidates of Eq.~(14)\maintag{} to the independent pair $\mathbf G^1_{(2|3)}$ and $\mathbf G^2_{(2|3)}$ in \sref{si:example:rank3}, kept there as $p=1,2$.
The generator carries each of the two into the other, so Eq.~(29)\maintag{} gives
\begin{equation} \label{eq:si:piezo:M2}
   \mathbf M^1
   =\begin{pmatrix}
      0 & 1 \\
      1 & 0
   \end{pmatrix},
\end{equation}
and $(\mathbf M^1-\mathbf I)\mathbf c=\mathbf 0$ again reduces to $c_1=c_2$, this time with no free variable left over.
The single solution $\mathbf c^1=(1,1)^{\mathsf T}$ gives $N_2^{\mathcal S}=1$ and, through Eq.~(27)\maintag{},
\begin{equation} \label{eq:si:piezo:X2}
   \mathbf Q^1_{(2|3)}
   =\mathbf G^1_{(2|3)}+\mathbf G^2_{(2|3)},
   \quad
   g^{Q}=6,
   \quad
   \widetilde{\mathbf Q}^1_{(2|3)}=\tfrac16\mathbf Q^1_{(2|3)},
\end{equation}
with $g^{Q}=g_{11}+2g_{12}+g_{22}=2+2+2$ from the weight-2 Gram matrix of \tref{tab:n3} and the dual from Eq.~(16)\maintag{}, and $\mathbf X_2=\widetilde{\mathbf Q}^1_{(2|3)}\odot^3\mathbf T$.

At weight three \tref{tab:n3} gives the one candidate $\mathbf G^1_{(3|3)}=\mathbf E_{(3|3)}$, with $g=1$, and it is symmetric in all its Cartesian indices, so $\mathbf M^1=(+1)$ and the condition reads $0\,c_1=0$, leaving $c_1$ free and $N_3^{\mathcal S}=1$,
\begin{equation} \label{eq:si:piezo:X3}
   \mathbf Q^1_{(3|3)}=\mathbf E_{(3|3)},
   \quad g^{Q}=1,
   \quad \mathbf X_3=\mathbf E_{(3|3)}\odot^3\mathbf T.
\end{equation}

\subsection{Second-order elastic tensor}
\label{si:elastic:operators}

The elastic tensor belongs to the class $((i_1i_2)(i_3i_4))$ and has the spectrum $\mathbf C=2\,\mathbf X_0\oplus2\,\mathbf X_2\oplus\mathbf X_4$ of Eq.~(33)\maintag{}, with $21=2\times1+2\times5+9$ components.

Its two generators both have $\eta_a=+1$.
The route is that of \sref{si:piezo}; the mixing matrices are $6\times6$ at weight two and are not reproduced, only the solutions of Eq.~(31)\maintag{} they lead to.

\subsubsection{Weight 0}
The two symmetry-adapted mappings, their Gram matrix and its inverse, and the two extracted scalars are given in Sec.~V\,A\maintag{}.

\subsubsection{Weight 2}
Write $\mathbf G^{(ab)}_{(2|4)}$ for the candidate of Eq.~(C1)\maintag{} that traces positions $a$ and $b$, the label standing in for the multiplicity index $p$,
\begin{equation}
   G^{(ab)}_{(2|4)} = E_{(\alpha_1\alpha_2|i_ci_d)}\,\delta_{i_ai_b},
\end{equation}
where $c,d$ are the untraced positions.
There are $N_2^{\mathrm c}=6$ such candidates, by Eq.~(B2)\maintag{}, and Eq.~(31)\maintag{} has the two independent solutions
\begin{equation}
   \mathbf Q^1_{(2|4)} = \mathbf G^{(12)}_{(2|4)}+\mathbf G^{(34)}_{(2|4)},
   \quad
   \mathbf Q^2_{(2|4)} = \mathbf G^{(13)}_{(2|4)}+\mathbf G^{(14)}_{(2|4)}+\mathbf G^{(23)}_{(2|4)}+\mathbf G^{(24)}_{(2|4)} .
\end{equation}
Their Gram matrix of Eq.~(15)\maintag{} and its inverse are
\begin{equation}
   \mathbf g^{Q} =
   \begin{pmatrix} 6 & 8\\ 8 & 20\end{pmatrix},
   \quad
   (\mathbf g^{Q})^{-1} =
   \begin{pmatrix} \tfrac{5}{14} & -\tfrac{1}{7}\\[2pt] -\tfrac{1}{7} & \tfrac{3}{28}\end{pmatrix}.
\end{equation}
Contracting with a general $\mathbf C$ collapses each of the two to a single second-order contraction,
\begin{equation}
   \mathbf Q^1_{(2|4)}\odot^4\mathbf C = 2\,\mathbf E_{(2|2)}\odot^2\mathbf A,
   \quad
   \mathbf Q^2_{(2|4)}\odot^4\mathbf C = 4\,\mathbf E_{(2|2)}\odot^2\mathbf B,
   \quad
   A_{kl}=C_{iikl},
   \quad
   B_{kl}=C_{ikil},
\end{equation}
the projector removing the trace as in Eq.~(12)\maintag{}.
The duals of Eq.~(16)\maintag{} then give the two weight-2 ICTs,
\begin{equation}
   \mathbf X_2^1 = \mathbf E_{(2|2)}\odot^2\left(\tfrac{5}{7}\mathbf A-\tfrac{4}{7}\mathbf B\right),
   \quad
   \mathbf X_2^2 = \mathbf E_{(2|2)}\odot^2\left(-\tfrac{2}{7}\mathbf A+\tfrac{3}{7}\mathbf B\right).
\end{equation}

\subsubsection{Weight 4}
The single mapping is
\begin{equation}
   \mathbf Q^1_{(4|4)}=\mathbf E_{(4|4)},
   \quad g^{Q}=1,
   \quad
   \mathbf X_4=\mathbf E_{(4|4)}\odot^4\mathbf C.
\end{equation}
The top-weight mapping is the only candidate and already satisfies the condition, and $g^{Q}=1$ by Eq.~(11)\maintag{} makes it self-dual.

\subsection{Third-order elastic tensor}
\label{si:toec:operators}

The third-order elastic tensor $C_{i_1\dots i_6}$ belongs to the class $((i_1i_2)(i_3i_4)(i_5i_6))$, with symmetry within each strain pair and under permutation of the three pairs.
A minimal generating set has $N_{\Pi}=3$: the exchange within the first pair and the exchanges of the first pair with the second and of the second pair with the third.
All have $\eta_a=+1$ and generate a group $\Gamma$ of order $48$.
At this rank the algebra has previously become prohibitive: symmetry conditions~\cite{norris1991symmetry} and explicit matrix representations~\cite{auffray2013matrix} are available for sixth-order constitutive tensors, and explicit harmonic decompositions have been obtained in two dimensions~\cite{auffray2021explicit}, but in three dimensions they have required a derivation tailored to each symmetry class.
What follows costs nothing beyond evaluating the same operators on this class.

Positions are labeled $a,b,c,\dots\in\{1,2,\dots,6\}$, so that $i_a$ is the Cartesian index in position $a$.
There are too many candidates here to carry a running multiplicity index, so each is named instead by the label that specifies it: $(ab)(cd)$ records the positions assigned to the $\bm\delta$ factors and $(uv)_{\epsilon}$ the pair assigned to $\bm\epsilon$, so that $\mathbf G^{(12)(34)}_{(2|6)}=E_{(\alpha_1\alpha_2|i_5i_6)}\delta_{i_1i_2}\delta_{i_3i_4}$, for instance.
The label plays the part of the index $p$ in $\mathbf G^p_{(\ell|n)}$ and nothing else changes.

As in \sref{si:elastic:operators}, only the solutions of Eq.~(31)\maintag{} are reported.
Each is a sum of candidates over a set of labels, written $\mathcal L^\ell_q$ for the $q$th solution at weight $\ell$.
A candidate carrying $\bm\epsilon$ may enter such a sum with either sign, since reversing its two $\bm\epsilon$ slots reverses the candidate.
Each weight below gives its candidates, its symmetry-adapted mappings $\mathbf Q^q$, and their Gram matrix $\mathbf g^{Q}$; the duals are not displayed, since each follows from the $\mathbf g^{Q}$ reported there through Eq.~(16)\maintag{}, and the ICTs are then
\begin{equation} \label{eq:si:toec:X}
   \mathbf X_\ell^p
   =\widetilde{\mathbf Q}^p_{(\ell|6)}\odot^6\mathbf C,
   \quad
   \ell=0,2,3,4,6.
\end{equation}

The candidate count $N_\ell^{\mathrm c}$ of Eq.~(B2)\maintag{} and Eq.~(B4)\maintag{}, the multiplicity $N_\ell$ that Algorithm~1\maintag{} leaves, and the symmetry-restricted multiplicity $N_\ell^{\mathcal S}$ solving Eq.~(31)\maintag{} are
\begin{equation} \label{eq:si:toec:multiplicities}
   \begin{array}{c|rrrrrrr}
      \ell                & 0  & 1  & 2  & 3  & 4  & 5  & 6 \\ \hline
      N_\ell^{\mathrm c}              & 15 & 45 & 45 & 90 & 15 & 15 & 1 \\
      N_\ell              & 15 & 36 & 40 & 29 & 15 & 5  & 1 \\
      N_\ell^{\mathcal S} & 3  & 0  & 3  & 1  & 2  & 0  & 1
   \end{array}.
\end{equation}
The last row recovers the component count,
\begin{equation}
   \sum_{\ell=0}^{6}N_\ell^{\mathcal S}(2\ell+1)
   =3\times1+3\times5+1\times7+2\times9+1\times13=56.
\end{equation}

\subsubsection{Weight 0}
The candidate mapping tensors are
\begin{equation}
   G^{(ab)(cd)(ef)}_{(0|6)}
   =\delta_{i_ai_b}\delta_{i_ci_d}\delta_{i_ei_f}.
\end{equation}
The three solutions sum them over the label sets
\begin{equation}
   \begin{aligned}
      \mathcal L^0_1
       & = \{(12)(34)(56)\},                                          \\
      \mathcal L^0_2
       & = \{(12)(35)(46),(12)(36)(45),(34)(15)(26),(34)(16)(25),     \\
       & \quad (56)(13)(24),(56)(14)(23)\},                           \\
      \mathcal L^0_3
       & = \{(13)(25)(46),(13)(26)(45),(14)(25)(36),(14)(26)(35),     \\
       & \quad (15)(23)(46),(15)(24)(36),(16)(23)(45),(16)(24)(35)\}.
   \end{aligned}
\end{equation}
These have $1$, $6$, and $8$ labels.
With
\begin{equation}
   \mathbf Q^q_{(0|6)}
   =\sum_{p\in\mathcal L^0_q}\mathbf G^p_{(0|6)},
\end{equation}
the Gram matrix and its inverse, ordered by $q=1,2,3$, are
\begin{equation}
   \mathbf g^{Q}
   =\begin{pmatrix}
      27 & 54  & 24  \\
      54 & 288 & 288 \\
      24 & 288 & 528
   \end{pmatrix},
   \quad
   (\mathbf g^{Q})^{-1}
   =\begin{pmatrix}
      \tfrac8{105} & -\tfrac1{42}     & \tfrac1{105}  \\
      -\tfrac1{42} & \tfrac{19}{1260} & -\tfrac1{140} \\
      \tfrac1{105} & -\tfrac1{140}    & \tfrac3{560}
   \end{pmatrix}.
\end{equation}

\subsubsection{Weight 1}
The character count of \eref{eq:symmetry:character:count}, averaging $\operatorname{tr}\mathbf M(\pi)$ over the $48$ elements of $\Gamma$, gives $N_1^{\mathcal S}=0$.
No combination of the $36$ independent candidates is invariant, so Eq.~(31)\maintag{} has only the zero solution and the weight is extinct.

\subsubsection{Weight 2}
The candidate mapping tensors are
\begin{equation}
   G^{(ab)(cd)}_{(2|6)}
   =E_{(\alpha_1\alpha_2|i_ei_f)}
   \delta_{i_ai_b}\delta_{i_ci_d},
\end{equation}
where $e$ and $f$ are the unpaired positions.
Four label sets $\mathcal L^2_q$ arise, containing $(12)(34)$, $(12)(35)$, $(13)(24)$, and $(13)(25)$ and having $3$, $12$, $6$, and $24$ labels respectively.
Let
\begin{equation}
   \mathbf R^q_{(2|6)}
   =\sum_{p\in\mathcal L^2_q}\mathbf G^p_{(2|6)},
   \quad q=1,2,3,4.
\end{equation}
These four are not independent, satisfying
\begin{equation}
   4\mathbf R^1_{(2|6)}-2\mathbf R^2_{(2|6)}
   -2\mathbf R^3_{(2|6)}+\mathbf R^4_{(2|6)}=\mathbf 0,
\end{equation}
so they span a three-dimensional space, and the conventions of \sref{si:conventions:symmetry} choose the basis
\begin{equation}
   \mathbf Q^1_{(2|6)}=\mathbf R^1_{(2|6)},
   \quad
   \mathbf Q^2_{(2|6)}=\tfrac12\mathbf R^3_{(2|6)}-\mathbf R^1_{(2|6)},
   \quad
   \mathbf Q^3_{(2|6)}=\mathbf R^2_{(2|6)},
\end{equation}
the scale and order being set by the independent candidates each contains rather than by the sums $\mathbf R^q$.
Then
\begin{equation}
   \mathbf g^{Q}
   =\begin{pmatrix}
      27  & -18 & 72  \\
      -18 & 33  & -48 \\
      72  & -48 & 276
   \end{pmatrix},
   \quad
   (\mathbf g^{Q})^{-1}
   =\begin{pmatrix}
      \tfrac17      & \tfrac2{63} & -\tfrac2{63} \\
      \tfrac2{63}   & \tfrac1{21} & 0            \\
      -\tfrac2{63}  & 0           & \tfrac1{84}
   \end{pmatrix}.
\end{equation}

\subsubsection{Weight 3}
The candidate mapping tensors are
\begin{equation}
   G^{(uv)_{\epsilon}(ab)_{\delta}}_{(3|6)}
   =E_{(\alpha_1\alpha_2\alpha_3|i_ri_sj)}
   \epsilon_{j\,i_ui_v}\delta_{i_ai_b},
\end{equation}
where $(u,v)$ and $(a,b)$ are disjoint and $r,s$ are the unpaired positions.
The single solution is proportional to the sum of the images of the candidate labeled $(13)_{\epsilon}(25)_{\delta}$ under $\Gamma$,
\begin{equation} \label{eq:toec:Q3}
   \mathbf Q^1_{(3|6)}
   =-\tfrac14\sum_{\pi\in\Gamma}
   \Pi_{\pi}\mathbf G^{(13)_{\epsilon}(25)_{\delta}}_{(3|6)},
\end{equation}
the factor $-\tfrac14$ being what the basis of \sref{si:conventions:symmetry} gives.
After the two $\bm\epsilon$ positions in each label are written in ascending order, the sum contains $24$ candidates with coefficient $+1$ and $24$ with coefficient $-1$, and
\begin{equation}
   g^{Q}=60 .
\end{equation}
The alternating coefficients are the $\bm\epsilon$ signs noted above; they do not indicate an antisymmetric intrinsic relation, since all generators have $\eta_a=+1$.

\subsubsection{Weight 4}
The candidate mapping tensors are
\begin{equation}
   G^{(ab)}_{(4|6)}
   =E_{(\alpha_1\dots\alpha_4|i_ci_di_ei_f)}\delta_{i_ai_b}.
\end{equation}
Two label sets arise, containing $(12)$ and $(13)$ and having $3$ and $12$ labels, and
\begin{equation}
   \mathbf Q^q_{(4|6)}
   =\sum_{p\in\mathcal L^4_q}\mathbf G^p_{(4|6)},
   \quad q=1,2,
\end{equation}
with
\begin{equation}
   \mathbf g^{Q}=\begin{pmatrix}9&24\\24&108\end{pmatrix},
   \quad
   (\mathbf g^{Q})^{-1}
   =\begin{pmatrix}\tfrac3{11}&-\tfrac2{33}\\-\tfrac2{33}&\tfrac1{44}\end{pmatrix}.
\end{equation}

\subsubsection{Weight 5}
The character count of \eref{eq:symmetry:character:count} likewise gives $N_5^{\mathcal S}=0$, so this weight is extinct as well.

\subsubsection{Weight 6}
The rank drop is zero, so there is one candidate and
\begin{equation}
   \mathbf Q^1_{(6|6)}
   =\mathbf E_{(6|6)},
   \quad g^{Q}=1 .
\end{equation}

\section{Character counts}
\label{si:character:counts}

Both counts below average a character over a group: the first over the index permutations of an intrinsic symmetry class, giving how many copies of a weight survive, the second over the point group of a crystal, giving how many components a copy retains.
Neither constructs an operator, and the two are independent of each other, which is why their counts multiply in \eref{eq:count:factorization}.

\subsection{Multiplicities of a symmetry class}
\label{si:character:count}

When only $N_\ell^{\mathcal S}$ is required, it can be obtained without solving for the coefficient vectors.
For the group $\Gamma$ generated by the prescribed index permutations, the mixing matrices form a representation: applying $\Pi_a$ and then $\Pi_b$ to Eq.~(29)\maintag{} gives the composed permutation the mixing matrix $\mathbf M^b\mathbf M^a$.
Write $\mathbf M(\pi)$ for the product of generator matrices along any word for $\pi\in\Gamma$, and $\eta_\pi$ for the product of the corresponding signs $\eta_a$.
If the $\eta_\pi$ form a consistent one-dimensional character of $\Gamma$, so that every word for $\pi$ gives the same sign, the multiplicity is the standard character projection~\cite{zee2016group},
\begin{equation} \label{eq:symmetry:character:count}
   N_\ell^{\mathcal S}
   =\frac{1}{|\Gamma|}\sum_{\pi\in\Gamma}
   \eta_\pi\operatorname{tr}\mathbf M(\pi).
\end{equation}
The sum is over the full group, not only its generators, and its value must be a non-negative integer.
It is what identifies an extinct weight without any construction: weights one and five of \sref{si:toec:operators} are extinct because \eref{eq:symmetry:character:count} returns zero there.
It gives only the multiplicity, however; the coefficient equations of Eq.~(31)\maintag{} are still needed to build the mappings themselves.

\subsection{Weight-resolved content by crystal system}
\label{si:crystal:counts}

\tref{tab:elastic:crystal} records how many independent components each weight-$\ell$ ICT of $\mathbf C$ retains under the proper point group (Laue class) of the crystal, together with the number of independent elastic constants that follows.
Each $a_\ell$ is the dimension of the $G$-invariant subspace of the weight-$\ell$ representation.
Averaging the representation over the group gives the projector onto that subspace, and the dimension of a subspace is the trace of its projector, so~\cite{zee2016group}
\begin{equation} \label{eq:crystal:invariants}
   a_\ell = \frac{1}{|G|}\sum_{R \in G} \chi_\ell(\theta_R) ,
\end{equation}
where $\theta_R$ is the rotation angle of $R$, the same character projection as \eref{eq:symmetry:character:count}.
There the trace was that of a mixing matrix $\mathbf M(\pi)$, built in Sec.~IV\maintag{}; here it is the character of the weight-$\ell$ representation of $\mathrm{SO}(3)$, the trace of a rotation acting on a weight-$\ell$ ICT.
Conjugate elements share a character, so only the rotation angle enters: a rotation by $\theta$ about any axis is conjugate to one by $\theta$ about $z$, which is diagonal in the basis of $L_z$ eigenstates with entries $e^{\mathrm{i}m\theta}$ for $m=-\ell,\dots,\ell$, and the trace is a geometric sum,
\begin{equation}
   \chi_\ell(\theta) = \sum_{m=-\ell}^{\ell} e^{\mathrm{i}m\theta}
   = \frac{\sin\bigl((\ell+\tfrac12)\theta\bigr)}{\sin(\theta/2)} ,
\end{equation}
with $\chi_\ell(0) = 2\ell+1$; this character, and its values for the crystallographic point groups, are tabulated in \olcite{bradley1972mathematical}.
Only proper rotations enter, and the rows of \tref{tab:elastic:crystal} are Laue classes rather than all thirty-two point groups, because $\mathbf C$ has even rank and is therefore invariant under inversion.

The two projections count different things: the index permutations $\Gamma$ act on the space of mapping tensors and count copies of each weight, while the point group $G$ acts within each ICT and counts the components a copy retains.
They are independent, so the counts multiply,
\begin{equation} \label{eq:count:factorization}
   n = \sum_\ell N_\ell^{\mathcal S}\, a_\ell ,
\end{equation}
which is how the last column of \tref{tab:elastic:crystal} is obtained.
The resulting $n$ agree with the numbers of independent elastic constants read off the classical stiffness matrices~\cite{nye1985physical}.
None of the mapping tensors of Sec.~IV\maintag{} enter any of this; the table is a property of the crystal class alone.
Sec.~V\,B\maintag{} uses two of its entries: the cubic classes have $a_2=0$, so that $f_2$ vanishes, and the isotropic limit has $a_2=a_4=0$.

\begin{table}[tbh!]
   \caption{Weight-resolved content of the elastic tensor across crystal systems.
      For each proper point group $G$, the column $a_\ell$ gives the number of components a weight-$\ell$ ICT retains under $G$, and $n=2\,a_0+2\,a_2+a_4$ the number of independent elastic constants.
   }
   \label{tab:elastic:crystal}
   \centering
   \begin{tabular}{lccccc}
      \hline
      Crystal system & Laue class       & $a_0$ & $a_2$ & $a_4$ & $n$ \\
      \hline
      Triclinic      & $C_1$            & 1     & 5     & 9     & 21 \\
      Monoclinic     & $C_2$            & 1     & 3     & 5     & 13 \\
      Orthorhombic   & $D_2$            & 1     & 2     & 3     & 9 \\
      Trigonal       & $C_3$            & 1     & 1     & 3     & 7 \\
                     & $D_3$            & 1     & 1     & 2     & 6   \\
      Tetragonal     & $C_4$            & 1     & 1     & 3     & 7 \\
                     & $D_4$            & 1     & 1     & 2     & 6   \\
      Hexagonal      & $C_6$, $D_6$     & 1     & 1     & 1     & 5 \\
      Cubic          & $T$, $O$         & 1     & 0     & 1     & 3 \\
      Isotropic      & $\mathrm{SO}(3)$ & 1     & 0     & 0     & 2 \\
      \hline
   \end{tabular}
\end{table}

\section{Conversion between operator conventions}
\label{si:brace}

\subsection{Coupling coefficient}
\label{si:coupling:coefficient}

This subsection derives $\kt$ of Eq.~(49)\maintag{} from the projector coefficient $c_t$ of Eq.~(8)\maintag{}.
The result is $\kt=c_t\,p_t$, where $p_t$ corrects for the terms of the projector that vanish on a product of two ICTs.

In the coupling, $\mathbf E_{(\ell_3|\ell_3)}$ acts on what $\mathbf F_{n\to \ell_3}$ of Eq.~(48)\maintag{} leaves of $\mathbf P=\mathbf X_{\ell_1}\otimes\mathbf Y_{\ell_2}$.
With its Greek indices written $\gamma$, each term of the projector is $c_t\avg{\delta_{\gamma i}^{\,\ell_3-2t}\delta_{\gamma\gamma}^{t}\delta_{ii}^{t}}$.
Its $\ell_3$ Roman indices $i$ land on $L_2$ indices $\alpha$ of $\mathbf X_{\ell_1}$ and $L_1$ indices $\beta$ of $\mathbf Y_{\ell_2}$, plus the free index of $\bm\epsilon$ when $L$ is odd.

The derivation rests on two facts about averages.
Grouping equal terms leaves an average unchanged, provided each distinct term occurs equally often, as noted after Eq.~(8)\maintag{}.
Terms that are zero add nothing, so if a fraction $p_t$ of the terms survives, the average over all of them is $p_t$ times the average over the survivors.
Carrying a term of the projector over to the coupling therefore replaces $c_t$ by $c_t\,p_t$.
For the harmonic operators of Sec.~VI\maintag{} no term vanishes, since the polyadic is not traceless, so $p_t=1$ and Eq.~(45)\maintag{} keeps $c_t$.

For the coupling, a term vanishes when one of its $t$ traces $\delta_{ii}$ joins two indices $\alpha$ or two indices $\beta$, taking a trace of the traceless $\mathbf X_{\ell_1}$ or $\mathbf Y_{\ell_2}$, or, for odd $L$, joins the free index of $\bm\epsilon$ to an $\alpha$ or a $\beta$, contracting $\bm\epsilon$ twice with a symmetric tensor.
A term survives only when every trace joins one $\alpha$ to one $\beta$.
Such a trace becomes one more factor $\delta_{\alpha\beta}$, which is why the exponent of $\delta_{\alpha\beta}$ in Eq.~(50)\maintag{} is $L_3+t$ rather than $L_3$.

The fraction $p_t$ is the chance that every trace joins an $\alpha$ to a $\beta$ when the Roman indices are placed at random.
The $2t$ slots of the $t$ traces can be filled in $\ell_3!/(\ell_3-2t)!$ ways.
In a surviving filling each trace takes one $\alpha$, one $\beta$, and one of their two orders, in $L_2!/(L_2-t)!$, $L_1!/(L_1-t)!$ and $2^t$ ways, so that
\begin{equation} \label{eq:si:pt}
   p_t = \frac{L_2!}{(L_2-t)!}\,\frac{L_1!}{(L_1-t)!}\,2^t\,\frac{(\ell_3-2t)!}{\ell_3!}
\end{equation}
in either parity.
For example, at $(\ell_1,\ell_2,\ell_3)=(2,2,4)$ the four Roman indices land on two $\alpha$ and two $\beta$.
With one trace, $4$ of the $6$ possible pairs join an $\alpha$ to a $\beta$, so $p_1=2/3$; with two traces, $2$ of the $3$ ways to pair all four indices do so twice, so $p_2=2/3$.

On a surviving term, the symmetry of $\mathbf X_{\ell_1}$ and $\mathbf Y_{\ell_2}$ leaves only the split of the $\gamma$ among $\delta_{\alpha\gamma}$, $\delta_{\beta\gamma}$ and $\delta_{\gamma\gamma}$ to matter.
Permuting the $\gamma$ maps surviving placements to surviving ones and carries any split to any other, so each split occurs equally often, and the average over the survivors is that of \eref{eq:si:avg:tp}:
\begin{equation}
   c_t\avg{\delta_{\gamma i}^{\,\ell_3-2t}\delta_{\gamma\gamma}^{t}\delta_{ii}^{t}}
   \cong c_t\,p_t\avg{\delta_{\alpha\gamma}^{\,L_2-t}\delta_{\beta\gamma}^{\,L_1-t}\delta_{\gamma\gamma}^{t}}\,\delta_{\alpha\beta}^{t},
\end{equation}
with $\epsilon_{\alpha\beta\gamma}$ inside the average when $L$ is odd, and $\cong$ meaning equality when acting on $\mathbf P$.

It remains to multiply.
Writing $c_t$ at weight $\ell_3$ in the double-factorial form of Appendix~G\maintag{}, the factors $\ell_3!$, $2^t$ and $(\ell_3-2t)!$ cancel against \eref{eq:si:pt}:
\begin{equation}
   \begin{aligned}
      c_t\,p_t
       & = (-1)^t\frac{(2\ell_3-2t-1)!!}{(2\ell_3-1)!!}\,\frac{\ell_3!}{2^t\,t!\,(\ell_3-2t)!}\;p_t \\
       & = (-1)^t\frac{(2\ell_3-2t-1)!!}{(2\ell_3-1)!!}\,\frac{L_2!\,L_1!}{(L_2-t)!\,(L_1-t)!\,t!}
       = \kt ,
   \end{aligned}
\end{equation}
which is Eq.~(49)\maintag{}.
In the example, $c_t=1,\,-6/7,\,3/35$ at $t=0,1,2$ becomes $\kt=1,\,-4/7,\,2/35$.

\subsection{Brace notation}
\label{si:brace:notation}

The harmonic and coupling operators of Sec.~VI\maintag{} and Sec.~VII\maintag{} are written throughout in the permutation average $\avg{\cdot}$, the notation of \olcite{coope1970irreducible2}.
Operators of the same two kinds appear in \olcite{lehman1989angular} in a different notation, a sum over the distinct terms rather than an average over all permutations, written $\{X\}$ as in \eref{eq:si:avg:brace}.
This subsection carries out the conversion between the two notations, and the converted operators are theirs.
A brace collects only the distinct terms: in $\{\delta_{i_1i_2}U_{i_3i_4}\}$ with $\mathbf U_2$ symmetric, $i_1$ and $i_2$ are interchangeable inside the $\bm\delta$ and so are $i_3$ and $i_4$ inside $\mathbf U_2$, leaving $4!/(2\times2)=6$ terms rather than $24$.
\sref{si:symmetrization} has already counted the terms for the two products that occur here, those of the harmonic and of the coupling operators, so this subsection supplies only the algebra that the counts set off.

For the harmonic operators the average is that of \eref{eq:si:avg:uv}, over the $n!$ Greek permutations of the summand $\Delta_t$ of Appendix~F\maintag{} with the $t$ factors $\delta_{ii}$ standing outside, and \eref{eq:si:count:delta:sym} counts $N=n!/[2^t\,t!\,(n-2t)!]$ distinct terms.
That class size cancels the binomials in $c_t$ and leaves a ratio of double factorials,
\begin{equation}
   c_t\,\frac{2^t\,t!\,(n-2t)!}{n!}
   = (-1)^t\frac{n!\,(2n-2t)!\,2^t}{(2n)!\,(n-t)!}
   = (-1)^t\frac{(2n-2t-1)!!}{(2n-1)!!},
\end{equation}
using $(2n-2t)!=2^{n-t}(n-t)!\,(2n-2t-1)!!$ and $(2n)!=2^n\,n!\,(2n-1)!!$.
Applied to Eq.~(45)\maintag{} and Eq.~(46)\maintag{}, this gives the compact forms
\begin{equation} \label{eq:E:symmetric:brace}
   E_{(n|n)} \cong
   \sum_{t=0}^{\lfloor n/2 \rfloor} (-1)^t
   \frac{(2n-2t-1)!!}{(2n-1)!!}
   \{\delta_{\alpha i}^{\,n-2t}\delta_{\alpha\alpha}^{t}\}\,
   \delta_{ii}^{t}
\end{equation}
and
\begin{equation} \label{eq:H:brace}
   H_{(n|n)} =
   C \sum_{t=0}^{\lfloor n/2 \rfloor} (-1)^t
   \frac{(2n-2t-1)!!}{(2n-1)!!}
   \{\delta_{\alpha i}^{\,n-2t}\delta_{\alpha\alpha}^{t}\}\,
   \delta_{ii}^{t}.
\end{equation}
In this notation the leading coefficient is unity at every $n$, which is why the literature prefers it, at the cost of requiring the distinct terms to be counted exactly.

The same conversion applies to the coupling operators of Sec.~VII\maintag{}, where the coefficient to convert is $\kt$ of Eq.~(49)\maintag{} rather than $c_t$, $\kt$ being what the composition of Eq.~(13)\maintag{} leaves in front of the average.
The average is that of \eref{eq:si:avg:tp}, and the brace at order $t$ carries the $\ell_3!/[(L_2-t)!\,(L_1-t)!\,2^t\,t!]$ terms of \eref{eq:si:count:tp}, which \eref{eq:si:count:tp:odd} shows to be the count in the odd parity as well.
Dividing $\kt$ by that count cancels its factorials, apart from $L_2!\,L_1!/\ell_3!$, and leaves a single power of two,
\begin{equation}
   \begin{aligned}
       & \kt\,\frac{(L_2-t)!\,(L_1-t)!\,2^t\,t!}{\ell_3!}          \\
       & = (-1)^t\,2^t\,\frac{(2\ell_3-2t-1)!!}{(2\ell_3-1)!!}\,\frac{L_2!\,L_1!}{\ell_3!}
       = (-2)^t\frac{(2\ell_3-2t-1)!!}{(2\ell_3-1)!!}\,\frac{L_2!\,L_1!}{\ell_3!},
   \end{aligned}
\end{equation}
the surviving $2^t$ being the $\bm\delta$ swap redundancy that the term count removes and $\kt$ does not carry.
The factor $L_2!\,L_1!/\ell_3!$ does not depend on $t$, so it moves into the constant, which is $C\,L_2!\,L_1!/\ell_3!$ in this notation.
With this, Eq.~(50)\maintag{} and Eq.~(51)\maintag{} become
\begin{equation} \label{eq:tp:even:brace}
   K_{(\ell_3|\ell_1,\ell_2)} =
   C\,\frac{L_2!\,L_1!}{\ell_3!}
   \sum_{t=0}^{\min(L_2,L_1)}
   (-2)^t
   \frac{(2\ell_3-2t-1)!!}{(2\ell_3-1)!!}
   \{\delta_{\alpha\gamma}^{\,L_2-t}
   \delta_{\beta\gamma}^{\,L_1-t}
   \delta_{\gamma\gamma}^{\,t}\}\,
   \delta_{\alpha\beta}^{\,L_3+t}
\end{equation}
and
\begin{equation} \label{eq:tp:odd:brace}
   K_{(\ell_3|\ell_1,\ell_2)} =
   C\,\frac{L_2!\,L_1!}{\ell_3!}
   \sum_{t=0}^{\min(L_2,L_1)}
   (-2)^t
   \frac{(2\ell_3-2t-1)!!}{(2\ell_3-1)!!}
   \{\epsilon_{\alpha\beta\gamma}\,
   \delta_{\alpha\gamma}^{\,L_2-t}
   \delta_{\beta\gamma}^{\,L_1-t}
   \delta_{\gamma\gamma}^{\,t}\}\,
   \delta_{\alpha\beta}^{\,L_3+t},
\end{equation}
which are the forms given by \olcite{lehman1989angular} in their Eqs.~(54) and (55).
Here the coefficient depends only on $\ell_3$ and $t$, whereas $\kt$ of Eq.~(49)\maintag{} depends on $L_2$ and $L_1$ as well, which set how many of the traces survive.

\section{From the coupling operator to the Cartesian \texorpdfstring{$3j$}{3j} tensor}
\label{si:tp:3j}

The coupling operator of Sec.~VII\maintag{} and the Cartesian $3j$ tensor of \olcite{coope1970irreducible3} carry the same coupling of three weights, and differ only in how that coupling is presented.
The coupling operator takes two ICTs and returns a third, so it singles out the output group and is written asymmetrically; the $3j$ tensor treats the three groups alike and changes at most by a sign when the weights are permuted.
Since the content is the same, the two convert into each other by a single scalar fixed by the three weights.
This section writes the $3j$ tensor down, derives that scalar, and relates the normalization conventions in use for the two objects.

The $3j$ tensor is Eqs.~(71) and (72) of \olcite{coope1970irreducible3}.
Throughout, $L=\ell_1+\ell_2+\ell_3$ and $L_1$, $L_2$ and $L_3$ are the triangle numbers of Sec.~VII\maintag{}, so that $L_2+L_1=\ell_3$ for even $L$ and $L_2+L_1=\ell_3-1$ for odd $L$; they are the exponents fixed by Eq.~(69) there, and the tensor is
\begin{equation} \label{eq:3j:tensor}
   \bm{\mathcal T}_{(\ell_1\ell_2\ell_3)} =
   \mathcal P
   \begin{cases}
      \delta_{\alpha\beta}^{\,L_3}
      \delta_{\beta\gamma}^{\,L_1}
      \delta_{\gamma\alpha}^{\,L_2},
       & L \text{ even}, \\[2pt]
      \epsilon_{\alpha\beta\gamma}\,
      \delta_{\alpha\beta}^{\,L_3}
      \delta_{\beta\gamma}^{\,L_1}
      \delta_{\gamma\alpha}^{\,L_2},
       & L \text{ odd},
   \end{cases}
\end{equation}
where $\mathcal P$ contracts the delta product with one natural projector per group, $\mathbf E_{(\ell_1|\ell_1)}$ on the $\alpha$ indices, $\mathbf E_{(\ell_2|\ell_2)}$ on the $\beta$ and $\mathbf E_{(\ell_3|\ell_3)}$ on the $\gamma$: each projector meets the product through its Roman slots and carries the free indices on its Greek ones, as in Eq.~(12)\maintag{}, so that $\bm{\mathcal T}$ is symmetric and traceless within each of the three groups separately.
Nothing distinguishes the three groups, and $\bm{\mathcal T}$ is accordingly invariant under permuting the weights when $L$ is even and changes sign under odd permutations when $L$ is odd, which is the rule obeyed by the columns of a $3j$ symbol; in the odd case the sign is carried by $\bm\epsilon$.

In symbols, the difference is which groups are projected.
$\bm{\mathcal T}$ projects all three through $\mathcal P$, which is why the permutation rule holds and why no sum over $t$ is needed, the $\mathbf E_{(\ell_3|\ell_3)}$ inside $\mathcal P$ having already removed the traces.
$\Kop$ projects only the output group, leaves the other two to meet $\mathbf X_{\ell_1}$ and $\mathbf Y_{\ell_2}$, and reaches a traceless output through its sum over $t$ instead.
\sref{si:3j:collapse} shows that this sum collapses when all three groups meet ICTs, \sref{si:3j:constant} evaluates the scalar in both directions, and \sref{si:3j:normalization} relates the conventions, including the symmetric one of \olcite{coope1970irreducible3}.

\subsection{Collapse of the series}
\label{si:3j:collapse}

The coupling operator of Eq.~(50)\maintag{} carries a sum over $t$, and its $t\geq1$ terms exist to make the output traceless: each carries $\delta_{\gamma\gamma}^{\,t}$, which acts on the $\gamma$ group alone.
When all three groups meet ICTs those terms contribute nothing, for the same reason that $\delta_{\alpha\alpha}$ and $\delta_{\beta\beta}$ never appear at all, namely that $\mathbf X_{\ell_1}$ and $\mathbf Y_{\ell_2}$ are traceless.
Only $t=0$ survives, leaving $\{\delta_{\alpha\gamma}^{\,L_2}\delta_{\beta\gamma}^{\,L_1}\}\,\delta_{\alpha\beta}^{\,L_3}$, whose exponents are those of \eref{eq:3j:tensor} in either parity.

The brace disappears as well.
It covers the $\gamma$ group alone, $\gamma$ being the free output group while $\alpha$ and $\beta$ meet the symmetric $\mathbf X_{\ell_1}$ and $\mathbf Y_{\ell_2}$, whereas $\mathcal P$ symmetrizes all three groups.
Under $\mathcal P$, then, a single representative product stands for every assignment of indices within a group, and \eref{eq:3j:tensor} needs no brace.
Written with braces instead, the symmetrized array has
\begin{equation} \label{eq:si:3j:count}
   N = \frac{\ell_1!\,\ell_2!\,\ell_3!}{L_2!\,L_1!\,L_3!}
\end{equation}
distinct terms in either parity, all equal after projection.

\subsection{Value of the constant}
\label{si:3j:constant}

At $t=0$ the coefficient $\kt$ of Eq.~(49)\maintag{} is unity, and after projection every term of its average equals $\bm{\mathcal T}_{(\ell_1\ell_2\ell_3)}$, so the average does too.
Restoring the normalization of Eq.~(50)\maintag{} therefore gives
\begin{equation} \label{eq:tp:3j:relation}
   \Kop \cong C\,\bm{\mathcal T}_{(\ell_1\ell_2\ell_3)},
\end{equation}
and inverting the scalar gives the conversion the other way,
\begin{equation} \label{eq:3j:tp:relation}
   \bm{\mathcal T}_{(\ell_1\ell_2\ell_3)} \cong
   \frac{1}{C}\,\Kop.
\end{equation}
In both, $C$ is the constant of Eq.~(53)\maintag{} for even $L$ and of Eq.~(54)\maintag{} for odd, and the condition on $\cong$ is that all three groups meet ICTs.
The relation is a rescaling only because $\mathcal P$ appears in \eref{eq:3j:tensor}: the bare product of deltas is not proportional to $\Kop$ under any scalar, since $\Kop$ is traceless in $\gamma$ and the bare product is not.
It also carries no phase, unlike the spherical Clebsch--Gordan to $3j$ conversion, whose phase comes from the spherical metric $(-1)^{\ell-m}\delta_{m,-m'}$ where the Cartesian metric is $\bm\delta$.

\subsection{Normalization conventions}
\label{si:3j:normalization}

Three conventions appear in the literature, related by known factors.
\olcite{lehman1989angular} fix the coupling by the Legendre property of Eq.~(47)\maintag{}, the choice adopted in Sec.~VII\maintag{}, which gives the constants of Appendix~G\maintag{}.
\olcite{coope1970irreducible3} instead normalizes the symmetric object to unit norm, $\widehat{\bm{\mathcal T}} = \Omega^{-1/2}\bm{\mathcal T}$ with $\Omega=\lVert\bm{\mathcal T}\rVert^2$, Eqs.~(73) and (74) there, and gives $\Omega$ in the closed form
\begin{equation}
   \Omega = \frac{(L+1)!\,(L-2\ell_1)!\,(L-2\ell_2)!\,(L-2\ell_3)!}{(2\ell_1)!\,(2\ell_2)!\,(2\ell_3)!},
\end{equation}
doubled when $L$ is odd, and gives the coupling itself at $([\ell_3]/\Omega)^{1/2}$ with $[\ell_3]=2\ell_3+1$.
Combining these with \eref{eq:tp:3j:relation},
\begin{equation}
   \Kop \cong C\,\sqrt{\Omega}\;\widehat{\bm{\mathcal T}},
\end{equation}
which converts between the convention used here and the symmetric one.

\section{Cartesian versus spherical formalisms}
\label{si:cart:vs:spherical}

The Cartesian ICT formalism and the spherical-tensor (Clebsch--Gordan) formalism carry the \emph{same} irreducible representations of $\mathrm{SO}(3)$: at each weight $\ell$ a $(2\ell+1)$-dimensional space, spanned by the ICTs of that weight in one and by the spherical harmonics of that degree in the other.
Neither basis has greater symmetry content.
What differs is the coupling data, which this section compares; the extra label that reflections require is common to both, and is taken first.

The comparison rests on one identity, \eref{eq:detid} of \sref{si:iso}: of the two elementary isotropic tensors, contracting $\bm\delta$ with copies of $\mathbf R$ gives $\bm\delta$ for every $\mathbf R\in\mathrm O(3)$, while contracting $\bm\epsilon$ gives $\det(\mathbf R)\,\bm\epsilon$.
An irreducible representation of $\mathrm{O}(3)$ is a weight-$\ell$ representation of $\mathrm{SO}(3)$ together with a parity $\pi\in\{+1,-1\}$, the eigenvalue under spatial inversion $\mathbf x\mapsto-\mathbf x$, so each weight occurs with both parities: $(1,-1)$ is a polar vector and $(1,+1)$ an axial one.
The product $\mathrm{O}(3)=\mathrm{SO}(3)\times\{\mathbf I,-\mathbf I\}$ is direct, so $\pi$ multiplies the rotational content rather than mixing with it, and in either formalism it is a label carried alongside the coupling data rather than something that data determines.
Being a one-dimensional character of $\{\mathbf I,-\mathbf I\}\cong\mathbb{Z}_2$, and characters of an abelian group multiplying, it obeys
\begin{equation} \label{eq:si:pilaw}
   \pi_z = \pi_x\,\pi_y
\end{equation}
in either formalism and with no extra factor.
Parity is sometimes tracked instead by the polar/axial flag $\sigma$, with $\sigma=+1$ for a polar tensor and $\sigma=-1$ for an axial tensor (a pseudo-tensor), related to $\pi$ for a rank-$n$ object by $\pi=\sigma\,(-1)^{n}$, where for an ICT the rank $n$ is its weight $\ell$.
\tref{tab:parity:objects} lists the objects used here.

\begin{table}[htb!]
   \caption{Parity $\pi$ and polar/axial flag $\sigma$ of vectors and of the isotropic tensors $\bm\delta$ and $\bm\epsilon$. The components of $\bm\delta$ and $\bm\epsilon$ do not change under inversion.}
   \label{tab:parity:objects}
   \centering
   \small
   \begin{tabular}{@{}l @{\hspace{1.5em}} c @{\hspace{1.5em}} c @{\hspace{1.5em}} c@{}}
      \hline
      object & rank $n$ & $\pi$ & $\sigma$ \\
      \hline
      polar vector & 1 & $-1$ & $+1$ \\
      axial vector & 1 & $+1$ & $-1$ \\
      $\bm\delta$ & 2 & $+1$ & $+1$ \\
      $\bm\epsilon$ & 3 & $+1$ & $-1$ \\
      \hline
   \end{tabular}
\end{table}

In that flag the same law reads $\sigma_z=\sigma_x\sigma_y(-1)^{\ell_1+\ell_2+\ell_3}$, where the factor $(-1)^{\ell_1+\ell_2+\ell_3}$ is only the $\pi\leftrightarrow\sigma$ conversion and must not be attached to \eref{eq:si:pilaw}.
Every operator built from $\bm\delta$ and $\bm\epsilon$ has $\pi=+1$ and so contributes nothing of its own, which covers extraction as well: a single input gives $\pi(\mathbf X_\ell)=\pi(\mathbf T_n)$ at every weight; the arrays are unchanged by inversion, and the factor $\det(\mathbf R)$ of \eref{eq:detid} enters only through $\sigma$.

What differs between the formalisms is where the relation between input and output types is displayed, not the relation itself.
The even-$L$ operator of Eq.~(50)\maintag{} is built from $\bm\delta$ alone, whereas the odd-$L$ operator of Eq.~(51)\maintag{} contains one $\bm\epsilon$, and by \eref{eq:detid} that single $\bm\epsilon$ is what turns a transformation of the inputs into a factor $\det(\mathbf R)$ in the output.
Counting it gives $\sigma_z=\sigma_x\sigma_y(-1)^{L}$ by inspection of the operator.
The spherical formalism gives the same relation, since $(-1)^{L}$ depends only on the three weights that label the Clebsch--Gordan coefficients; there it is read from those weights and the parity labels rather than off the array.
The change of basis does not replace $\bm\epsilon$ by a numerical sign: a global sign is independent of $\mathbf R$ and cannot reproduce the $\mathbf R$-dependent $\det(\mathbf R)$ response of an $\bm\epsilon$-containing operator.

Neither formalism escapes the label once a coupled ICT is kept only through its compact components.
Parity is not recoverable from those components, since a compact axial weight-1 feature and a polar one hold identical numbers, so a label is how the next operation knows the parity that the components dropped.
In a machine learning model the point is sharper: a learned linear combination of a polar and an axial feature has no definite parity and would break equivariance, so features must be segregated into parity channels, each carrying a label, kept consistent by \eref{eq:si:pilaw}.
Only a fully $\bm\epsilon$-explicit representation, storing every axial object as the polar tensor it is dual to, removes the label, at the cost of inflated feature dimension.
The other storage asymmetry is the component count itself: a weight-$\ell$ ICT occupies $3^\ell$ Cartesian components of which $2\ell+1$ are independent, where the spherical formalism carries the minimal $2\ell+1$, real or complex.
\tref{tab:cart:vs:spherical} collects the comparison.

\begin{table}[htb!]
   \caption{Comparison of the Cartesian and spherical-tensor formalisms.}
   \label{tab:cart:vs:spherical}
   \centering
   \small
   \begin{tabular}{@{}l @{\hspace{1.5em}} p{0.32\textwidth} @{\hspace{1.5em}} p{0.32\textwidth} @{}}
      \hline
                         & Cartesian / ICT                               & Spherical / Clebsch--Gordan                \\
      \hline
      Full symmetry
                         & $\mathrm{O}(3)$                               & $\mathrm{O}(3)$                            \\
      Rotational coupling
                         & explicit $\bm\delta$/$\bm\epsilon$ tensor     & Clebsch--Gordan coefficients               \\
      Parity law
                         & $\pi_z=\pi_x\pi_y$                           & $\pi_z=\pi_x\pi_y$                       \\
      Input--output type relation
                         & visible in the $\bm\delta$/$\bm\epsilon$ structure        & supplied by the parity labels              \\
      Inputs / outputs
                         & native Cartesian tensors                      & basis change required                      \\
      Component count
                         & $3^\ell$ stored ($2\ell+1$ independent)       & minimal $2\ell+1$ (real or complex)        \\
      \hline
   \end{tabular}
\end{table}

\section{Computational cost of \texttt{natto}}
\label{si:cost}

This section reports the time and storage of the operators of \texttt{natto}: the reduction operators of a general tensor (\sref{si:cost:reduction}) and of a tensor with an intrinsic symmetry (\sref{si:cost:symmetry}), and the coupling operators (\sref{si:cost:coupling}).

\subsection{Reduction of general tensors}
\label{si:cost:reduction}

\tref{tab:cost:reduction} gives the cost of building and applying the reduction operators of a rank-$n$ tensor, over all weights $\ell=0,\dots,n$ of one rank.
The operators are the orthonormal mapping tensors $\widehat{\mathbf G}^{p}_{(\ell|n)}$ of Eq.~(21)\maintag{}, built from the candidate mapping tensors of Eq.~(13)\maintag{} once the independent ones are selected over the rationals.
They are applied to a tensor $\mathbf T_n$ with \texttt{numpy.einsum}, by extraction, which gives the ICTs of every weight and channel by Eq.~(23)\maintag{}, and by reconstruction, which returns the tensor from them by Eq.~(24)\maintag{}.
At rank six, all 141 operators are built in 1.5~s, evaluated in 0.6~s and take 25~MB, and a tensor is extracted in 1.0~ms and reconstructed in 0.66~ms.

\begin{table*}[tbh!]
   \caption{\textbf{Construction, storage and application of the reduction operators of a general tensor.} Candidates and independent are the numbers of candidate and independent mapping tensors, summed over the weights of the rank. Symbolic is the time of constructing the operators exactly, numerical the time of evaluating them into arrays, and dense and sparse are the storage of these arrays, the sparse form keeping the non-zero entries with int8 indices and float64 values. Extraction and reconstruction are the times of applying the operators to one tensor. All arrays and operations are in double precision. Each time is the median of seven timed calls after three untimed warmup calls, averaged over five independent processes, on one core of an Apple M4 Pro.}
   \label{tab:cost:reduction}
   \vspace{4pt}
   \centering
   \footnotesize
   \begin{tabular}{@{\hspace{6pt}}c@{\hspace{7pt}}c@{\hspace{7pt}}c@{\hspace{7pt}}c@{\hspace{7pt}}c@{\hspace{7pt}}c@{\hspace{7pt}}c@{\hspace{7pt}}c@{\hspace{7pt}}c@{\hspace{6pt}}}
      \hline
      rank & candidates & independent & symbolic & numerical & dense & sparse & extraction & reconstruction \\
      $n$ & & & (ms) & (ms) & (MB) & (MB) & (ms) & (ms) \\
      \hline
      1 & 1 & 1 & 0.130 & 0.00845 & 6.87e-05 & 2.86e-05 & 0.00102 & 0.00153 \\
      2 & 3 & 3 & 0.504 & 0.0426 & 0.000893 & 0.000332 & 0.00284 & 0.00396 \\
      3 & 8 & 7 & 1.77 & 0.263 & 0.0113 & 0.00414 & 0.00682 & 0.00884 \\
      4 & 22 & 19 & 10.3 & 2.56 & 0.146 & 0.0587 & 0.0212 & 0.0270 \\
      5 & 76 & 51 & 108 & 34.8 & 1.90 & 0.824 & 0.101 & 0.104 \\
      6 & 226 & 141 & 1\,480 & 573 & 24.6 & 12.0 & 1.01 & 0.657 \\
      \hline
   \end{tabular}
\end{table*}

\subsection{Reduction with an intrinsic symmetry}
\label{si:cost:symmetry}

\tref{tab:cost:symmetry} gives the same for a symmetric rank-two tensor, the piezoelectric tensor, and the second- and third-order elastic tensors, $((ij)(kl))$ at rank four and $((ij)(kl)(mn))$ at rank six.
The operators are the orthonormal symmetry-adapted mapping tensors $\mathbf Q^{p}_{(\ell|n)}$ of Eq.~(28)\maintag{}, obtained from the independent mappings of the general tensor by the exact adaptation of Procedure~2, and applied as in Eq.~(23)\maintag{}.
The third-order elastic tensor is built in 1.6~s, of which the adaptation is 0.61~s, and extracted in 0.19~ms.

\begin{table*}[tbh!]
   \caption{\textbf{Construction, storage and application of the reduction operators of tensors with an intrinsic symmetry.} The tensors are the symmetric rank-two tensor $\sigma_{ij}$, the piezoelectric tensor $d_{ijk}$, and the second- and third-order elastic tensors $C_{ijkl}$ and $C_{ijklmn}$. Independent is the number of independent mapping tensors of the general tensor of the same rank, and adapted is the number of symmetry-adapted mapping tensors, which are the operators that are applied. Symbolic includes the adaptation, and the other columns and the timing are as in \tref{tab:cost:reduction}.}
   \label{tab:cost:symmetry}
   \vspace{4pt}
   \centering
   \footnotesize
   \begin{tabular}{@{\hspace{6pt}}c@{\hspace{10pt}}c@{\hspace{10pt}}c@{\hspace{10pt}}c@{\hspace{10pt}}c@{\hspace{10pt}}c@{\hspace{10pt}}c@{\hspace{10pt}}c@{\hspace{10pt}}c@{\hspace{6pt}}}
      \hline
      tensor & rank & independent & adapted & symbolic & numerical & dense & extraction & reconstruction \\
      & $n$ & & & (ms) & (ms) & (MB) & (ms) & (ms) \\
      \hline
      $\sigma_{ij}$ & 2 & 3 & 2 & 0.474 & 0.0301 & 0.000687 & 0.00193 & 0.00282 \\
      $d_{ijk}$ & 3 & 7 & 4 & 1.68 & 0.190 & 0.00865 & 0.00416 & 0.00564 \\
      $C_{ijkl}$ & 4 & 19 & 5 & 10.2 & 1.20 & 0.0624 & 0.00600 & 0.00797 \\
      $C_{ijklmn}$ & 6 & 141 & 10 & 1\,570 & 199 & 5.27 & 0.193 & 0.109 \\
      \hline
   \end{tabular}
\end{table*}

\subsection{Coupling}
\label{si:cost:coupling}

\tref{tab:tp:build} gives the cost of building the coupling operators of Eq.~(50)\maintag{} and Eq.~(51)\maintag{} for all weight triples up to $\ell_{\max}=4$.

\begin{table*}[tbh!]
   \caption{\textbf{Construction and storage of the coupling operators.} Each row covers every operator with weights up to $\ell_{\max}$, that is, every triple of weights obeying the triangle rule, of either parity. Symbolic is the time of constructing the operators as analytical closed forms, and numerical the time of evaluating them into arrays in double precision, both summed over the operators of the row. The last three columns refer to the largest array of the row: dense stores all $3^{\ell_1+\ell_2+\ell_3}$ entries in single precision, and sparse stores only the non-zero entries, their number being given, in coordinate form with int8 indices and float32 values. Each time is the median of seven timed calls after three untimed warmup calls, averaged over five independent processes, on one core of an Apple M4 Pro.}
   \label{tab:tp:build}
   \vspace{4pt}
   \centering
   \footnotesize
   \begin{tabular}{@{\hspace{6pt}}c@{\hspace{10pt}}c@{\hspace{10pt}}c@{\hspace{10pt}}c@{\hspace{10pt}}c@{\hspace{10pt}}c@{\hspace{10pt}}c@{\hspace{6pt}}}
      \hline
      $\ell_{\max}$ & number of operators & symbolic (ms) & numerical (ms) & dense (MB) & non-zeros & sparse (MB) \\
      \hline
      0 & 1 & 0.0148 & 0.00548 & 3.81e-06 & 1 & 3.81e-06 \\
      1 & 4 & 0.0582 & 0.0285 & 0.000103 & 6 & 4.01e-05 \\
      2 & 11 & 0.206 & 0.131 & 0.00278 & 63 & 0.000601 \\
      3 & 23 & 0.655 & 0.553 & 0.0751 & 810 & 0.0100 \\
      4 & 42 & 2.16 & 2.76 & 2.03 & 7\,209 & 0.110 \\
      \hline
   \end{tabular}
\end{table*}

\begin{figure}[tbh!]
   \centering
   \includegraphics[width=0.7\columnwidth]{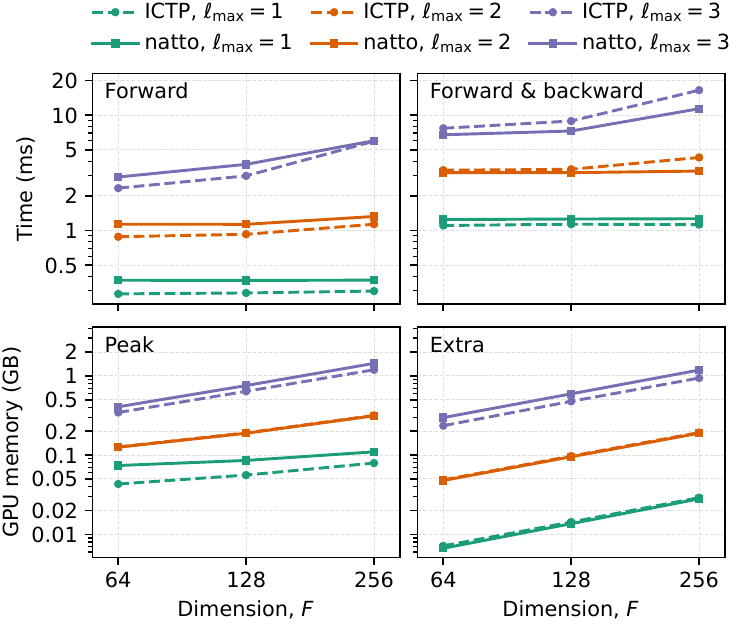}
   \caption{\textbf{Cost of applying the coupling operators.}
   Time (top) and GPU memory (bottom) for evaluating every even-parity coupling $(\ell_1,\ell_2)\to\ell_3$ with ranks up to $\ell_{\max}$, on $B=1000$ pairs of ICTs with $F$ features each, on one NVIDIA RTX~5090 in single precision.
   Forward is one evaluation, and forward \& backward includes the gradient with respect to the inputs.
   Peak is the largest GPU memory in use during the call, and extra is the part of it above the memory that holds the inputs.
   ICTP is the released implementation of Ref.~\cite{zaverkin2024higher}, as distributed, which covers the even-parity couplings up to rank three; natto applies the stored operators.
   Each run is the median of 20 timed calls after three untimed warmup calls, and each point is the mean of five independent runs.
   The times and the memory for all batch sizes ($B=100$, $10^3$ and $10^4$) and feature dimensions ($F=32$ to $256$) are in \tref{tab:tp:cost:time} and \tref{tab:tp:cost:memory}.
   }
   \label{fig:tp:cost:si}
\end{figure}

Applying all even-parity couplings $(\ell_1,\ell_2)\to\ell_3$ with $\ell_1,\ell_2,\ell_3\leq\ell_{\max}$ by Eq.~(52)\maintag{} with \texttt{torch.einsum}~\cite{paszke2019pytorch} to arrays of shape $(B,F,3^\ell)$ is compared with ICTP~\cite{zaverkin2024higher}, which implements \eref{eq:tp:even:brace}, in \fref{fig:tp:cost:si}, \tref{tab:tp:cost:time} and \tref{tab:tp:cost:memory}.
The operators take 1.3 to 1.4 times the forward time of ICTP at $B=100$ and 0.4 to 0.9 times at $B=10^4$ and $F\geq64$, and the same extra memory at $\ell_{\max}\leq2$ and 1.25 times at $\ell_{\max}=3$.

\begin{table*}[tbh!]
   \caption{\textbf{Time of applying the coupling operators, in ms.} The values for batch size $B$ and feature dimension $F$, for ICTP and natto, of which the rows of $B=1000$ are shown in \fref{fig:tp:cost:si}.}
   \label{tab:tp:cost:time}
   \vspace{4pt}
   \centering
   \footnotesize
   \setlength{\tabcolsep}{4pt}
   \begin{tabular}{@{}rr @{}p{20pt}@{} rr @{}p{8pt}@{} rr @{}p{20pt}@{} rr @{}p{8pt}@{} rr @{}p{20pt}@{} rr @{}p{8pt}@{} rr@{}}
      \hline
      \multicolumn{3}{c}{} & \multicolumn{5}{c}{$\ell_{\max}=1$} &  & \multicolumn{5}{c}{$\ell_{\max}=2$} &  & \multicolumn{5}{c}{$\ell_{\max}=3$} \\
      \cline{4-8} \cline{10-14} \cline{16-20}
      \multicolumn{3}{c}{} & \multicolumn{2}{c}{Forward} & & \multicolumn{2}{c}{\shortstack{Forward \&\\backward}} &  & \multicolumn{2}{c}{Forward} & & \multicolumn{2}{c}{\shortstack{Forward \&\\backward}} &  & \multicolumn{2}{c}{Forward} & & \multicolumn{2}{c}{\shortstack{Forward \&\\backward}} \\
      \cline{4-5} \cline{7-8} \cline{10-11} \cline{13-14} \cline{16-17} \cline{19-20}
      $B$ & $F$ & & ICTP & natto &  & ICTP & natto & & ICTP & natto &  & ICTP & natto & & ICTP & natto &  & ICTP & natto \\
      \hline
      100 & 32 & & 0.277 & 0.367 & & 1.08 & 1.21 & & 0.874 & 1.14 & & 3.27 & 3.21 & & 2.16 & 2.72 & & 7.65 & 6.69 \\
       & 64 & & 0.273 & 0.38 & & 1.07 & 1.26 & & 0.879 & 1.14 & & 3.29 & 3.2 & & 2.16 & 2.72 & & 7.66 & 6.7 \\
       & 128 & & 0.28 & 0.362 & & 1.09 & 1.21 & & 0.875 & 1.13 & & 3.28 & 3.16 & & 2.19 & 3.16 & & 7.74 & 6.66 \\
       & 256 & & 0.278 & 0.373 & & 1.1 & 1.26 & & 0.895 & 1.14 & & 3.35 & 3.23 & & 2.19 & 2.77 & & 7.7 & 6.81 \\
      \hline
      1000 & 32 & & 0.28 & 0.374 & & 1.1 & 1.26 & & 0.882 & 1.11 & & 3.31 & 3.19 & & 2.2 & 2.76 & & 7.69 & 6.74 \\
       & 64 & & 0.283 & 0.371 & & 1.11 & 1.25 & & 0.885 & 1.13 & & 3.34 & 3.18 & & 2.33 & 2.91 & & 7.71 & 6.77 \\
       & 128 & & 0.288 & 0.37 & & 1.14 & 1.26 & & 0.928 & 1.13 & & 3.4 & 3.18 & & 2.98 & 3.75 & & 8.92 & 7.3 \\
       & 256 & & 0.299 & 0.372 & & 1.13 & 1.27 & & 1.13 & 1.32 & & 4.3 & 3.28 & & 5.96 & 6.01 & & 16.5 & 11.4 \\
      \hline
      10000 & 32 & & 0.326 & 0.441 & & 1.13 & 1.39 & & 1.37 & 1.47 & & 4.91 & 3.34 & & 7.64 & 7.74 & & 21.1 & 14.1 \\
       & 64 & & 0.505 & 0.373 & & 1.49 & 1.31 & & 3.01 & 2.74 & & 9.72 & 5.27 & & 18.3 & 15.9 & & 47.6 & 28.4 \\
       & 128 & & 0.872 & 0.418 & & 2.4 & 1.41 & & 6.67 & 5.44 & & 20.8 & 10.3 & & 39 & 32.9 & & 100 & 58.3 \\
       & 256 & & 1.81 & 0.791 & & 4.89 & 1.8 & & 14.6 & 11.5 & & 44.5 & 21.3 & & 80 & 67.4 & & 205 & 119 \\
      \hline
   \end{tabular}
\end{table*}

\begin{table*}[tbh!]
   \caption{\textbf{GPU memory of applying the coupling operators, in MB.} The values for batch size $B$ and feature dimension $F$, for ICTP and natto, of which the rows of $B=1000$ are shown in \fref{fig:tp:cost:si}. Peak is the largest memory in use during the call and extra is the part above the memory holding the inputs.}
   \label{tab:tp:cost:memory}
   \vspace{4pt}
   \centering
   \footnotesize
   \setlength{\tabcolsep}{4pt}
   \begin{tabular}{@{}rr @{}p{20pt}@{} rr @{}p{8pt}@{} rr @{}p{20pt}@{} rr @{}p{8pt}@{} rr @{}p{20pt}@{} rr @{}p{8pt}@{} rr@{}}
      \hline
      \multicolumn{3}{c}{} & \multicolumn{5}{c}{$\ell_{\max}=1$} &  & \multicolumn{5}{c}{$\ell_{\max}=2$} &  & \multicolumn{5}{c}{$\ell_{\max}=3$} \\
      \cline{4-8} \cline{10-14} \cline{16-20}
      \multicolumn{3}{c}{} & \multicolumn{2}{c}{Peak} & & \multicolumn{2}{c}{Extra} &  & \multicolumn{2}{c}{Peak} & & \multicolumn{2}{c}{Extra} &  & \multicolumn{2}{c}{Peak} & & \multicolumn{2}{c}{Extra} \\
      \cline{4-5} \cline{7-8} \cline{10-11} \cline{13-14} \cline{16-17} \cline{19-20}
      $B$ & $F$ & & ICTP & natto &  & ICTP & natto & & ICTP & natto &  & ICTP & natto & & ICTP & natto &  & ICTP & natto \\
      \hline
      100 & 32 & & 32.6 & 64.6 & & 0.366 & 0.342 & & 67.2 & 67.2 & & 2.45 & 2.42 & & 78.5 & 81.7 & & 12 & 15.2 \\
       & 64 & & 33.2 & 65.2 & & 0.732 & 0.684 & & 70.5 & 70.4 & & 4.91 & 4.83 & & 92.9 & 99.3 & & 24 & 30.3 \\
       & 128 & & 34.4 & 66.3 & & 1.46 & 1.37 & & 77 & 76.9 & & 9.81 & 9.67 & & 123 & 135 & & 49.4 & 61 \\
       & 256 & & 36.9 & 68.7 & & 2.93 & 2.73 & & 90 & 89.7 & & 19.6 & 19.3 & & 181 & 206 & & 97.1 & 123 \\
      \hline
      1000 & 32 & & 38.1 & 69.9 & & 3.66 & 3.42 & & 97.1 & 96.1 & & 25.1 & 24.2 & & 210 & 241 & & 121 & 152 \\
       & 64 & & 44.2 & 75.7 & & 7.32 & 6.84 & & 130 & 129 & & 50 & 49 & & 354 & 417 & & 241 & 305 \\
       & 128 & & 57.4 & 87.7 & & 14.6 & 14 & & 196 & 193 & & 99.5 & 97.6 & & 654 & 773 & & 489 & 608 \\
       & 256 & & 81.3 & 113 & & 29.8 & 28.7 & & 325 & 321 & & 197 & 193 & & 1223 & 1475 & & 962 & 1213 \\
      \hline
      10000 & 32 & & 94 & 123 & & 37.1 & 34.2 & & 390 & 386 & & 246 & 242 & & 1511 & 1824 & & 1203 & 1515 \\
       & 64 & & 156 & 183 & & 74.2 & 68.7 & & 716 & 708 & & 492 & 484 & & 2959 & 3586 & & 2405 & 3032 \\
       & 128 & & 279 & 302 & & 147 & 138 & & 1366 & 1350 & & 983 & 967 & & 5851 & 7104 & & 4808 & 6061 \\
       & 256 & & 527 & 538 & & 296 & 275 & & 2667 & 2637 & & 1966 & 1936 & & 11635 & 14144 & & 9615 & 12123 \\
      \hline
   \end{tabular}
\end{table*}

\clearpage
%\bibliography{si}
%aipnum4-2.bst 2019-01-14 (MD) hand-edited version of apsrev4-1.bst